\documentclass[12pt]{article}

\usepackage{amsmath}
\usepackage{graphicx,psfrag,epsf}
\usepackage{enumerate}
\usepackage{natbib}
\setcitestyle{authoryear,open={(},close={)}}
\newcommand{\blind}{0}
\usepackage[export]{adjustbox}
\usepackage{titlesec}
\titlespacing*{\section}{0pt}{0pt}{0pt}
\titlespacing*{\subsection}{0pt}{0pt}{0pt}
\titlespacing*{\subsubsection}{0pt}{0pt}{0pt}

\usepackage[linkcolor=blue,urlcolor=blue,citecolor=blue, colorlinks,linktocpage,bookmarks,breaklinks,pdfstartview=FitH]{hyperref}
\usepackage{graphicx,kantlipsum,setspace}
\usepackage{caption}
\usepackage{paralist}
\usepackage{longtable}
\usepackage{multirow}
\usepackage{spverbatim}
\usepackage[section]{placeins}
\usepackage{fancyhdr}
\usepackage{mathtools}  
\usepackage{makecell}   
\usepackage{amsopn}
\usepackage{textcomp}
\usepackage{breqn} 
\usepackage[titletoc]{appendix}
\DeclareMathOperator{\sgn}{sign}

\RequirePackage[OT1]{fontenc}
\usepackage{amsthm,amsmath}
\RequirePackage[colorlinks,citecolor=blue,urlcolor=blue]{hyperref}
\usepackage{multicol}

\usepackage{verbatim} 
\usepackage{amsmath,mathtools}
\usepackage{graphicx,psfrag,epsf}
\usepackage{enumerate}

\usepackage{rotating}
\usepackage{pdflscape}
\usepackage{float}
\usepackage[export]{adjustbox}
\usepackage{moreverb}
\usepackage{color}
\usepackage{tikz}

\usepackage{graphicx}
\usepackage{caption}
\usepackage{subcaption}
\usepackage{paralist}
\usepackage{longtable}
\usepackage{multirow}
\usepackage[section]{placeins}
\usepackage{fancyhdr}
\usepackage{makecell} 
\usepackage{amsopn}
\usepackage{textcomp}
\numberwithin{equation}{section}
\theoremstyle{plain}
\newtheorem{thm}{Theorem}[section]
\newcommand{\fast}{F^{\ast}}

\newcommand{\hmi}{\hat{\mu}_i}

\newcommand{\hti}{\hat{\theta}_i}
\newcommand{\tti}{\tilde{\theta}_i}

\newcommand{\pin}{\Phi^{-1}}

\newcommand{\logfitted}{l\left( \boldsymbol y ; \boldsymbol{\hat{\mu}} \right)}
\newcommand{\logsaturated}{l\left( \boldsymbol y ; \boldsymbol{\tilde{\mu}} \right)}

\def \ddeviance {D}
\def\mb#1{\mbox{\boldmath $#1$}}

\def\rep{^{\mbox{\scriptsize rep}}}
\def\sc#1{^{\mbox{\scriptsize #1}}}

\usetikzlibrary{shapes}
\newcommand{\solidline}{\raisebox{2pt}{\tikz{\draw[-,black,solid,line width = 0.9pt](0,0) -- (5mm,0);}}}
\newcommand{\dashedline}{\raisebox{2pt}{\tikz{\draw[-,black,dashed,line width = 0.9pt](0,0) -- (5mm,0);}}}
\newcommand{\dottedline}{\raisebox{2pt}{\tikz{\draw[-,black,dotted,line width = 0.9pt](0,0) -- (5mm,0);}}}

\newcommand{\markcircle}{\raisebox{0.5pt}{\tikz{\node[draw,scale=0.4,circle](){};}}}
\newcommand{\marktriangle}{\raisebox{0pt}{\tikz{\node[draw,scale=0.3,regular polygon, regular polygon sides=3, rotate=0](){};}}}

\begin{document}

\def\spacingset#1{\renewcommand{\baselinestretch}%
{#1}\small\normalsize} \spacingset{1}

\if0\blind
{
  \title{\bf Randomized Predictive P-values: A Versatile Model Diagnostic Tool with Unified Reference Distribution}
  \author{
    Cindy Feng\thanks{The author gratefully acknowledges NSERC for financial support.}\hspace{.2cm}\\
    School of Public Health, University of Saskatchewan \\
     %and \\
   Alireza Sadeghpour \\
    Department of Mathematics and Statistics, University of Saskatchewan \\
      %  and \\
    Longhai Li\thanks{
    Correspondence should be sent to Longhai Li. The author gratefully acknowledges NSERC and CFI for financial support.}\hspace{.2cm}\\
    Department of Mathematics and Statistics, University of Saskatchewan 
 }
  \maketitle
} \fi

\if1\blind
{
  \bigskip
  \bigskip
  \bigskip
  \begin{center}\doublespacing
    {\Large \bf Randomized Predictive P-values: A Versatile Model Diagnostic Tool with Unified Reference Distribution}
\end{center}
  \medskip
} \fi

\begin{abstract}
%\vspace{-1in}
%=============================================================================
Residual analysis is a standard tool for assessing normal regression. However, for a discrete response, the traditional Pearson and deviance residuals cluster on lines and their distributions are far from normality.  Graphical and quantitative inspection of these residuals provides little information for model diagnosis. Marshall and Spiegelhalter (2007) defined cross-validatory predictive p-values which are uniformly distributed for a continuous response but not for a discrete response.  Randomized predictive p-values (RPP) are uniformly distributed for discrete responses.  Normally-transformed RPPs (NRPPs) can be used to diagnose a regression model with a discrete response in the same way as diagnosing normal regression with Pearson residuals.  The NRPPs are nearly the same as the randomized quantile residuals (RQR) proposed by Dunn and Smyth (1996) but remain little known by statisticians. This paper provides an exposition of RQR using the RPP perspective. The contributions of this paper include: (1) we give a rigorous proof and illustrative examples of the uniformity of RPPs;  (2) we conduct extensive simulation studies and a real data example to demonstrate the normality of NRPPs; (3) we show that the NRPP method is a versatile diagnostic tool for detecting many kinds of model inadequacies. 
\end{abstract}

\noindent%
\spacingset{1.45} % DON'T change the spacing!

\section{Introduction} \label{sec:intro}

Model diagnosis/checking via residual analysis is a standard practice in normal regression modeling based on the theory that Pearson residuals are normally distributed.  First,  the discrepancy nature between a model and data (e.g., non-linear effects and heavy tails) can be revealed by looking at residual plots. Second, the overall goodness-of-fit (GOF) of the model can be checked graphically and quantitatively by examining the normality of residuals using QQ plots and statistical tests. Third, residuals can identify outliers with the so-called ``empirical rules'' for normal distribution.   However, when a response is discrete,  traditional Pearson and deviance residuals cluster on lines corresponding to distinct response values, hence these residuals are far from being normally distributed. As a result, graphical and quantitative examination of these residuals provides little information for diagnosing non-normal models.    According to \citet{lin_model-checking_2002}, ``although model misspecification can seriously affect the validity and efficiency of regression analysis, model checking has not become a routine practice, partly because of the lack of suitable tools.''  Many alternative residuals have been proposed for specific problems in the literature, see \citet{lin_model-checking_2002, arbogast_model-checking_2005, leon_model_2012, yuan_goodness_fit_2012}; among others.  However,  the methods for analyzing these residuals are quite dissimilar to those for Pearson residuals in normal regression. To the best of our knowledge, statistical inferences without serious model checking remain common in today's statistical practice. 

In Bayesian statistics,  \cite{gelman1996posterior} proposed to use the posterior predictive distribution of a discrepancy measure, such as $\chi^{2}$ statistic to the posterior mean,  to check the overall GOF of a model.  A posterior predictive p-value is defined as the probability that the replicated discrepancy measure is greater than the observed discrepancy measure. This posterior predictive p-values are not uniformly distributed but are more concentrated around 0.5, because the data is used twice in both training and validating process in machine learning language. Nevertheless,  the posterior p-values are still very informative and are widely used to check the GOF of Bayesian models \citep{gelman_two_2013-2}.  However, such an overall GOF checking method cannot reveal the discrepancy nature which residual diagnosis can do.    \cite{Marshall03,marshall2007identifying} defined a cross-validatory (CV) predictive p-value for each observed $y_{i}$ as follows: 
\begin{equation}
P(Y\rep_{i}>y_{i}|y_{-i})+0.5 P(Y\rep_{i}=y_{i}|y_{-i}), \label{eqn:cvpp}
\end{equation}
where $Y\rep_{i}$ is distributed as the CV predictive distribution of $y_{i}$ given the observations except $y_i$. \cite{marshall2007identifying} shows that when $y_{i}$ is continuous the CV predictive p-values are uniformly distributed on $(0,1)$. For a discrete response, in order to obtain uniformly distributed predictive p-values, the 0.5 in \eqref{eqn:cvpp} needs to be modified to a random number on $(0,1)$, that is, the probability of $y_i$ is \textit{randomly} split into the left and right tails. This modification results in randomized predictive p-values (RPPs).  

Due to the uniformity,  the RPPs can be transformed to quantities with any desired distribution with its quantile function. For example, using normal quantile function, RPPs can be transformed to residuals with the normal distribution, which we will call normally-transformed RPP (NRPP). If we do not use cross-validation to eliminate the bias of using the data twice, the NRPPs are the same as the randomized quantile residuals(RQR) introduced by   \citet{dunn1996randomized}. Unfortunately, the  RQR method has not been widely embraced as a standard model diagnostic tool for regression models, although it has been used in a few statistical packages \citep{benjamin2003generalized, rigby2005generalized, ospina2012general}.  The lack of awareness and application of RQR is due to two reasons: (1) The definition of RQR is quite different from Pearson residuals;  (2) There is a shortage of empirical and theoretical studies for RQR in the literature; \cite{dunn1996randomized} provided neither a proof of the normality nor an investigation of its statistical properties with simulated datasets. %We have not seen further theoretical or empirical studies on RQR by others either. 

This article provides an exposition to RQR using the RPP perspective. In what follows, we will use NRPP instead of RQR because the essential technique in this method is the randomization on predictive p-values, and the fact that p-values of test statistics are uniformly distributed under the null hypothesis is well-known to statisticians.  Our contributions are summarized as follows. (1) We provide a rigorous proof of the uniformity of RPP, and illustrative examples for explaining why the randomization  is necessary in order to obtain truly uniformly distributed predictive p-values for discrete response variables. (2) We use extensive simulation studies to demonstrate that the NRPPs are normally distributed under the true model, and the overall GOF tests by applying Shapiro-Wilk normality test (as opposed to other tests) to the NRPPs are well-calibrated. (3) We show that the NRPP method is a versatile model diagnostic tool for detecting many kinds of mis-specifications due to lack of necessary complexity, such as non-linearity, zero-inflation, and over-dispersion. Identification of these modeling complexities is of great interest in contemporary application areas such as epidemiology, ecology, and bioinformatics \citep{feng2012joint, xu2015assessment, brilleman2016joint, zhang2017negative}. %Considering its unification in computing and interpretation, and capability in detecting many types of model inadequacies in various models, the NRPP is an excellent model diagnostic tool for statistical modeling in many contemporary application areas. 

This article is organized as follows. In Section \ref{sec:TraditionalResiduals}, we review the traditional residuals, followed by a discussion of the problems Pearson and deviance residuals. Then, we define the RPPs and NRPPs in Section \ref{sec:NRPP}, where we provide illustrative examples for explaining the uniformity of RPPs. In Section \ref{sec:simulation}, simulation studies in three scenarios (non-linearity, zero-inflation, and over-dispersion) are conducted to demonstrate that NRPPs have the normality under the true model and great power in detecting these model complexities. In Section \ref{sec:application}, we further demonstrate the advantage of the NRPPs with a health care utilization dataset. Concluding remarks are given in Section \ref{sec:conclusion}.

\section{Review of Traditional Residuals}\label{sec:TraditionalResiduals}
\subsection{Common Non-Normal Regression Models} \label{sec:reviewmod}\label{sec:zip}

%In this section, we review some commonly used non-normal regression models, including generalized linear models as well as zero-inflated models. 
%\subsubsection{Generalized Linear Model (GLM)}
The GLM framework  generalizes the ordinary linear regression to allow the response variable to follow a non-normal distribution, such as the Poisson and negative binomial. These non-normal distributions belong to a broad family called the \textit{exponential family} with its probability density function (PDF) or probability mass function (PMF) defined by 
\begin{equation}
f(y_i; \theta_i , \phi ) = \exp \left\{ \frac{ y_i \theta_i - b( \theta_i )}{a(\phi)} + c(y_i, \phi) \right\}
\end{equation} 
for some functions $a(\cdot)$, $b(\cdot)$ and $c(\cdot)$. The expected value  of the response variable $\mu_i=E(y_i|x_i)$ is linked to a set of covariates with a linear function  $g(\mu_i) = x_i\beta$.

%\begin{equation}
%y_i \sim 
%\begin{cases}
%\delta_{0} & \quad \text{with probability } \, p_i \\
%\text{Poisson}(\lambda_i) & \quad \text{with probability } 1-p_i 
%\end{cases}
%\end{equation}
%where $\delta_{0}$ is a point mass at $0$, $\lambda_i$ is the mean of the Poisson for counts in at-risk group and

%\subsubsection{Zero-Inflated Model}
In many contemporary application areas, count data often contains excessive zeros that may not be captured by a conventional Poisson or negative binomial (NB) model; these data are commonly known as \textit{zero-inflated} data. One popular approach to model such data is a mixture model of degenerate zeros from the non-risk group (i.e., structural zeros) and responses with random zeros or positive values from the at-risk group \citep{lambert1992zero, Yu2013}.  The zero-inflated Poisson (ZIP) model is denoted by $\mathit{ZIP}(\lambda_i,p_i)$, with two components  $\lambda_i$ and $p_i$:  $p_i$ is the probability that the $i$th observation belongs to the non-risk group, and $\lambda_{i}$ is the mean of counts from at-risk group. The $p_{i}$ and $\lambda_{i}$ are typically linked to the covariates by
%\begin{align}
$\text{logit}(p_i) =z_i \gamma \, \, \hbox{and} \, \, 
\log (\lambda_i) = x_i \beta,
$
where $z_i$ and $x_i$ are vectors of covariates.  The PMF of ZIP, $p\sc{zip}(y_i)$, is written as $p_{i}\sc{zip}(y_i) = p_i +(1-p_i)e^{-\lambda_i}$ for $y_{i}=0$ and $p_{i}\sc{zip}(y_i) = (1-p_i)\frac{e^{-\lambda_i}\lambda_i^{y_{i}}}{y_{i}!}, \, \, \mbox{for } y_{i}>0.$
With $F\sc{pois}(y; \lambda_i)$ denoting the CDF of a Poisson distribution with mean parameter $\lambda_{i}$, the CDF of ZIP can be written as
%\begin{equation}
$
F_{i}\sc{zip}(y_i)=p_i+(1-p_i)F\sc{pois}(y_{i}; \lambda_i), \mbox{for } y_{i} \geq 0; =0 \mbox{ otherwise}.
$
%\end{equation}
The zero-inflated negative binomial (ZINB) model is similarly defined with Poisson distribution replaced by NB distribution.

\iffalse
The mean and variance of a ZIP distribution can be derived as 
\begin{align}
E(y_i)&= \mu_i = (1-p_i)\lambda_i \label{E}\\
V(y_i)&= (1-p_i)\lambda_i \left(1+p_i \lambda_i \right).
\label{V}
\end{align}
According to (\ref{E}) and (\ref{V}), since $V(y_i) > E(y_i)$, the ZIP distribution is another form of an overdispersed Poisson distribution.
\fi

\subsection{Pearson an Deviance Residuals}\label{sec:pearsonresid}
\textit{Pearson residual} is defined as the raw residual scaled by the standard deviation of the response variable,  denoted as $r_i = \frac{y_i-\hmi}{\sqrt{\widehat{V}(y_i)}}$
where $\hmi$ is the fitted value for $y_i$ and $\widehat{V}(y_i)$ is the estimated variance for $y_i$. Calculating the Pearson residuals for non-normal models is straightforward once we can compute the mean $\mu_{i}$ and $V(y_{i})$. 

%Table \ref{tab:pearson} contains the derivation of the Pearson residuals for some commonly used non-normal regression models. 
%\begin{table}[h]
%\centering
%\caption{Pearson residuals for commonly used non-normal regression models.\label{tab:pearson}}
%\setcellgapes{5pt}
%\makegapedcells
%\begin{tabular}{|c|c|}
%\hline
%Model & Pearson Residuals \\
%\hline
%Poisson & $r_i = \frac{y_i-\hmi}{\sqrt{\hmi}}$ \\
%\hline
%Negative Binomial & $r_i = \frac{y_i-\hmi}{\sqrt{\hmi+\hmi^2/k}}$ \\
%%\hline
%%Gamma & $r_i = \frac{y_i-\hmi}{\hmi/ \sqrt{k}} $ \\
%\hline
%ZIP & $ r_i = \frac{y_i-\hzipa}{\sqrt{\hzipa \left[ 1+\hzipb \right]}}$ \\
%\hline
%\end{tabular}
%\vspace*{-15pt}
%\end{table}

%\subsection{Deviance Residuals}
%\label{sec:devianceresid}
Let $l(\boldsymbol y;\boldsymbol\mu)$ be the log-likelihood function based on a fitted model. A \textit{saturated model} is the model which there are as many estimated parameters as data points \citep{agresti2015foundations}. By definition, this will lead to a perfect fit and has the highest log-likelihood among all models. For example, for Poisson and negative binomial regression models, $l(\boldsymbol y, \boldsymbol y)$ yields the highest achievable log-likelihood. Let $\logsaturated$ and $\logfitted$ denote the log-likelihoods for the saturated and the fitted model, respectively. The deviance is then defined by  
%\begin{equation}
$
\ddeviance=2 \left\{ \logsaturated - \logfitted \right\}.
$
%\end{equation}
\noindent The \textit{deviance residual} is defined as the signed square root of the $i$th term in $\ddeviance$, possibly rescaled by a factor that is free of $i$. For the exponential family, the deviance residual is 
$%\begin{equation}\label{deviance}
d_i=\sgn(y_i-\hmi)\sqrt{2 \left[y_i ( \tti-\hti )-b( \tti)+b( \hti)\right] },
$
where $\tti$ and $\hti$ denote the parameters in the saturated and the fitted model. Finding a saturated model for the deviance residuals may be ambiguous when the model is not in the exponential family. For the ZIP model, \citet{lee2001analysis} showed that $\text{Poisson}(y_i)$ is a saturated model for ZIP; hence, the deviance residual for the ZIP model is defined as the signed square root of the $i$th term in the log likelihood difference between the ZIP and the $\text{Poisson}(y_i)$ model; see \citet{lee2001analysis}.

\subsection{Problems with Pearson and Deviance Residuals}
\label{sec:problemstradresiduals}
For normal regression, the Pearson and deviance residuals are identical and have an exact normal distribution under the true model. However, their distributions are often skewed and non-normally distributed for non-normal regression models. The residuals cluster on separated lines according to distinct response values, imposing great challenges for visual inspection. Therefore, Pearson and deviance residuals are difficult to use for graphically diagnosing non-normal regression models. Quantitative assessment of the overall GOF with Pearson and deviance residuals are also challenging. The {\it Pearson $\chi^{2}$ statistic} is written as
$%\begin{equation}
X^2=\sum_{i=1}^n r_{i}^{2}.$  The asymptotic distribution of $X^2$ and $\ddeviance$ under the true model is often assumed to be $\chi^2_{n-p}$, where $n$ is the sample size and $p$ is the number of parameters. However, the use of this asymptotic distribution for both $X^{2}$ and $\ddeviance$ lacks theoretical underpinning. To justify a $\chi^{2}$ distribution as the asymptotic distribution for $X^{2}$, the number of squares must be fixed  as $n\to\infty$; this scenario obviously does not occur in $X^{2}$ because the number of squares approaches infinity as $n\to\infty$. The general theory (Wilk's theorem) for the likelihood ratio test (LRT) is often used to justify the $\chi^{2}_{n-p}$ as the asymptotic distribution for the deviance $\ddeviance$. However, this argument is flawed \citep{wood2006generalized}.  Wilk's theorem  assumes that the numbers of parameters in two nested models being fixed as $n\to\infty$. However, the number of parameters in the saturated model increases linearly with $n$. 
%To remedy this deficiency within the context of logistic regression, \cite{Hosmer1980} proposed a method by grouping the predicted probabilities into a small number of intervals, resulting in a fixed number of cells and reasonable expected number of observations in each interval. However, the value of the test statistic may vary based on the number of chosen groups \citep{Hosmer1997, Pigeon1999, lin_model-checking_2002, arbogast_model-checking_2005}. %Another evidence to illustrate this flaw is that the log LRT statistic for normal regression is the $D$ in \eqref{lrtD} not the $D^{*}$ in \eqref{scaledD}.  

%%%%%%%%%%%%%%%%%%%%%%%%%%%%%%%%%%%%%%%%%%%%%%%%%%%%%
\section{Randomized Predictive P-value}\label{sec:NRPP}
%%%%%%%%%%%%%%%%%%%%%%%%%%%%%%%%%%%%%%%%%%%%%%%%%%%%
\subsection{Definition of Randomized Predictive P-values}
%When a response variable is continuous, a residual can be defined as a standard normal quantile by inverting the predictive p-values (i.e., CDF) of each response observation. These residuals are known as quantile residuals. A normal probability plot of the quantile residuals can validate the assumption of the true model. When a response is discrete, the corresponding tail probabilities are discrete; therefore, we draw a random ``probability'' between two consecutive tail probabilities to find a standard normal quantile. This random draw for the quantile residual is the Randomized Predictive P-values (NRPPs) introduced by \citep{dunn1996randomized}. 

We will now define the randomized predictive p-values in technical terms. Let $F_i(y_{i})$ be the cumulative distribution function (CDF) for a response variable $y_i$ given a set of covariates $x_{i}$ in an assumed regression model, and  $p_{i}(y_{i})$ be the corresponding PMF. The randomized predictive p-value (RPP) for an observed $y_{i}$ is defined as 
\begin{eqnarray}
\fast (y_{i}, u_i) &=& P(Y_i\rep < y_i|x_i) + u_i\cdot P(Y_i\rep = y_i|x_i)
= F_{i}(y_{i}-)+u_{i} \cdot \, p_{i}(y_{i}) \label{pvi} 
\end{eqnarray}
where $Y_i\rep$ represents a random variable with the same distribution as observed $y_i$ given $x_{i}$, $F_{i}(y_{i}-)$ is the lower limit of $F_{i}$ at $y_{i}$ (i.e., $\sup_{y < y_{i}} F_{i}(y)$) and $u_{i}$ is a random number from a uniform distribution on $(0,1)$.  A less intuitive expression given by  \cite{dunn1996randomized} for the RPP is that $\fast (y_{i}, u_i)$ is a uniform random number between $a_{i}= F_{i}(y_i-)$ and $b_{i} = F_{i}(y_{i})$.  We can easily see that they are the same. 

The RPPs in \eqref{pvi} are uniformly distributed on $(0,1)$ under the true model, which we will prove in Section \ref{sec:proof}. Therefore, we can transform RPPs to quantities with any desired distribution using its quantile function, and assess the transformed RPPs by comparing to this distribution. The normal distribution is  well-understood by statisticians with the so-called ``empirical rules''. To obtain normally distributed residuals, we can transform RPPs with normal quantile function, resulting in normally-transformed RPP (NRPP):
\begin{equation}\label{quantile}
q_{i} = \pin(\fast (y_i, u_i))%= q(y_i;\hmi,{\phi_i},u_i)
\end{equation}
where $\Phi^{-1}$ is the quantile function of the standard normal distribution.

When $F_{i}$ is continuous at $y_{i}$, the $p_{i}(y_{i}) = 0$ (note that, $p_{i}(y_{i})$ is the probability \textit{mass} function); that is, there is no actual ``randomness'' in $\fast (y_{i}, u_i)$. The formula given in \eqref{pvi} encompasses this situation. \textit{Particularly, one can easily see that for normal regression, the NRPP in \eqref{quantile} is the same as Pearson residual}. Additionally, the variability in $F^{*}(y_{i}, u_{i})$  will be smaller when the probability at $y_{i}$ is smaller. This scenario typically occurs when the mean of $y_{i}$  is large. However, when the probability at $y_{i}$ is large, the randomization with $u_{i}$ is necessary to produce uniform p-values. 

In \cite{Marshall03,marshall2007identifying}, a predictive p-value with $u_{i}=0.5$  is defined for identifying outliers.   We will refer this predictive value as ``middle-point predictive p-values'' (MPPs), and refer the corresponding normally-transformed MPPs as NMPPs. NMPP will be compared to NRPP in our examples to see the necessity of the randomization for obtaining uniformly distributed predictive p-values. 

%NRPPs can be widely applied to diagnose regression models for scalar $y_i$ if the CDF and PMF exist. 

%Table \ref{tab:quanitle} provides the derivation of the NRPPs for the Poisson, NB,  and ZIP regression models, where $ppois$ and $dpois$, $pnbinom$ and $dnbinom$, and $pzip$ and $dzip$ denote their respective CDFs and PMFs. 
%
%\begin{table}[h]
%\centering
%\caption{NRPPs for commmonly used non-normal regression models. \label{tab:quanitle}}
%\setcellgapes{5pt}
%\makegapedcells
%\begin{tabular}{|c|c|}
%\hline
%Model & NRPPs \\
%\hline
%Poisson & $q_i = \Phi^{-1} \Bigl( ppois(y_i-1;\hmi)+u_i \cdot dpois(y_i;\hmi) \Bigr)$ \\
%\hline
%Negative Binomial & $q_i = \Phi^{-1} \Bigl( pnbinom(y_i-1;\hmi, \hat{k})+u_i \cdot dnbinom(y_i;\hmi,\hat{k}) \Bigr)$ \\
%%\hline
%%Gamma & $q_i = \Phi^{-1} \Bigl( pgamma(y_i;\hmi, \hat{k}) \Bigr)$ \\
%\hline
%ZIP & $q_i = \Phi^{-1} \Bigl( pzip(y_i-1;\hmi,\hat{p_i})+u_i \cdot dzip(y_i;\hmi,\hat{p_i}) \Bigr)$ \\
%\hline
%\end{tabular}
%\end{table}
%
\subsection{Illustrative Examples}\label{sec:ill}

\subsubsection{An Example with No Covariate}\label{sec:ill1}
\begin{figure}[htp]
    \centering
    \begin{subfigure}[t]{0.45\textwidth}
    \centering
    \includegraphics[width=\textwidth, height = 0.2\textheight]{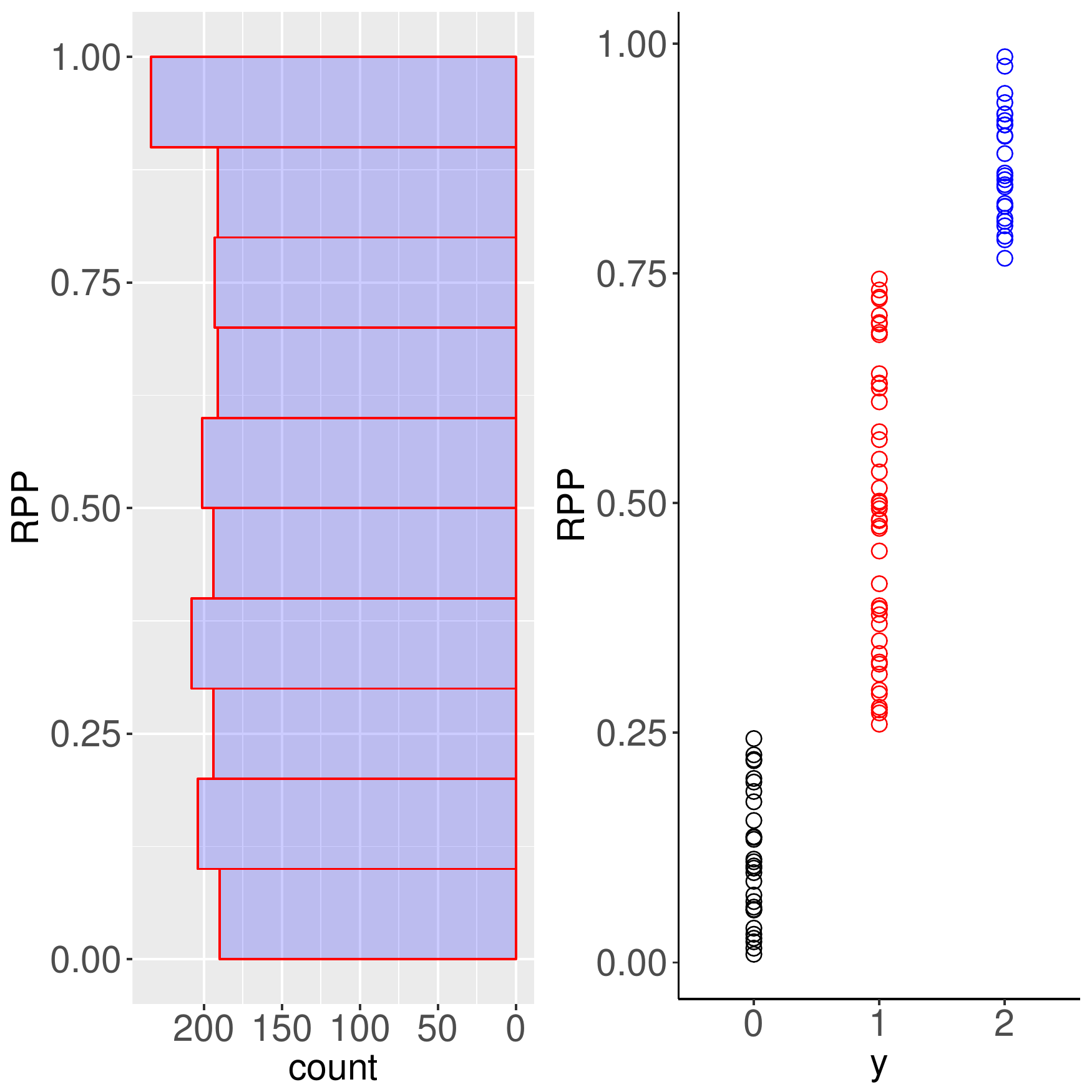}
   \caption{(a) RPPs for the true model} \label{fig:pvr}
    \end{subfigure}
     \hspace{1em}
     \begin{subfigure}[t]{0.38\textwidth}
     \centering
    \includegraphics[width=\textwidth, height = 0.2\textheight]{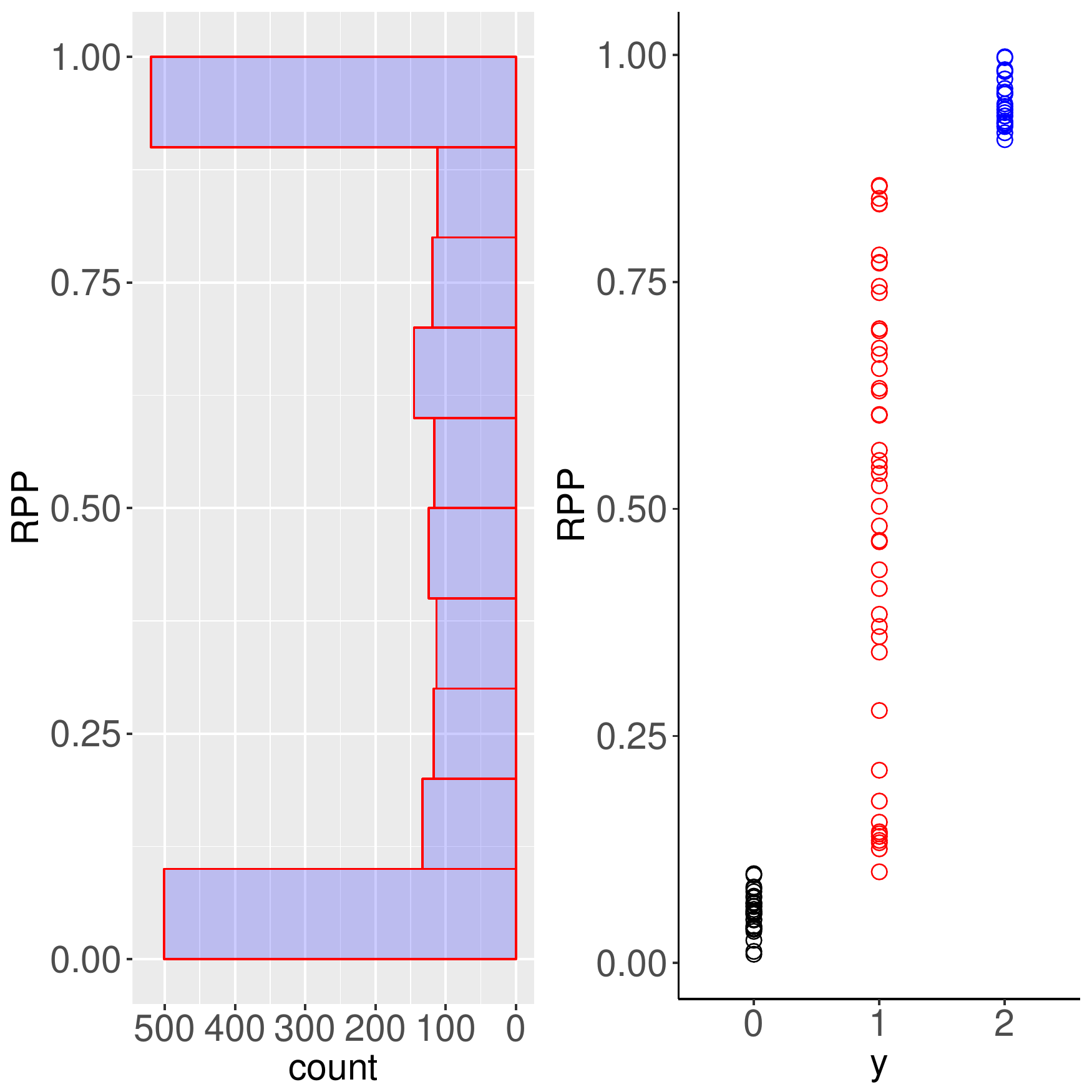}  
     \caption{(b) RPPs for a wrong model} \label{fig:pvw} 
    \end{subfigure}
    \caption{The RPPs for the true model (left panel) and a wrong model (right panel) for the first example given in Section~\ref{sec:ill1}. For each subfigure, the left plot shows the histogram of 2000 RPPs, and the right plot shows the scatterplots of  100 RPPs against observed $y_{i}$.}
    \vspace*{-10pt}
\end{figure}

We first consider a scenario without covariate. Suppose that the true distribution for $y_i$ has the following PMF: $p_{0}(y_{i})=0.25$ for $y_{i}=0$ or $2$; $=0.5$ for $y_{i}=1$.
%\begin{equation*}\label{table:truebinom}
%\begin{tabular}{c|ccc}
%$y_i$ & 0 & 1 & 2 \\
%\hline
%$p_0(y_i)$ & 0.25 & 0.5 & 0.25
%\end{tabular}
%\end{equation*}
For a dataset generated from $p_0$, we expect that a quarter of $y_i$ are 0, half of $y_i$ are 1, and a quarter of $y_i$ are 2. The RPP function $F^*$ with $p_0$ as the considered model converts $y_i=0$ into a uniform random number on $(0, 0.25)$, $y_i=1$ into a uniform random number on $(0.25, 0.75)$ and $y_i=2$ into a uniform random number on $(0.75, 1)$. Overall, the random numbers converted with $F^*$ are uniformly distributed on $(0,1)$. To illustrate, we simulate a sample of size $2000$ from this distribution and compute $\fast (y_i,u_{i})$ based on $p_{0}$. As depicted in Figure \ref{fig:pvr}, $\fast (y_i, u_{i})$ is uniformly distributed between 0 and 1. 

Suppose we compute RPPs based on a wrong model with PMF given by: $p_{1}(y_{i})=0.1$ for $y_{i}=0$ or $2$; $=0.8$ for $y_{i}=1$.
%\begin{equation*}\label{table:wrongbinom}
%\begin{tabular}{c|ccc}
%$y_i$ & 0 & 1 & 2 \\
%\hline
%$p_1(y_i)$ & 0.1 & 0.8 & 0.1
%\end{tabular}
%\end{equation*}
All zeros (around $40\%$ of data) will be scattered uniformly to the interval (0, 0.1), all ones (around $50\%$ data) will be uniformly scattered to the interval (0.1, 0.9), and all two's (around $25\%$ of data) will be scattered uniformly to the interval (0.9, 1). As such,  the distribution of RPPs based on $p_{1}$ is more dense on both left and right tails than the middle interval (0.1, 0.9); see Figure \ref{fig:pvw}. The non-uniformity of  RPPs indicates that the model $p_{1}$ is wrong for the data. The fat tails of the histogram of RPPs reveal that the model $p_{1}$ has shorter tails than the true model ($p_{0}$). The non-uniformity is indeed caused by the mis-matching of the observed frequencies (0.25,0.5, and 0.25) of RPPs on the three intervals and the theoretical frequencies (0.1,0.8, and 0.1) postulated by the model $p_{1}$.

\subsubsection{An Example with Covariate}\label{sec:ill2}

%\begin{landscape}
\begin{figure}[p]
\centering
\includegraphics[width=\textwidth, height =7.5in]{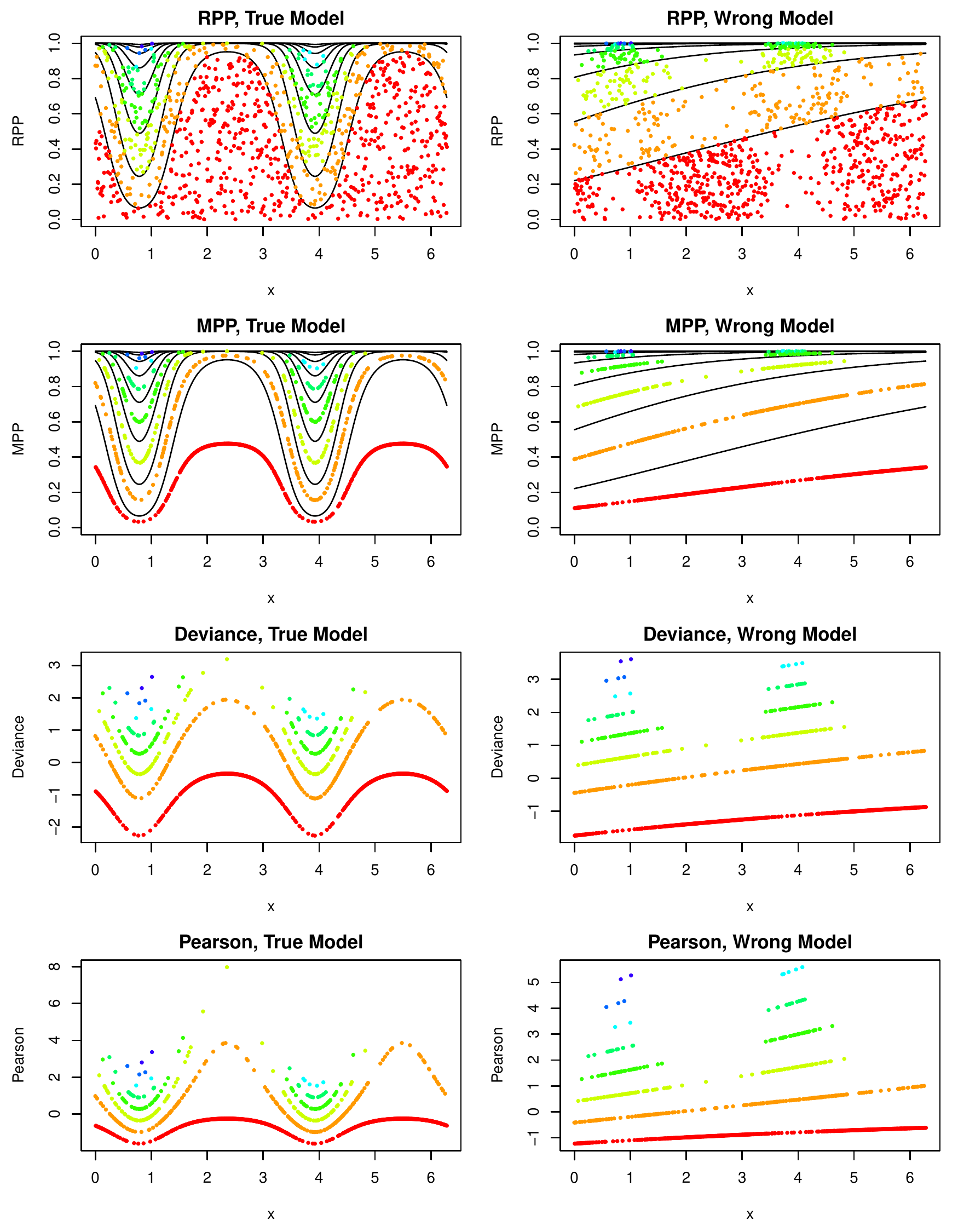}
\caption{An example of the RPPs in comparison with other residuals. Panels in the left column present the residuals or predictive p-values under the true model and panels in the right column present the corresponding values under a wrong model. For the RPPs and MPPs, each black line is a CDF curve of $F(k|x_i) \mbox{ versus } x_{i}$ associated with a value of $k$. The colours of points represent values of $y_{i}$.}
\label{fig:ill2}
\end{figure}
%\end{landscape}

To further demonstrate the idea of RPP in regression settings, we simulate 1000 observations from a Poisson model with mean $\mu_{i}$ given by $\hbox{log}(\mu_i)=-1+ 2\hbox{sin}(2 x_i)$,  and $x_i \sim \mathit{Uniform}(0, 2\pi)$. Then, we fitted the Poisson regression with the true mean structure  and a wrong Poisson model with mean structure $\hbox{log}(\mu_i)=\beta_0+\beta_1 x_i$. 

The CDF of the response variable $Y_i$ given $x_i$  is denoted by $F(k|x_i)= P(Y_{i}\leq k|x_{i})$, for $k=0, 1, \ldots$. Figure \ref{fig:ill2} shows $F(k|x_i)$ as a function of $x_{i}$, with each black line representing a CDF curve associated with a value $k$ in $\{0,1,2,\ldots,\}$. The distance between two curves, $F(y_{i}-1|x_i)$ and $F(y_{i}|x_i)$, is the ``theoretical'' (model-based) probability of observing $y_{i}$. The $\fast (y_i, u_i)$ for each observed $y_{i}$ is a random point between the CDF lines $F(y_{i}-1|x_i)$ and $F(y_{i}|x_{i})$. This random scattering of  $y_{i}$ facilitates the comparison of the ``observed'' frequency (fraction of points between two curves) and ``theoretical'' frequency (distance of two curves). If the ``observed'' and ``theoretical'' frequencies agree well, the $\fast (y_i, u_i)$ should be uniformly distributed on $(0,1)$ in each neighbourhood of $x_{i}$; otherwise, they are not. Under the true model, the top-left plot in Figure \ref{fig:ill2} depicts that the RPPs are uniformly distributed on (0, 1) given each neighbourhood of $x_{i}$. By contrast, under the wrong model, the top-right plot of Figure \ref{fig:ill2} shows that the RPPs given each $x_{i}$ are not uniformly distributed. For example, given $x_{i}=2$, the wrong model postulates that $P(y_{i}=1)$ is about 0.7-0.4=0.3.  However, the actual probability of $y_{i}=1$ is near 0.1 in the true model, hence, we see very few RPPs in the interval (0.4,0.7) in RRP plot for the wrong model. The mis-matching in ``observed'' and ``theoretical'' frequencies results in non-uniformity of RPPs under the wrong model.  Without randomization, the mid-point predictive p-values (MPP) cluster on separated lines with each associated a distinct value of $y$. Therefore, we cannot graphically assess the matching of the observed and frequencies, and the MPPs are not uniformly distributed in any neighbourhood of $x_{i}$ when the true model is fitted; indeed they show a clear non-linear trend. Similarly, the Pearson and deviance residuals cluster on lines, hence, are not normally distributed under the true model; they also show a clear non-linear trend under the true model. 

In summary, due to the clustering on lines, MPPs, Pearson and deviance residuals cannot confirm the good fit by the true model,  and hence cannot be used to identify the non-linearity in the wrong model due to lack of a unified reference distribution under the true model for comparison. The randomization in RPP is necessary to produce truly uniform predictive p-values for discrete response under the true model so that the model mis-specification can be revealed by comparing the RPPs to this reference distribution.

\subsection{The Uniformity of RPPs}\label{normality proof}
The RPPs are uniformly distributed, and correspondingly the NRPPs are normally distributed, under the true model. First, let us recall the well-known property of p-value for a continuous response variable written as 
\begin{thm}\label{thm:0}
Suppose a continuous random variable $Y$ has the CDF given by $F(y)$, then $F(Y)$ is uniformly distributed on (0,1). 
\end{thm}
Theorem \ref{thm:0} leads to the well-known fact that the p-values of a test statistic are uniformly distributed on $(0,1)$ when the null distribution is true. This uniformity is used to validate the well-calibration of computed p-values.  Another equivalent way to express Theorem \ref{thm:0} is that: suppose $F(y)$ is a continuous CDF, let $F^{-1} (u)$ denote the inverse function defined as $F^{-1} (u)=\inf \{x|F(x)>u\}$, and $U$ is uniformly distributed on (0, 1), then $F^{-1}(U)$ is distributed as $F(y)$. That is, we can transform uniform random numbers to random numbers with any desired continuous distribution, such as normal.  When the $y_{i}$ is discrete, Theorem \ref{thm:0} can be extended to: 

\begin{thm}\label{thm:1}
Suppose the true distribution of $Y_{i}$ given $X_{i}=x_{i}$ has the CDF $F_{i}(y_{i})$ and PMF $p_{i}(y_{i})$, where the subscript $i$ indicates that $F_{i}$ and $p_{i}$ depends on a covariate $x_i$. The randomized predictive p-values $F^{*}(y_i, u_i)$ is defined as $F_{i}(y_{i}-)+u_{i} \,p_{i}(y_{i})$ \eqref{pvi}. Suppose $U_i$ is uniformly distributed on (0,1). Then, we have 
\begin{equation}\label{eqn:uniform}
F^*(Y_i,U_i) \sim \mbox{Uniform}((0,1)), 
\mbox {and }
%\begin{equation}\label{eqn:normality}
\phi^{-1} (F^*(Y_i,U_i)) \sim N(0,1).
\end{equation}
\end{thm}

\def \sj {^{(j)}}
\def \Fstar {F^*(Y_i,U_i)}
\allowdisplaybreaks

The proof of  Theorem  \ref{thm:1} can be found in Section~\ref{sec:proof}. Next, we make a few remarks to clarify the applications of  Theorem  \ref{thm:1}. 
\begin{enumerate}

\item Since the conditional distribution of $\fast (Y_{i}, U_{i})$ given $X_{i}=x_i$ is uniformly distributed on $(0,1)$,  and this distribution is free of $x_{i}$, the marginal distribution of $\fast (Y_{i}, U_{i})$ with $x_{i}$ marginalized away is still  uniformly distributed on $(0,1)$. This justifies that the overall distribution of RPPs is uniform on $(0,1)$. 
\item In frequentist paradigm, the $F_{i}(y_{i})$ is the CDF of the true model with the true parameters that have generated the dataset. In practice, the parameters may be estimated with the sample data including $y_{i}$ itself. The use of estimated parameters that have learned from $y_{i}$ itself will introduce conservatism in the predictive p-values due to using $y_{i}$ twice. As a result, the predictive p-values may be more concentrated around 0.5 than the uniform distribution on (0,1); correspondingly, the NRPPs tend to be more concentrated around 0 than distributed as $N(0,1)$. This conservatism is minor when the sample size is much larger than the number of parameters. Our empirical studies (not shown in this paper) also indicated that the conservatism affects less in the following overall GOF test applied to NRPP if we use Shapiro-Wilk normality test compared to other tests, such as Kolmogorov-Smirnov test. For very complex models with a high risk of overfitting, it is necessary to eliminate this conservatism by computing cross-validatory NRPP.  In this paper, we focus on discussing the necessity of using ``randomization'' to obtain truly uniform predictive p-values, hence, we ignore this conservatism by considering relatively simple models. %However, note that this conservatism will be a serious problem for complex models   \cite{li_estimating_2017}. 

\item In Bayesian paradigm, the $F_{i}(y_{i})$ is the CDF of the CV predictive distribution of $Y_{i}\rep$ given $y_{-i}$ (and covariates $x_{1},\ldots,x_{n}$) with model parameters $\mb\theta$ marginalized away with respect to the posterior based on $y_{-i}$, as given below:
$$F_{i}(y_{i})=P(Y_{i}\rep\leq y_{i}|y_{-i})=\int P(Y_{i}\rep\leq y_{i}|\mb\theta)P(\mb\theta|y_{-i})d\mb\theta.
$$
Therefore, the $F^{*}(y_{i}, u_{i})$ is the cross-validatory randomized predictive p-values (CVRPP).  Theorem \ref{thm:1} is an extension of the theorem proved by \cite{marshall2007identifying} about the uniformity of  CV predictive p-values for continuous response variable under the true model to the uniformity of CVRPP for discrete response variable.  In Bayesian sense, the uniformity of CVRPP holds when both of the prior and the likelihood are correctly specified.  Therefore, the non-uniformity of CVRPPs may result from mis-specification in either prior or likelihood, or both, and hence,  reveals the discrepancies in both prior and likelihood.  Last, we note that, although NRPPs without CV is the same as RQRs in simple regression,  CVRPPs can be applied to diagnose hierarchical Bayesian models for data with complex correlation structure; see \citet{marshall2007identifying,  li_estimating_2017}.  Theorem \ref{thm:1} provides theoretical foundation for model diagnosis with CVRPPs for models with discrete response.  Further discussion of computing CVRPPs for Bayesian models is given in Section~\ref{sec:conclusion}. 

\end{enumerate}

\vspace*{-10pt}

%%%%%%%%%%%%%%%%%%%%%%%%%%%%%%%%%%%%%%%%%%%%%%%%%%%%%
\section{Simulation Studies}\label{sec:simulation}
%%%%%%%%%%%%%%%%%%%%%%%%%%%%%%%%%%%%%%%%%%%%%%%%%%%%%
In this section, we investigate via simulation the performance of the NRPPs and compare this approach with the NMPPs, deviance and Pearson residuals. The simulations consist of testing non-linearity in the covariate effect, over-dispersion, and zero-inflation. For each simulation, we first present the performance of NRPP under one simulation scenario. Furthermore, in order to gain more insight of the finite-sample performance, we perform power analysis by setting the sample sizes to $n=20, 50, 100, 200, 400, 600, 800$ and $1000$ with varying degrees of model complexity. In overall GOF, we tested the specified hypotheses: $H_0$: the model fits the data well versus $H_a$: the model does not fit the data well. Shapiro-Wilk (SW) test was used for evaluating the normality of residuals. Under each simulation scenario, we randomly generated 500 datasets from a true model for examining the type I error rates and statistical powers using significance level 0.05. 
%The type I error rate is defined as the probability of rejecting the true model, estimated by the proportion of SW test p-values $<0.05$ when the true model is fitted. The statistical power is the probability of rejecting the wrong model, estimated by the proportion of SW test p-values $< 0.05$ when the wrong model is fitted. Ideally, a desirable GOF test should have type I error rate close to the nominal level $0.05$ and have high statistical power in rejecting the wrong model. 

\subsection{Detection of Non-linearity}\label{sec:4.1}
The performance of the NRPPs for detecting non-linearity in the covariate effect is evaluated with a count response variable following a NB distribution based on a single dataset. We first simulate the covariate $x_{i} \sim \mathit{Uniform}(-1.5, 1.5)$ of size $n=1000$. The response variable is then simulated from a NB regression model $\hbox{log}(\mu_i)=\beta_0+\beta_1 x_i^2$, where $\mu_i$ is the expected count for the $i$th subject. A wrong model $\hbox{log}(\mu_i)=\beta_0+\beta_1 x_i$ is considered for fitting. The regression parameters were set as $\beta_0=0$ and $\beta_1=1$ while the reciprocal for the dispersion parameter associated with the NB distribution was set as $k=2$. 

\begin{figure}[htp]
\begin{subfigure}[t]{0.30\textwidth}
\subcaption{\centering \textbf{Residual plot}}
\includegraphics[width=1.8in, height=1.3in, trim=0in 1.3in 0in 2.1in,clip]{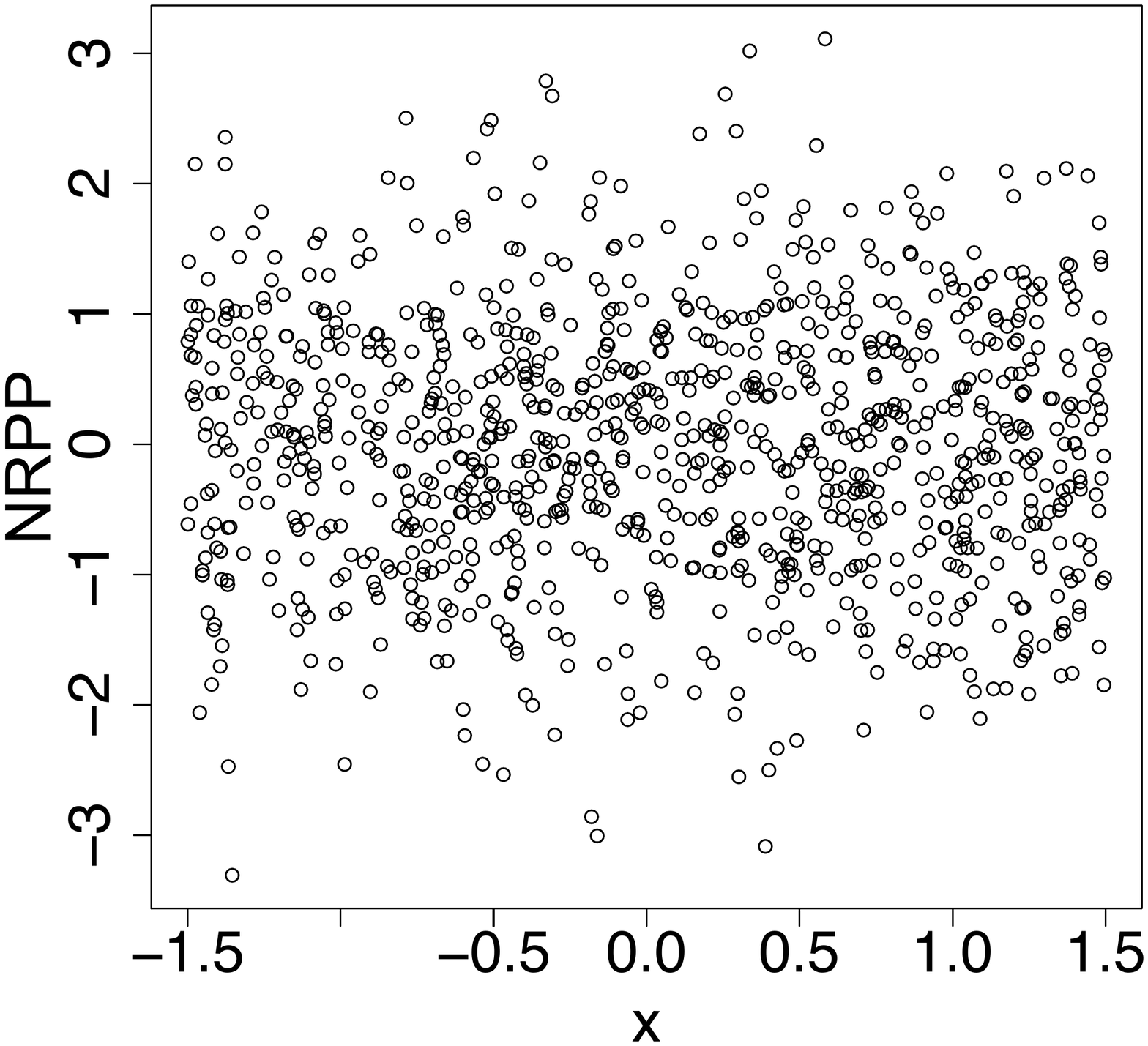}
\end{subfigure}\hspace{1em}
\begin{subfigure}[t]{0.30\textwidth}
\subcaption{\centering \textbf{QQ plot}}
\includegraphics[width=1.8in, height=1.3in,trim=0 0 0 0.6in,clip]{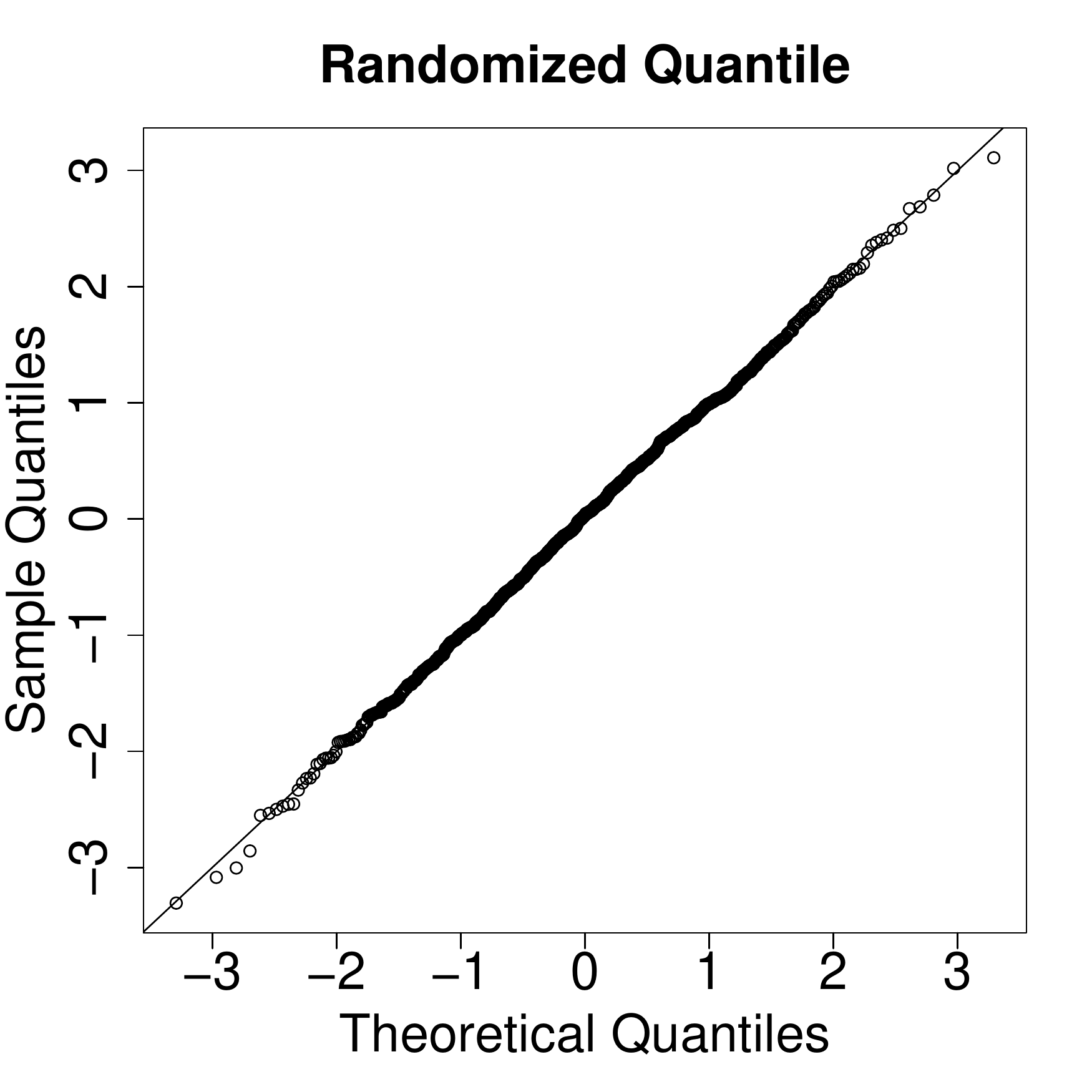}
\end{subfigure}\hspace{1em}
\begin{subfigure}[t]{0.30\textwidth}
\subcaption{\centering \textbf{SW p-values}}
\includegraphics[width=1.8in, height=1.3in,trim=0 0 0 0.6in,clip]{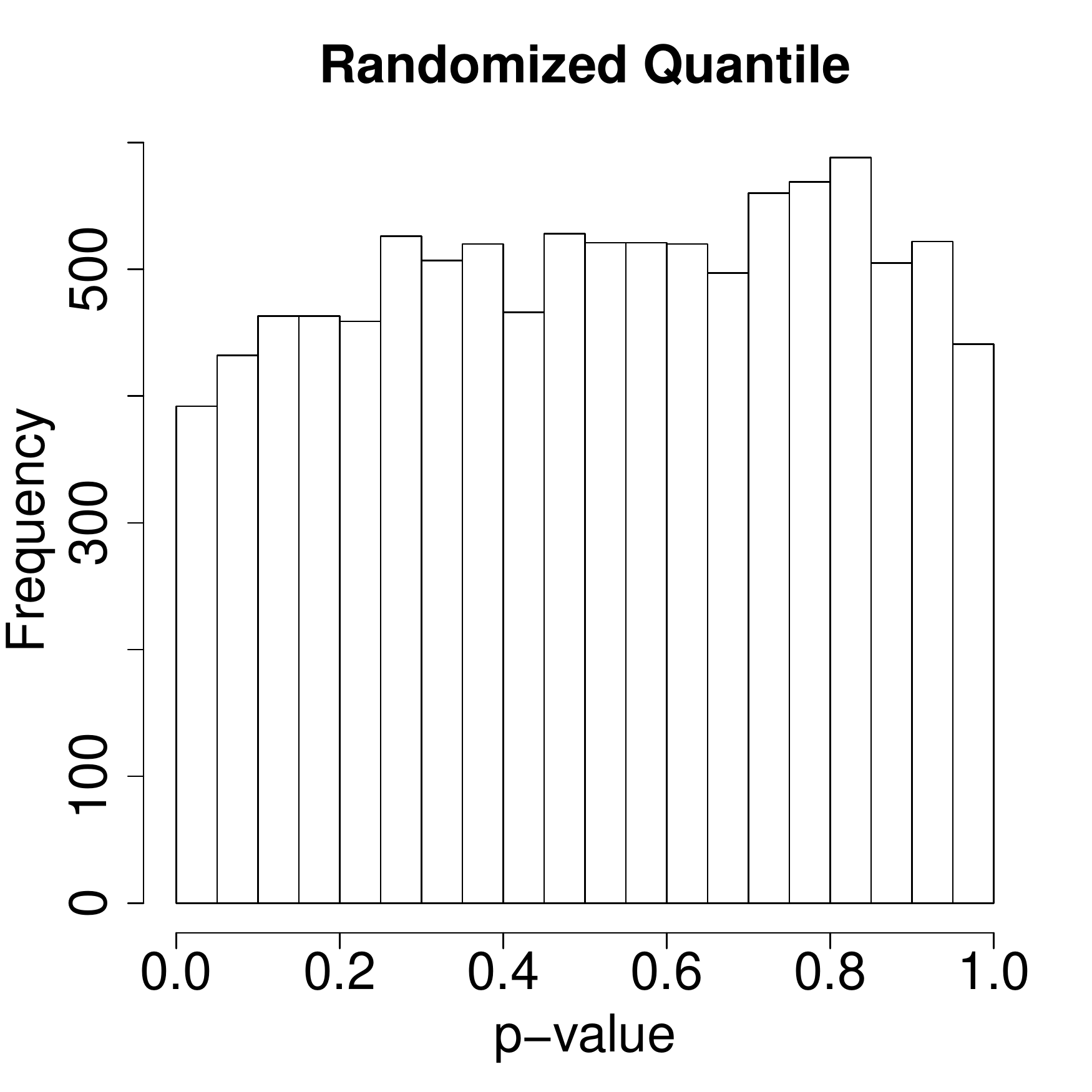}
\end{subfigure}\hspace{1em}

\begin{subfigure}[t]{0.3\textwidth}
\includegraphics[width=1.8in, height=1.3in,trim=0in 1.3in 0in 2.1in,clip]{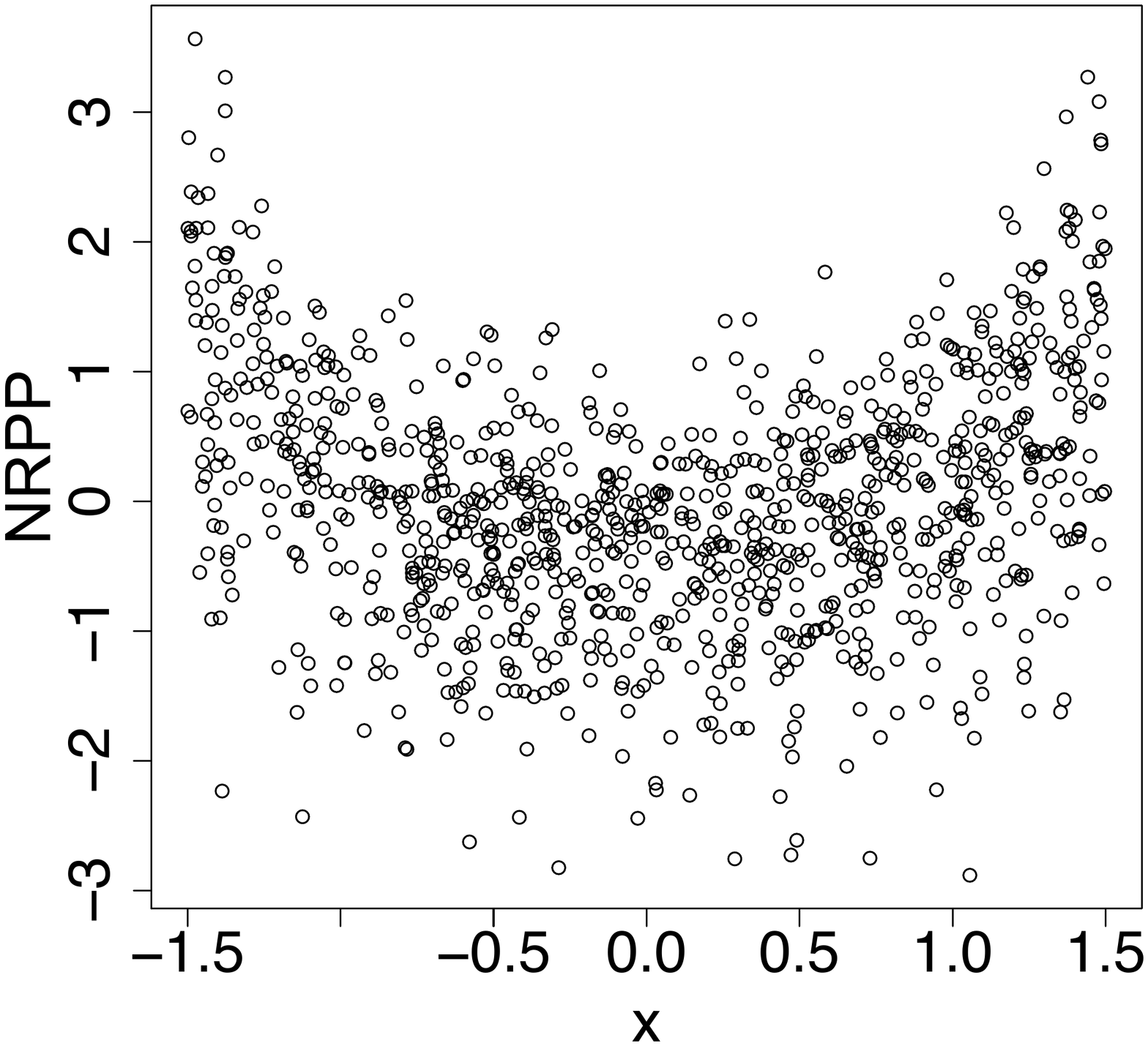}
\end{subfigure}\hspace{1em}
\begin{subfigure}[t]{0.3\textwidth}
\includegraphics[width=1.8in, height=1.3in,trim=0 0 0 0.6in,clip]{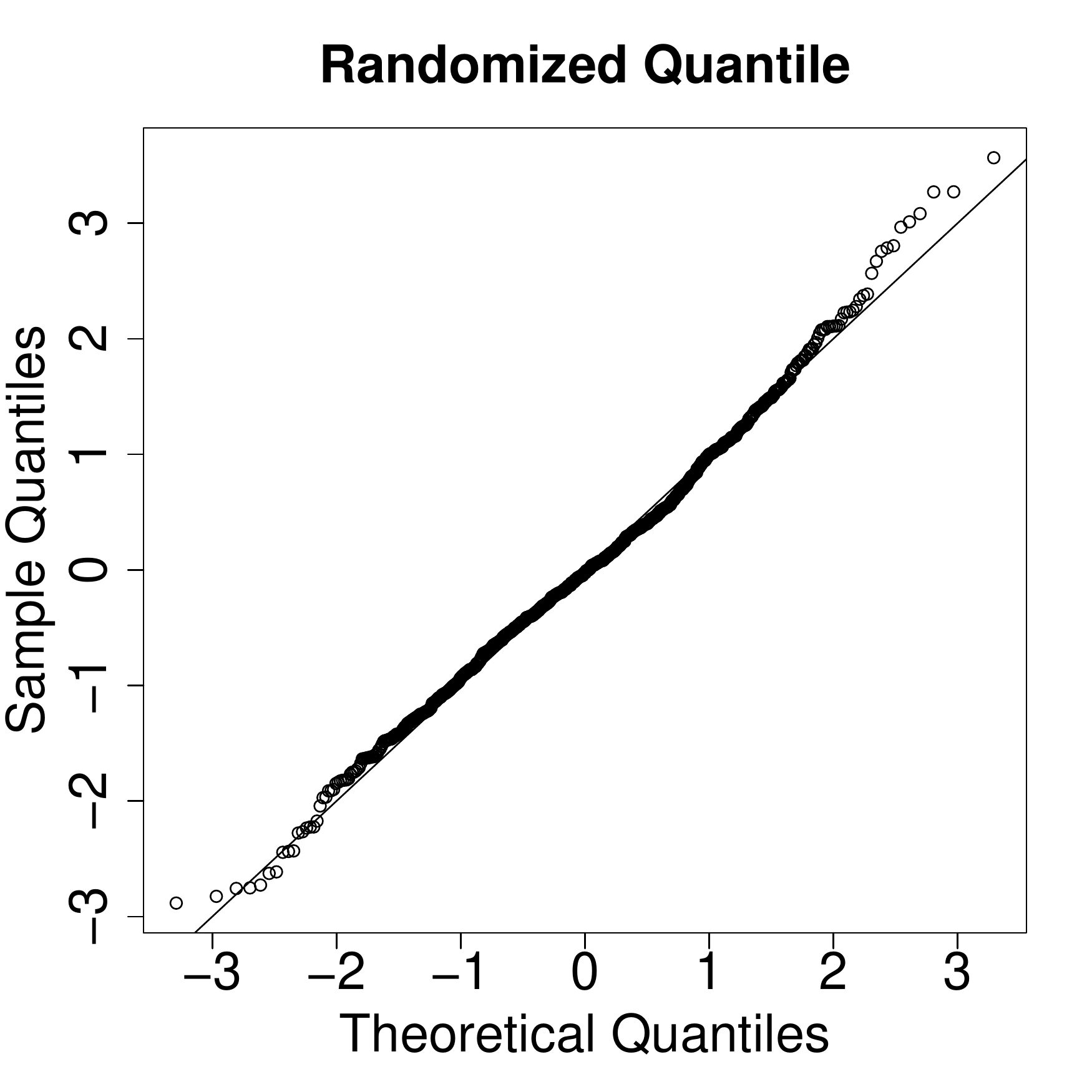} 
\end{subfigure}\hspace{1em}
\begin{subfigure}[t]{0.3\textwidth}
\includegraphics[width=1.8in, height=1.3in,trim=0 0 0 0.6in,clip]{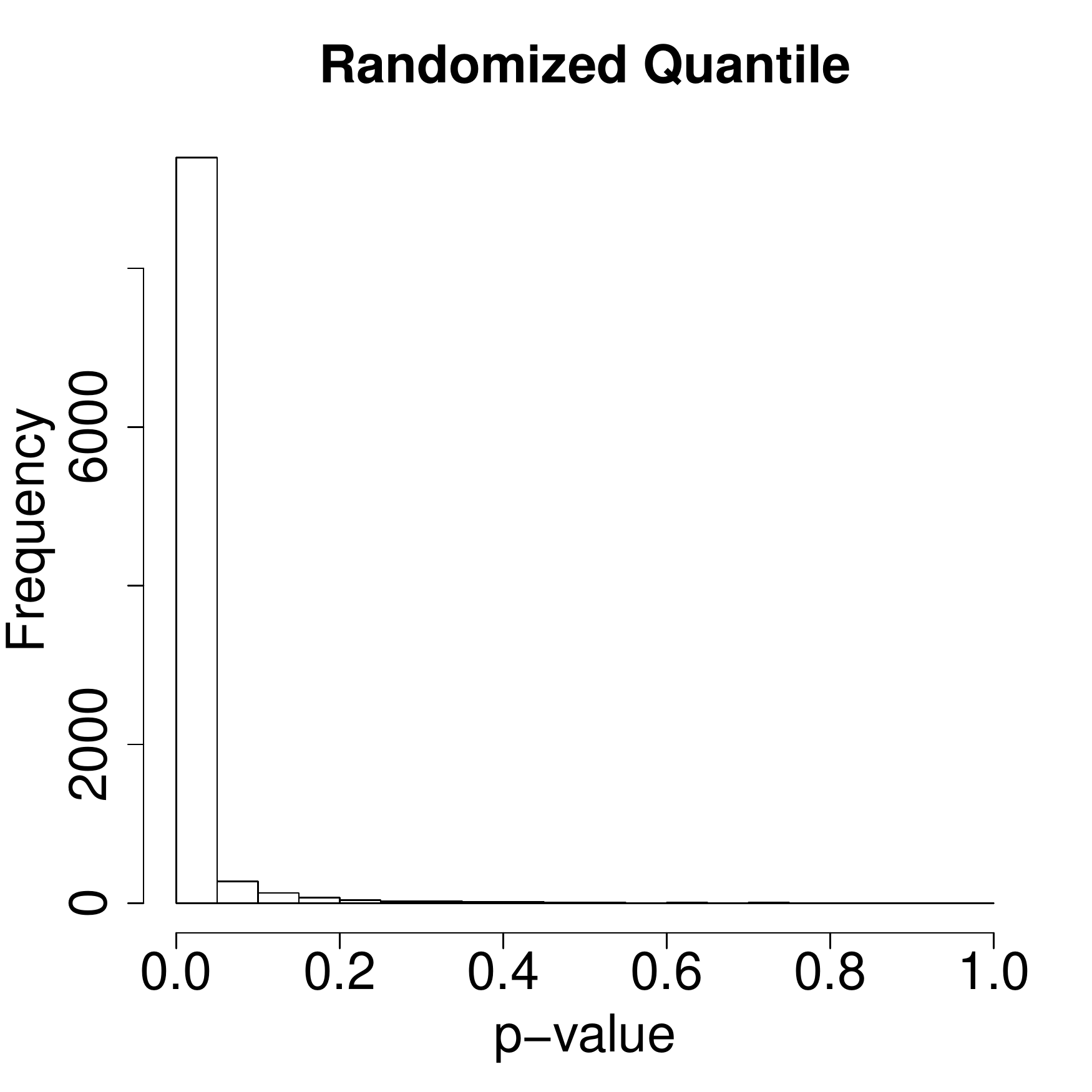} 
\end{subfigure}\hspace{1em}

\caption{\footnotesize Performance of the NRPPs in detecting covariate non-linearity effect of a sample dataset of size $n=1000$. The panels in the first row present the NRPPs for the fitted true model: NB model with quadratic covariate effect (i.e., $\exp(\beta_1 x^2)$). The panels in the second row present the NRPPs for the fitted wrong model: NB model with linearity covariate effect (i.e.,$\exp(\beta_1 x)$). The first two columns display the scatter plots and QQ plots of the NRPPs, respectively. The third column presents the histograms of the SW p-values for the NRPPs over 500 simulated datasets from the true model.  
\label{fig:nbresid}}
\end{figure}

The panels of the first column of Figure \ref{fig:nbresid} display the NRPPs against the covariate under the true and wrong models. Under the true model, NRPPs are randomly scattered without exhibiting any pattern and most being within -3 and 3 as standard normal variates; conversely, under the wrong model, the NRPPs clearly indicate a quadratic trend. The panels of the second column of Figure \ref{fig:nbresid} present the quantile-quantile (QQ) plots of the NRPPs under the true and wrong models. Under the true model, the QQ plot almost perfectly aligns with the diagonal line, whereas under the wrong model, the QQ plot deviates from the diagonal line in both the upper and lower tails. The deviation in the tails is minor for this simulation due to the simulated covariate $x_{i}$ being symmetric about 0; larger deviations will be clearly observed if $x_{i}$ is asymmetric about 0.

In order to quantitatively assess the overall GOF, we applied the SW test to evaluate the normality of the NRPPs resulted from fitting the true and wrong models to the 500 datasets. The panels in the third column of Figure \ref{fig:nbresid} present the histograms of 500 SW p-values under the true and wrong models. The SW p-values under the true model are nearly uniform, indicating the well-calibration of this overall GOF test. In contrast, under the wrong model, the SW p-values are highly distributed near 0, implying that the wrong model will be rejected most of the times at a small nominal threshold. Thus, the overall GOF test via the SW test for the NRPPs has a probability of type I error close to 0.05 and great power in detecting the non-linear relationship in the simulation. 

\begin{figure}[htp]
\begin{subfigure}[t]{0.24\textwidth}
\subcaption{\centering \textbf{~~~~~NRPP}}
\includegraphics[width=\textwidth,trim=0in 0.1in 0.4in 0.6in,clip]{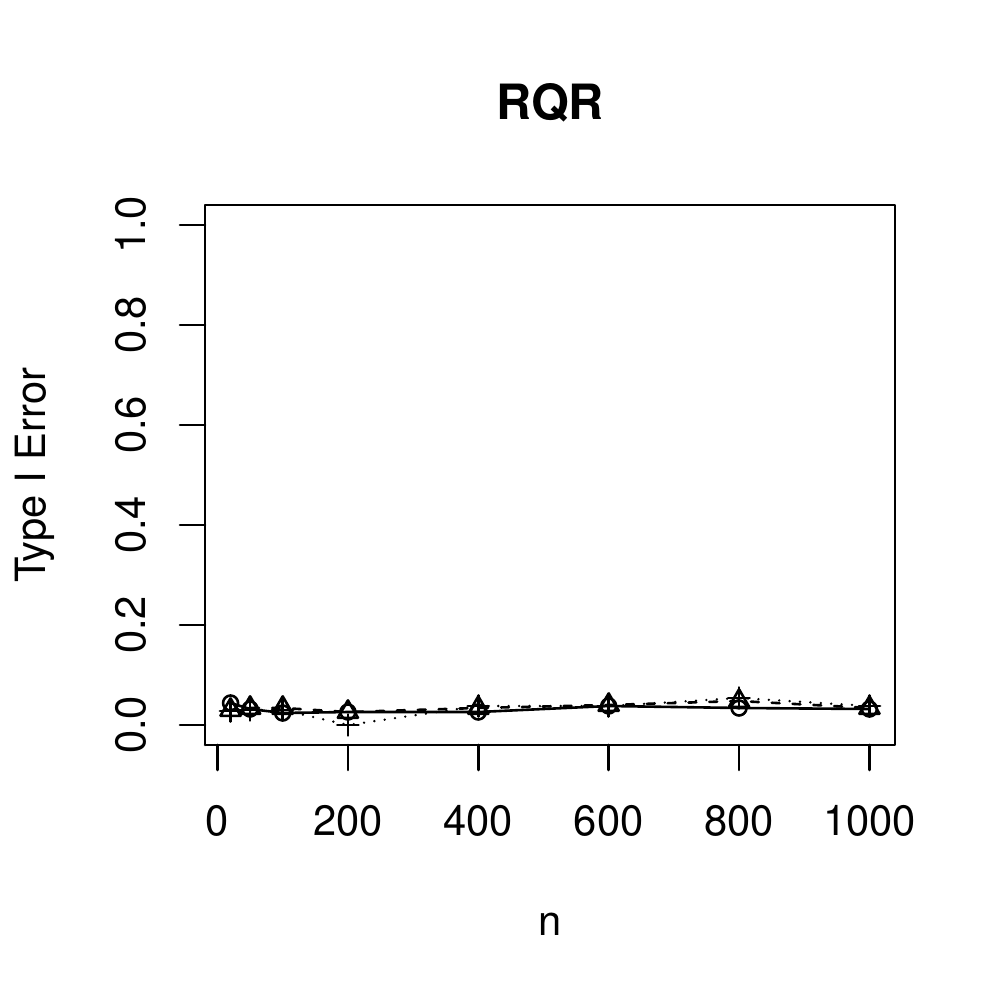}
\end{subfigure}
\begin{subfigure}[t]{0.24\textwidth}
\subcaption{\centering \textbf{~~~~~NMPP}}
\includegraphics[width=\textwidth,trim=0in 0.1in 0.4in 0.6in,clip]{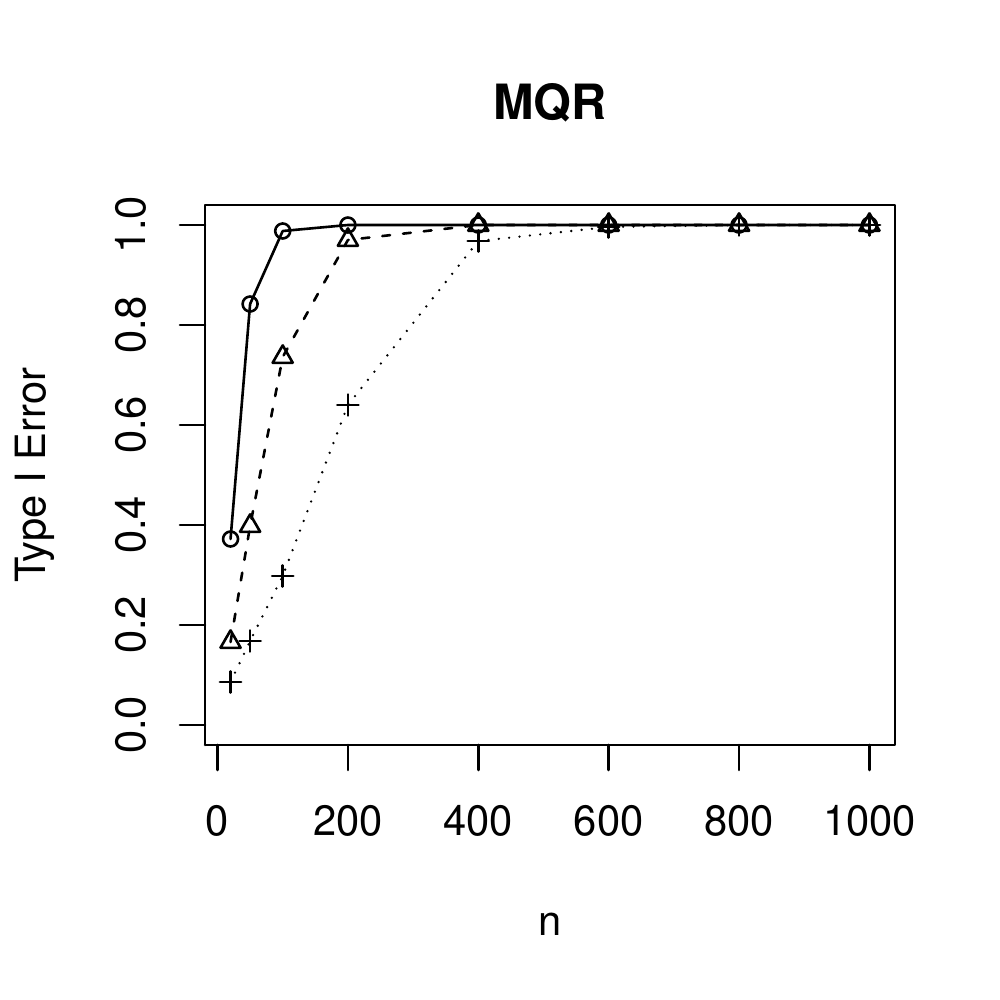}
\end{subfigure}
\begin{subfigure}[t]{0.24\textwidth}
\subcaption{\centering \textbf{~~~~~Deviance}}
\includegraphics[width=\textwidth,trim=0in 0.1in 0.4in 0.6in,clip]{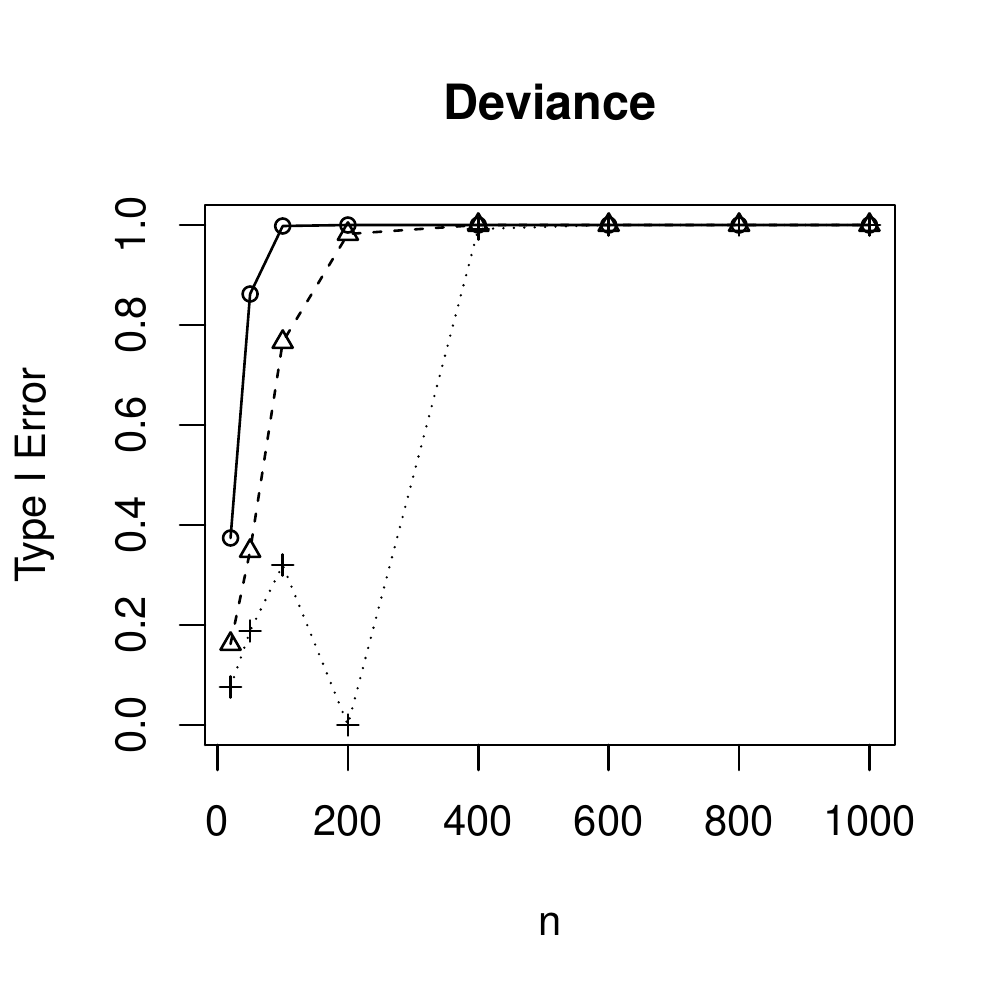}
\end{subfigure}
\begin{subfigure}[t]{0.24\textwidth}
 \subcaption{\centering \textbf{~~~~~Pearson}}
\includegraphics[width=\textwidth,trim=0in 0.1in 0.4in 0.6in,clip]{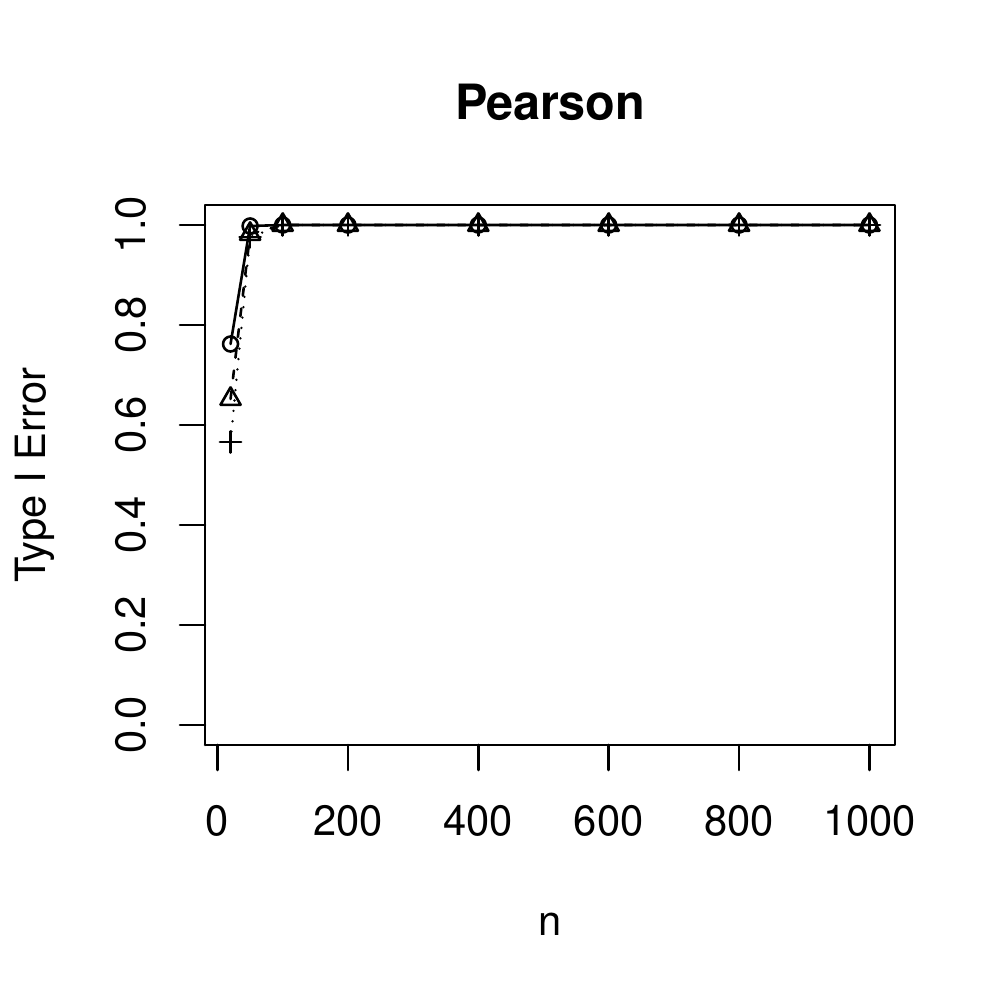}
\end{subfigure}

\begin{subfigure}[t]{0.24\textwidth}
\includegraphics[width=\textwidth,trim=0in 0.1in 0.4in 0.6in,clip]{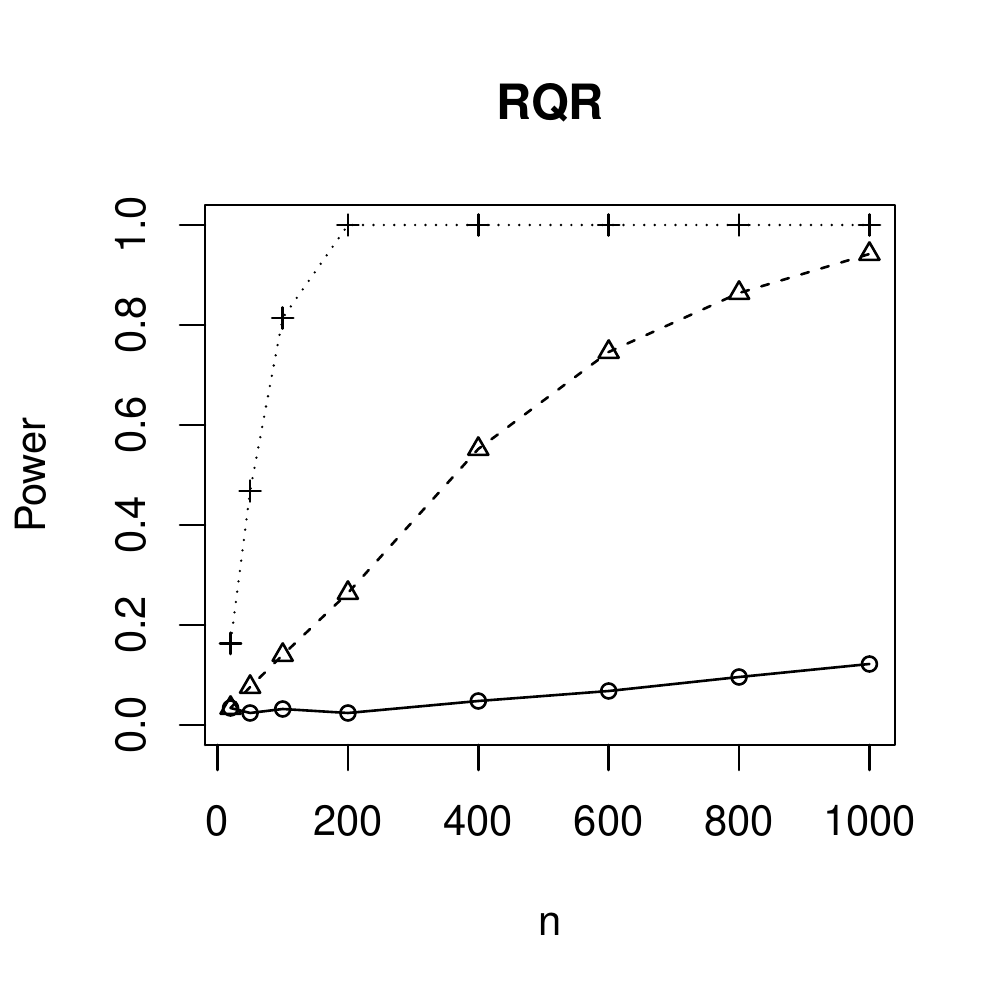}
\end{subfigure}
\begin{subfigure}[t]{0.24\textwidth}
\includegraphics[width=\textwidth,trim=0in 0.1in 0.4in 0.6in,clip]{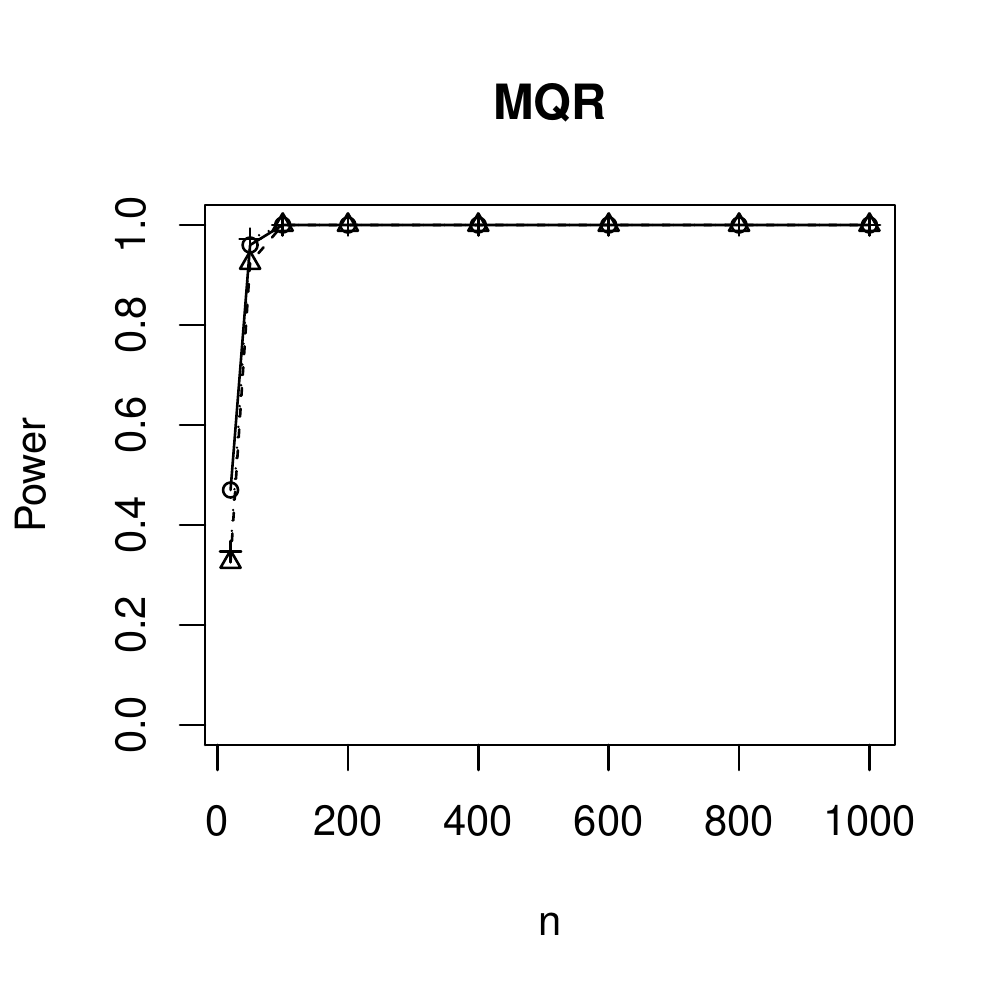} 
\end{subfigure}
\begin{subfigure}[t]{0.24\textwidth}
\includegraphics[width=\textwidth,trim=0in 0.1in 0.4in 0.6in,clip]{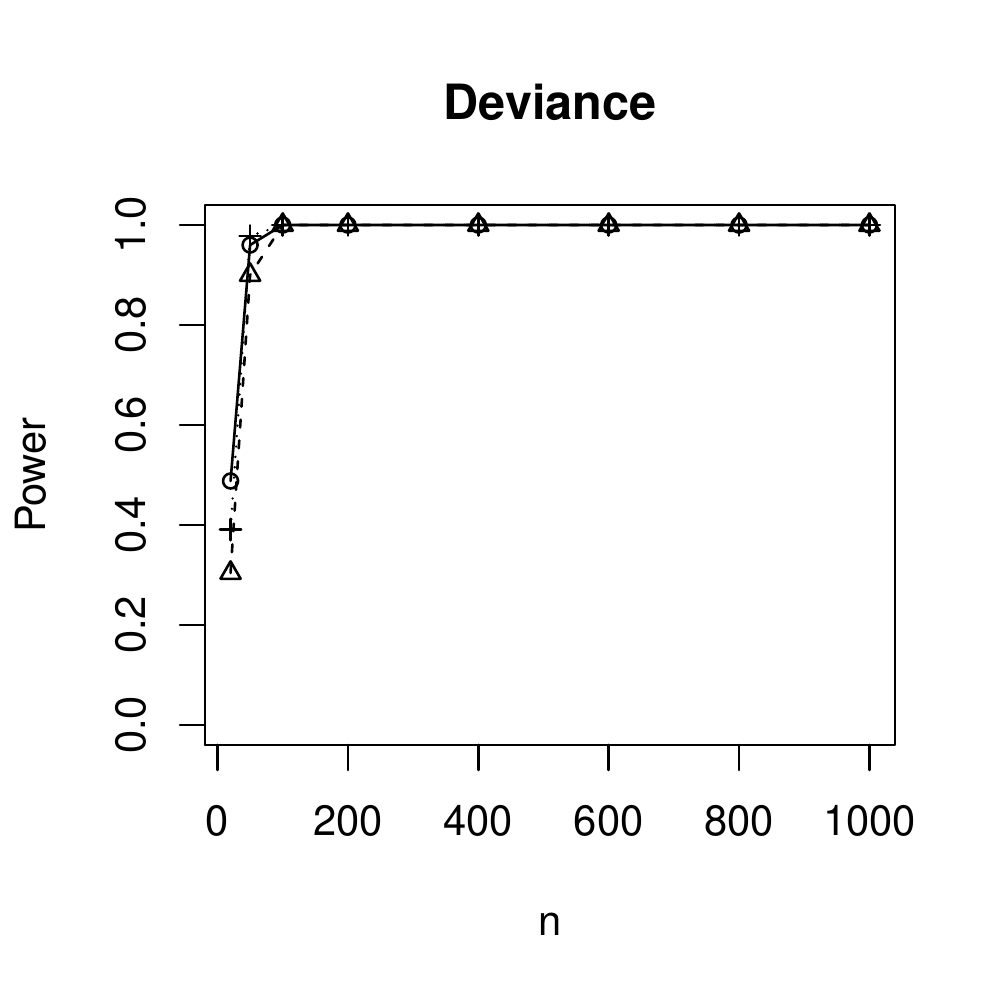} 
\end{subfigure}
\begin{subfigure}[t]{0.24\textwidth}
\includegraphics[width=\textwidth,trim=0in 0.1in 0.4in 0.6in,clip]{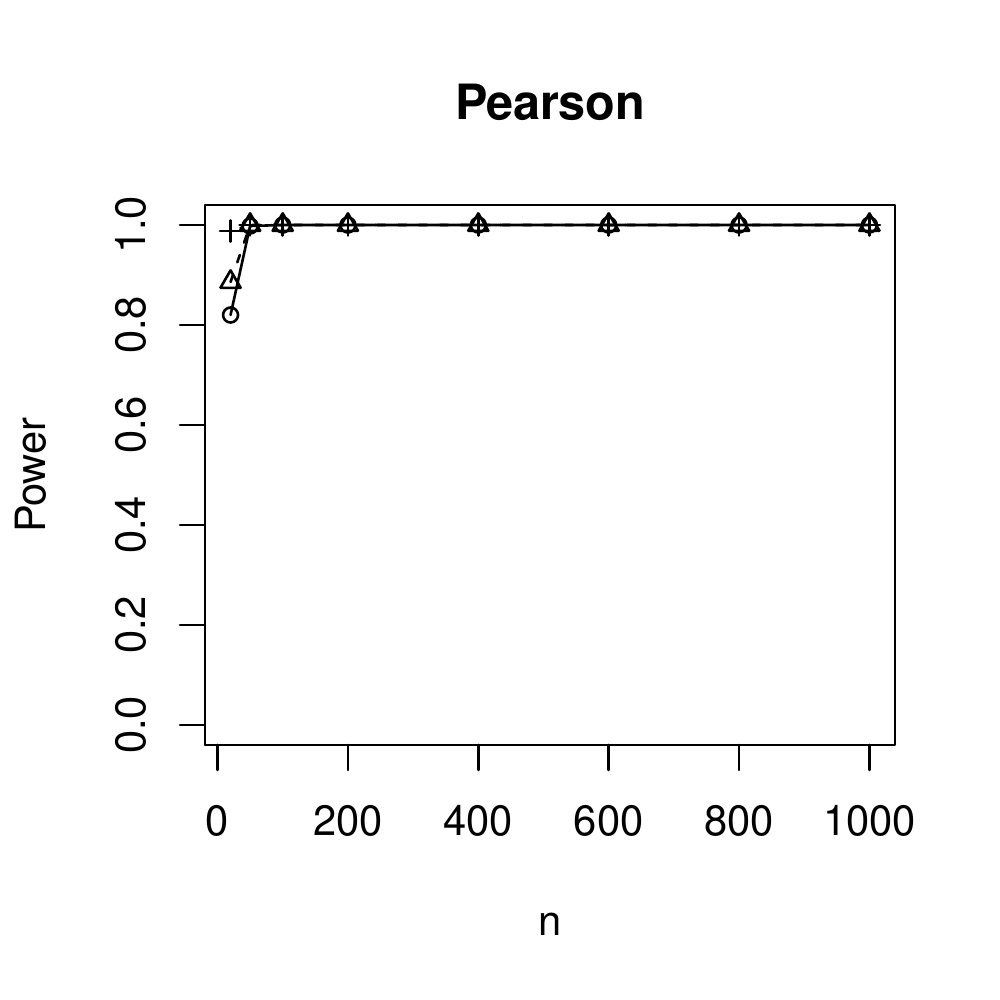} 
\end{subfigure}
\caption{\footnotesize Comparison of the type I errors and powers of the SW tests for the NRPPs, NMPPs, and deviance and Pearson residuals. Response variable is simulated from the true model at varying sample sizes, and nonlinear covariate effects of $\beta_1=0.5$ (\protect\solidline \protect\markcircle), 1 (\protect\dashedline \protect\marktriangle) and 2 (\protect\dottedline $+$). True model: NB model with $\mu_{i}=\exp(\beta_1 x^2_{i})$. Wrong model: NB model with $\mu_{i}=\exp(\beta_1 x_{i})$.} \label{fig:poweranalysis_NB}
\end{figure}

To further investigate the finite-sample performance of the SW test for the NRPPs, Figure \ref{fig:poweranalysis_NB} presents the results of the power analysis by generated 500 datasets for each $\beta_1=0.5, 1, 2$ and with varying sample sizes. The probabilities of the type I errors are consistently low at nominal level 0.05 for all scenarios due to the SW p-values for the NRPPs being nearly uniformly distributed. In contrast, the probabilities of type I errors for the SW tests applied to the NMPPs, and deviance and Pearson residuals are significantly above 0.05, as their SW p-values are incorrectly distributed near 0 when the true model is fitted. Thus, these results show that evaluating GOF with the SW test is well-calibrated for the NRPPs, but is unsuitable for the NMPPs, and deviance and Pearson residuals. Figure \ref{fig:poweranalysis_NB} also show that the power of this GOF test for the NRPPs are reasonably high as long as a large difference between the true and wrong model is present. Although high power results are obtained for the NMPPs, and deviance and Pearson residuals, the substantially high probabilities of type I errors make the GOF tests with these residuals useless. 

%\begin{landscape}
%\end{landscape}

\subsection{Detection of Over-dispersion}\label{sec:4.2}
As in the previous section, the same approach is implemented to investigate the performance of the NRPPs in detecting over-dispersion in the data. We first simulate a covariate $x\sim \mathit{Uniform}(-1, 2)$ of size $n=1000$. Then, the response variable is simulated from a NB model with $\hbox{log}(\mu_i) = \beta_0+\beta_1 x_i$. We set the regression parameters as $\beta_0=1$ and $\beta_1=2$ and reciprocal for the dispersion parameter as $k=2$. To examine the performance of the various types of residuals in diagnosing over-dispersion, we considered fitting a wrong model: Poisson model with the same mean function as in the true NB model.  

\begin{figure}[htp]
%\vspace{2em}
\begin{subfigure}[t]{0.30\textwidth}
\subcaption{\centering \textbf{Residual plot}}
\includegraphics[width=1.8in, height=1.3in,trim=0in 1.3in 0in 2in,clip]{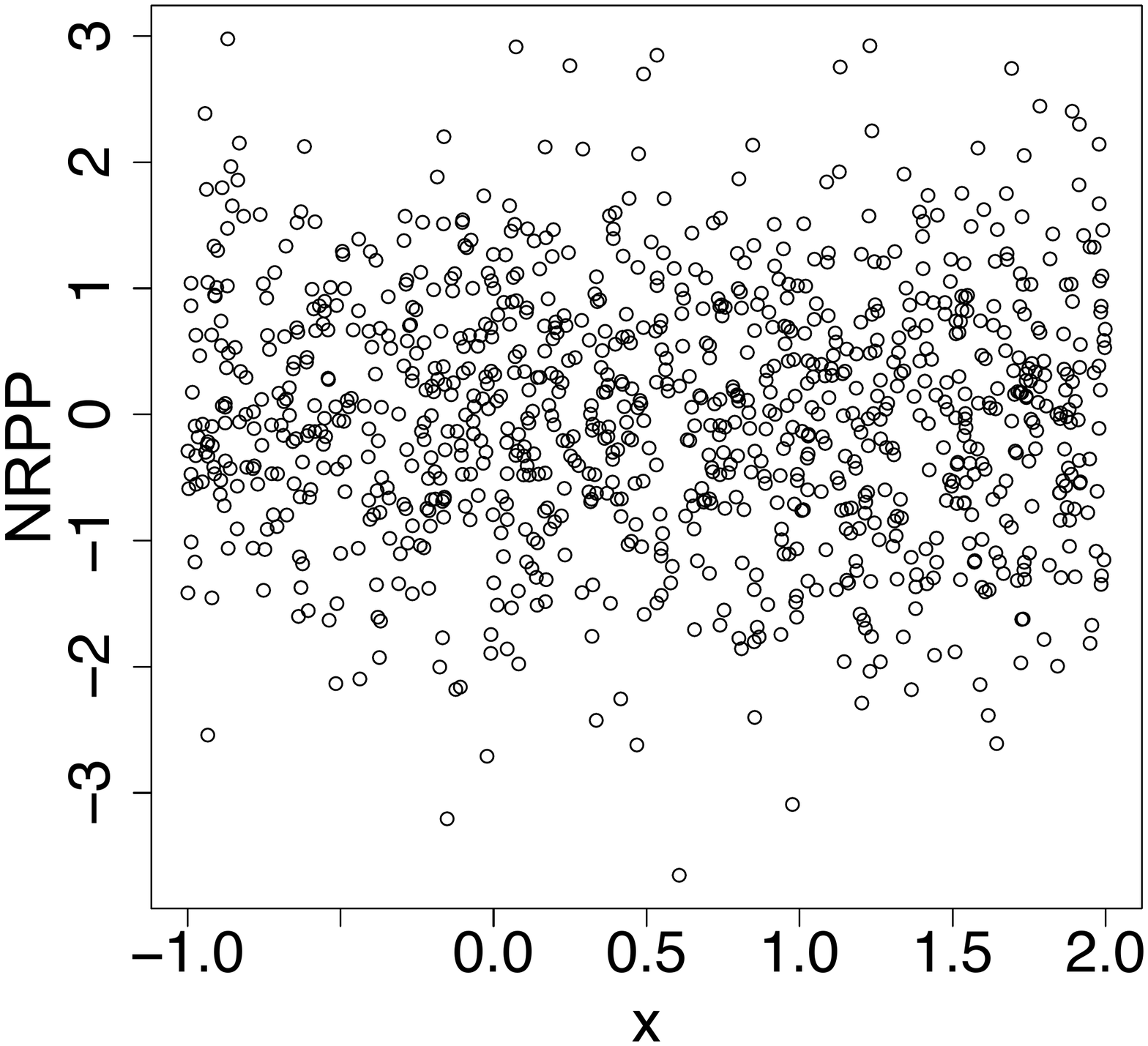}
\end{subfigure}\hspace{1em}
\begin{subfigure}[t]{0.30\textwidth}
\subcaption{\centering \textbf{QQ plot}}
\includegraphics[width=1.8in, height=1.3in,trim=0.00in 0.1in 0.1in 0.6in,clip]{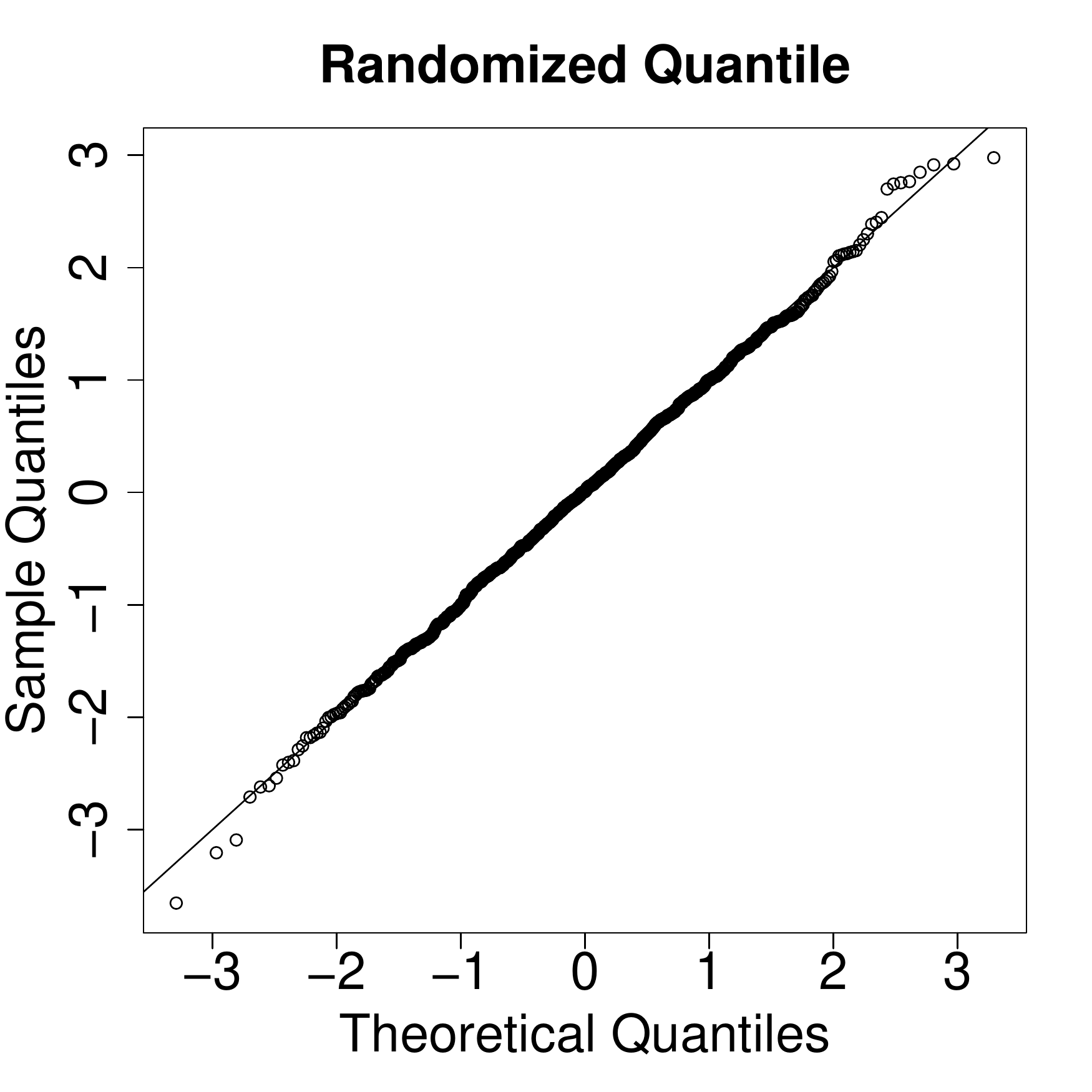}
\end{subfigure}\hspace{1em}
\begin{subfigure}[t]{0.30\textwidth}
\subcaption{\centering \textbf{SW p-values}}
\includegraphics[width=1.8in, height=1.3in,trim=0.00in 0.1in 0.1in 0.6in,clip]{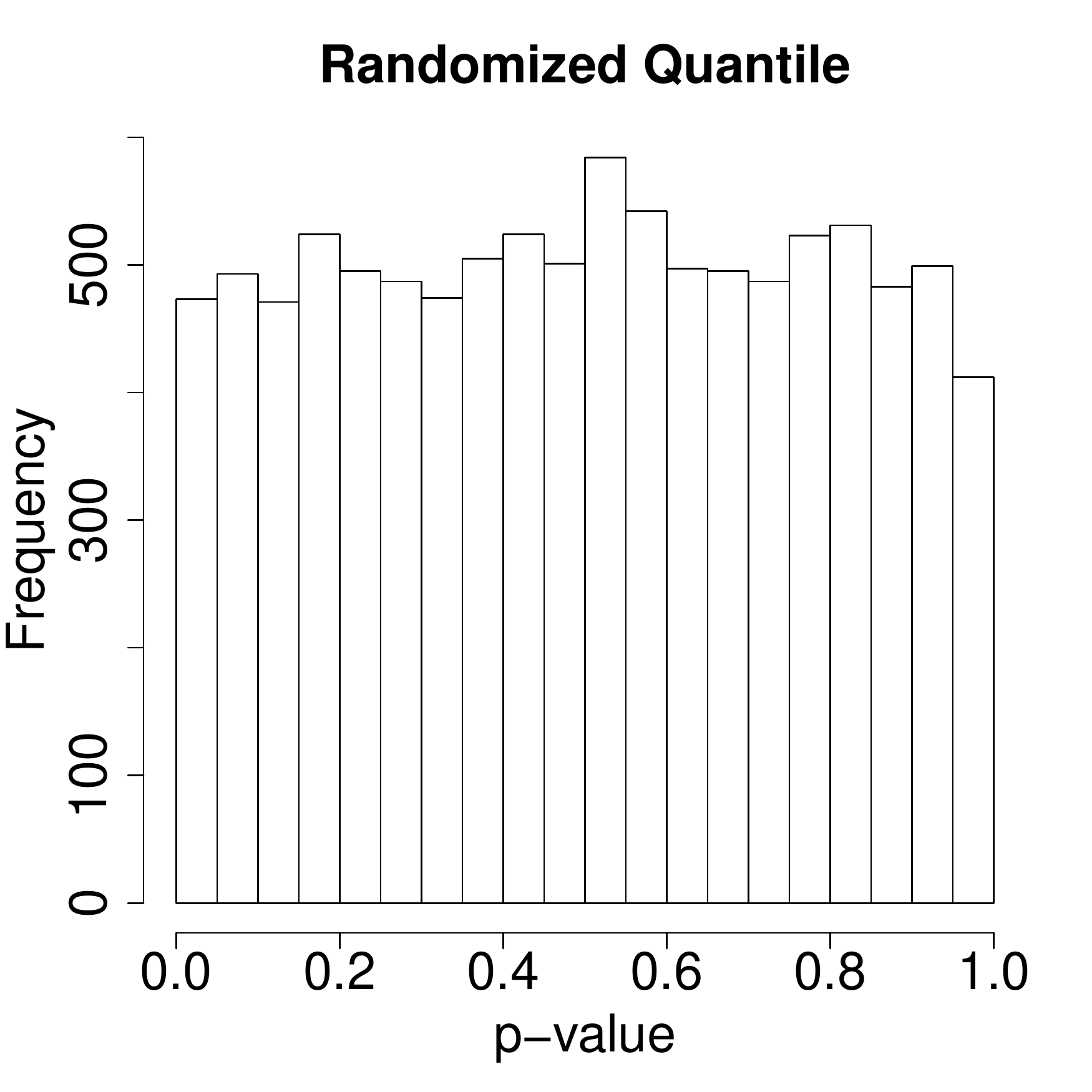}
\end{subfigure}\hspace{1em}

\begin{subfigure}[t]{0.30\textwidth}
\includegraphics[width=1.8in, height=1.3in,trim=0in 1.3in 0in 2.1in,clip]{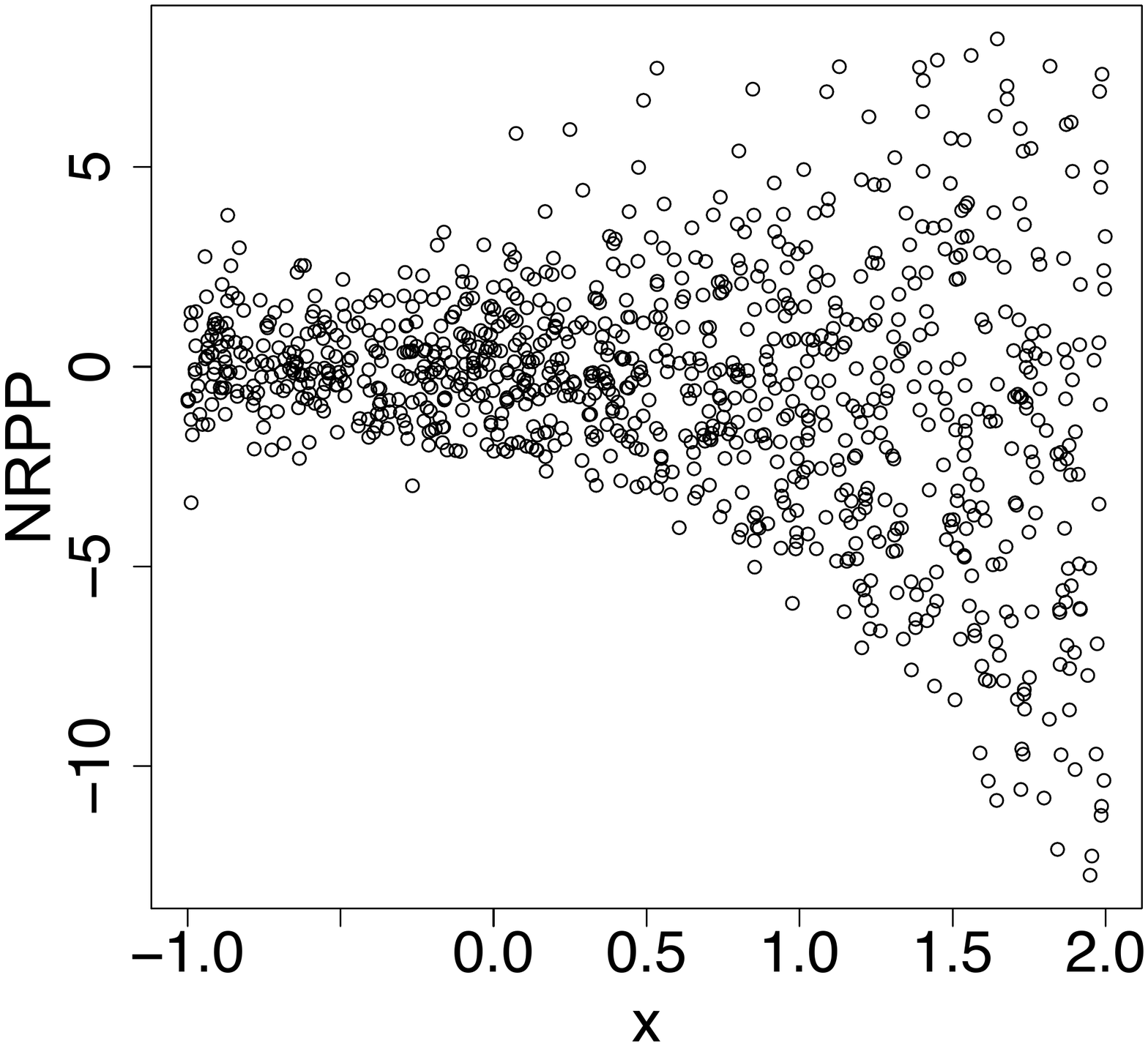}
\end{subfigure}\hspace{1em}
\begin{subfigure}[t]{0.30\textwidth}
\includegraphics[width=1.8in, height=1.3in,trim=0in 0.1in 0.1in 0.6in,clip]{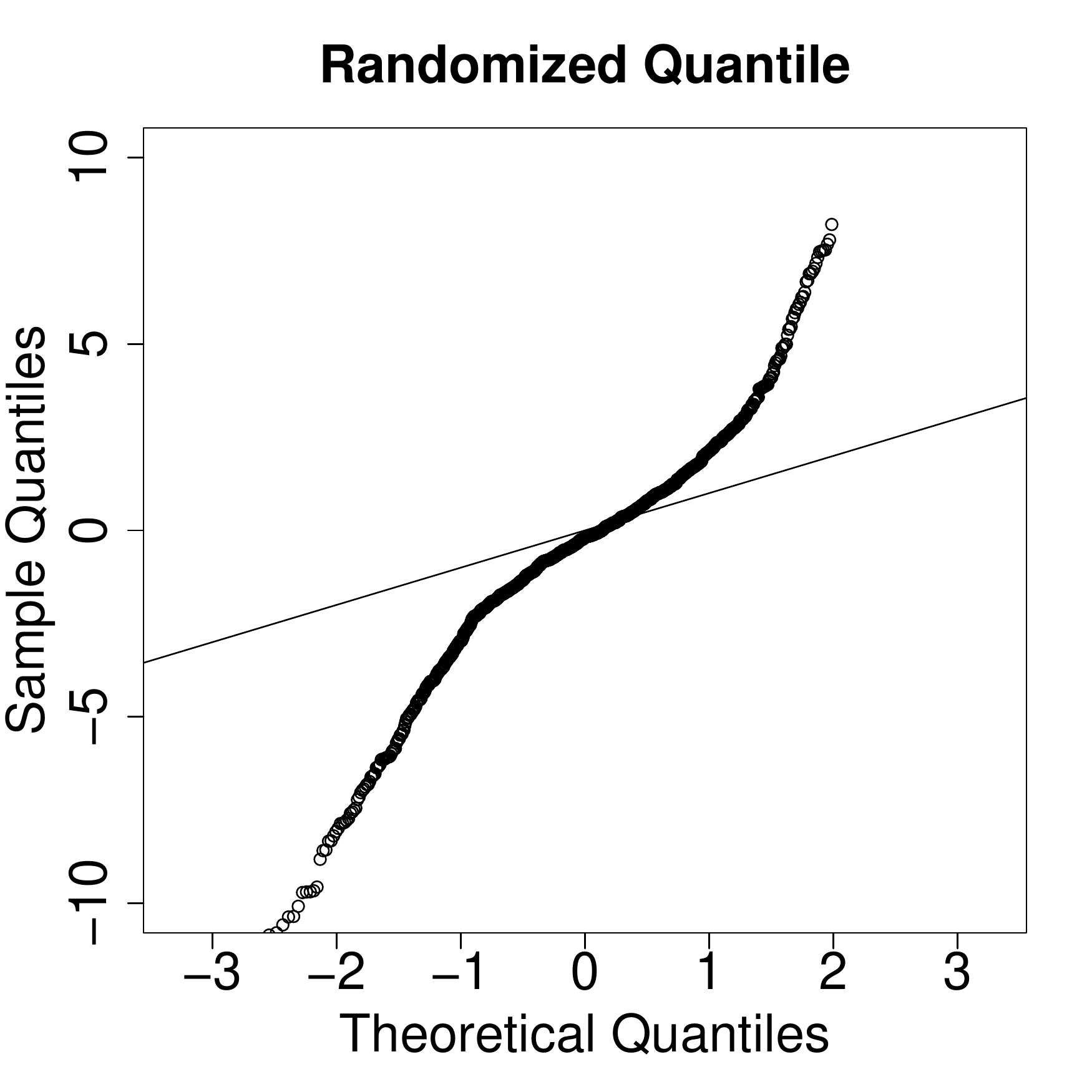} 
\end{subfigure}\hspace{1em}
\begin{subfigure}[t]{0.30\textwidth}
\includegraphics[width=1.8in, height=1.3in, trim=0in 0.1in 0.1in 0.6in, clip]{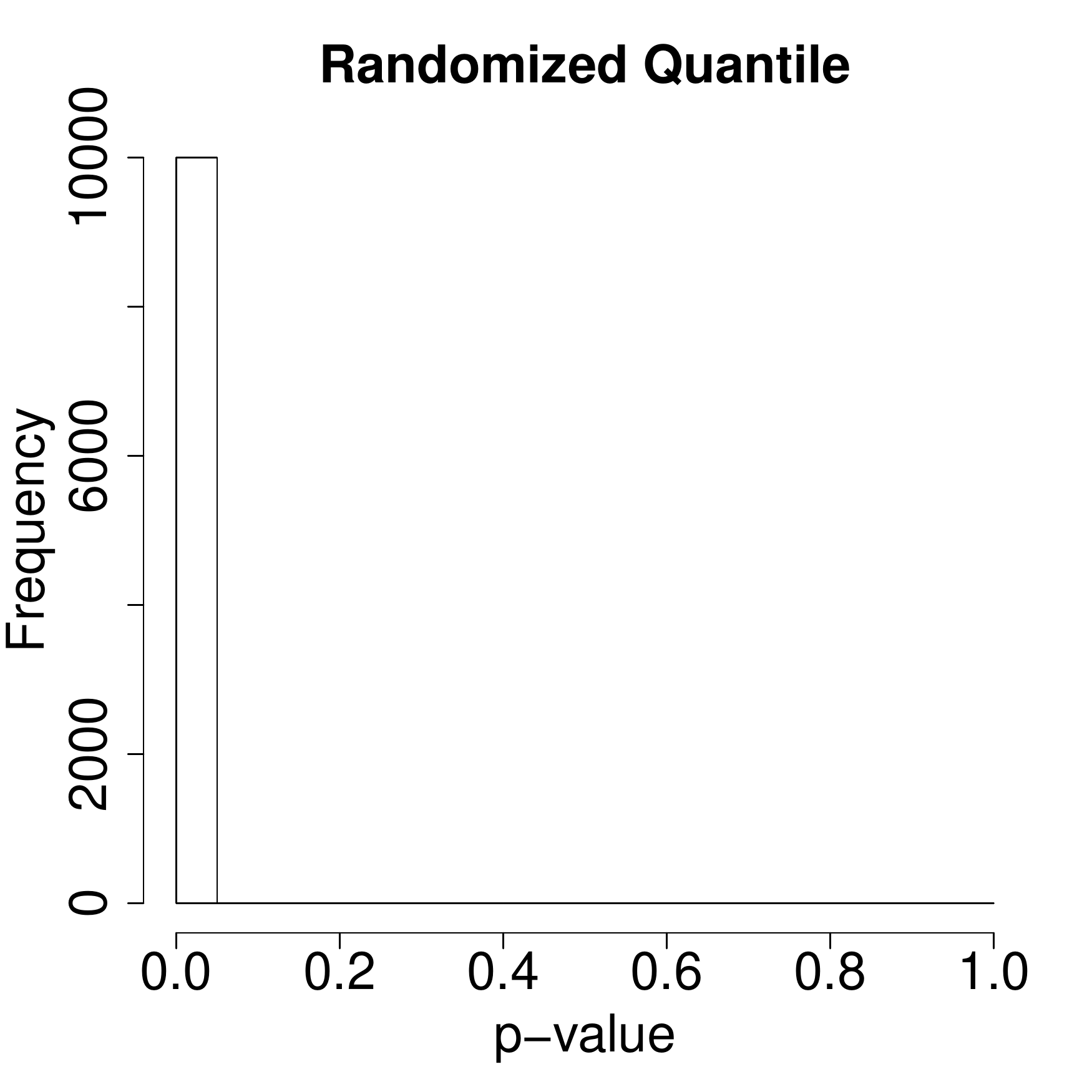} 
\end{subfigure}\hspace{1em}

\caption{\footnotesize Performance of the NRPPs in detecting over-dispersion of a sample dataset of size $n=1000$. The panels in the first row present the NRPPs for the true NB model. The panels in the second row present the NRPPs for the wrong Poisson model with the same mean structure as the true model. 
%The first two columns display the scatter plots and QQ plots of the NRPPs, respectively. The third column presents the histograms of the SW p-values for the NRPPs over 500 simulated datasets from the true model. 
\label{fig:nbpoissonresid}}
\end{figure}

The panels in the first column of Figure \ref{fig:nbpoissonresid} present the scatter plots of the NRPPs against the covariate under the true and wrong models. Under the true model, the residual plot is mostly bounded between -3 and 3 as standard normal variates without any visible trends; on the other hand, under the wrong model, all the residuals ``fan out'' from left to right, suggesting the presence of over-dispersion for increasing values of $x_{i}$.  The panels in the second column of Figure \ref{fig:nbpoissonresid} provide the QQ plots of NRPP residuals under the true and wrong models. Under the true model, the QQ plot aligns with the diagonal line, whereas under the wrong model, the QQ plot significantly deviates from the diagonal line, exhibiting a non-linear trend. The panels in the third column of Figure \ref{fig:nbpoissonresid} present the histograms of 500 SW p-values for testing the normality of the NRPPs under the true and wrong models; the SW p-values under the true model are nearly uniform while the SW p-values under the wrong model are highly distributed around 0. These results demonstrate that the overall GOF test by using the SW test for the NRPPs has a nominal-level probability of type I error and high statistical power for detecting over-dispersion.

\begin{figure}[htp]
\begin{subfigure}[t]{0.24\textwidth}
\subcaption{\centering \textbf{~~~~~NRPP}}
\includegraphics[width=\textwidth,trim=0in 0.1in 0.4in 0.6in,clip]{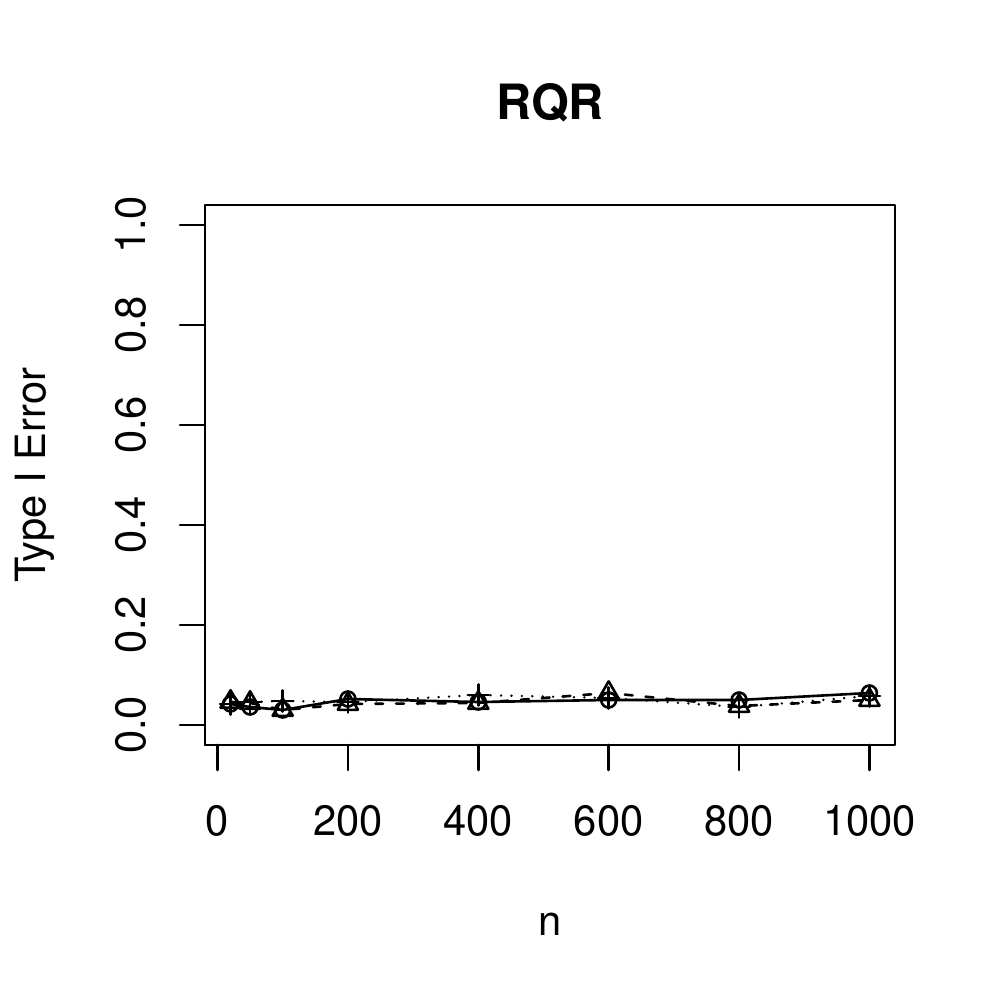}
\end{subfigure}
\begin{subfigure}[t]{0.24\textwidth}
\subcaption{\centering \textbf{~~~~~NMPP}}
\includegraphics[width=\textwidth,trim=0in 0.1in 0.4in 0.6in,clip]{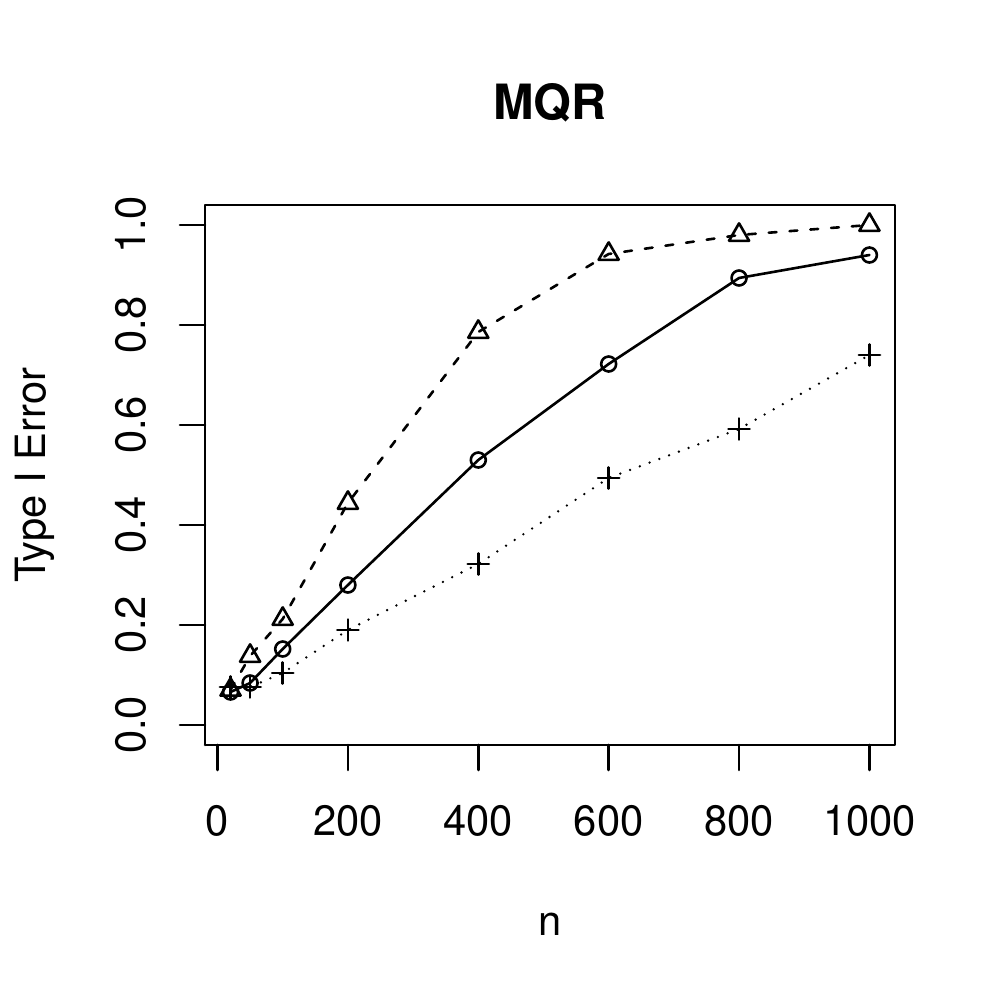}
\end{subfigure}
\begin{subfigure}[t]{0.24\textwidth}
\subcaption{\centering \textbf{~~~~~~Deviance}}
\includegraphics[width=\textwidth,trim=0in 0.1in 0.4in 0.6in,clip]{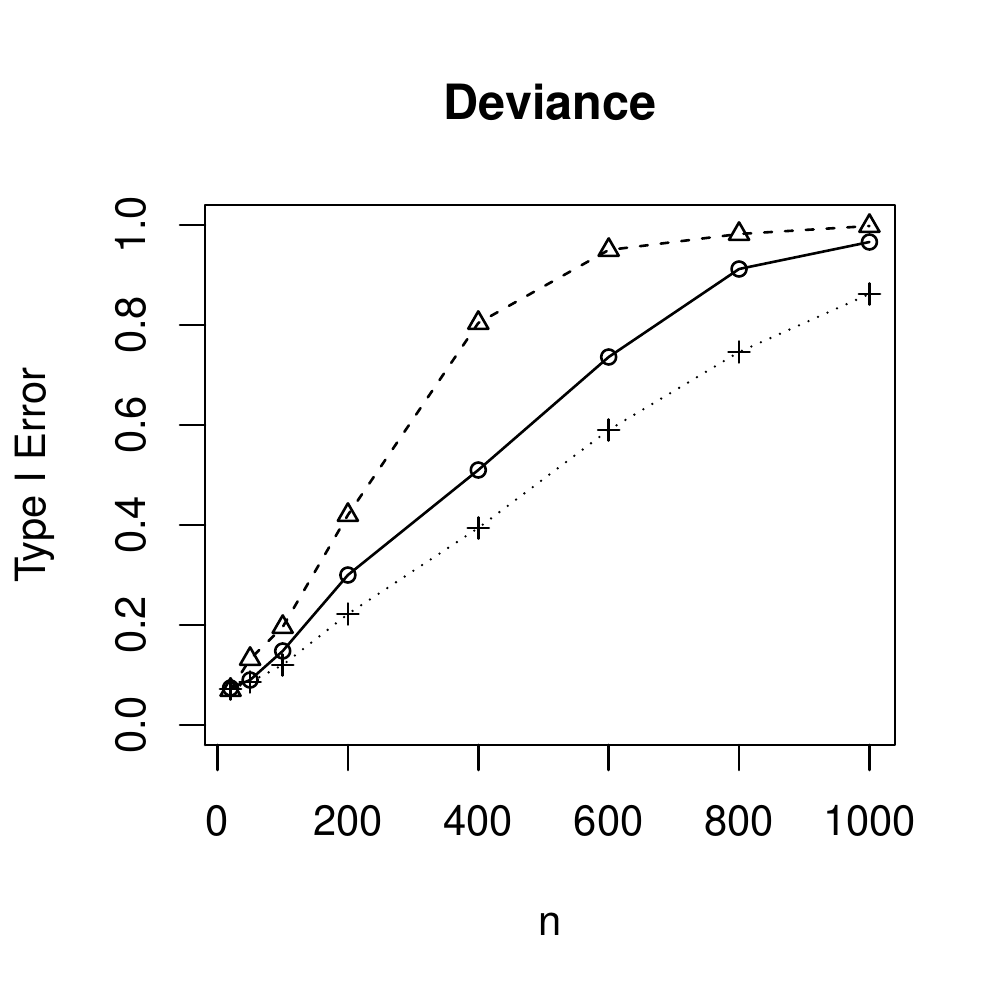}
\end{subfigure}
\begin{subfigure}[t]{0.24\textwidth}
\subcaption{\centering \textbf{~~~~~~Pearson}}
\includegraphics[width=\textwidth,trim=0in 0.1in 0.4in 0.6in,clip]{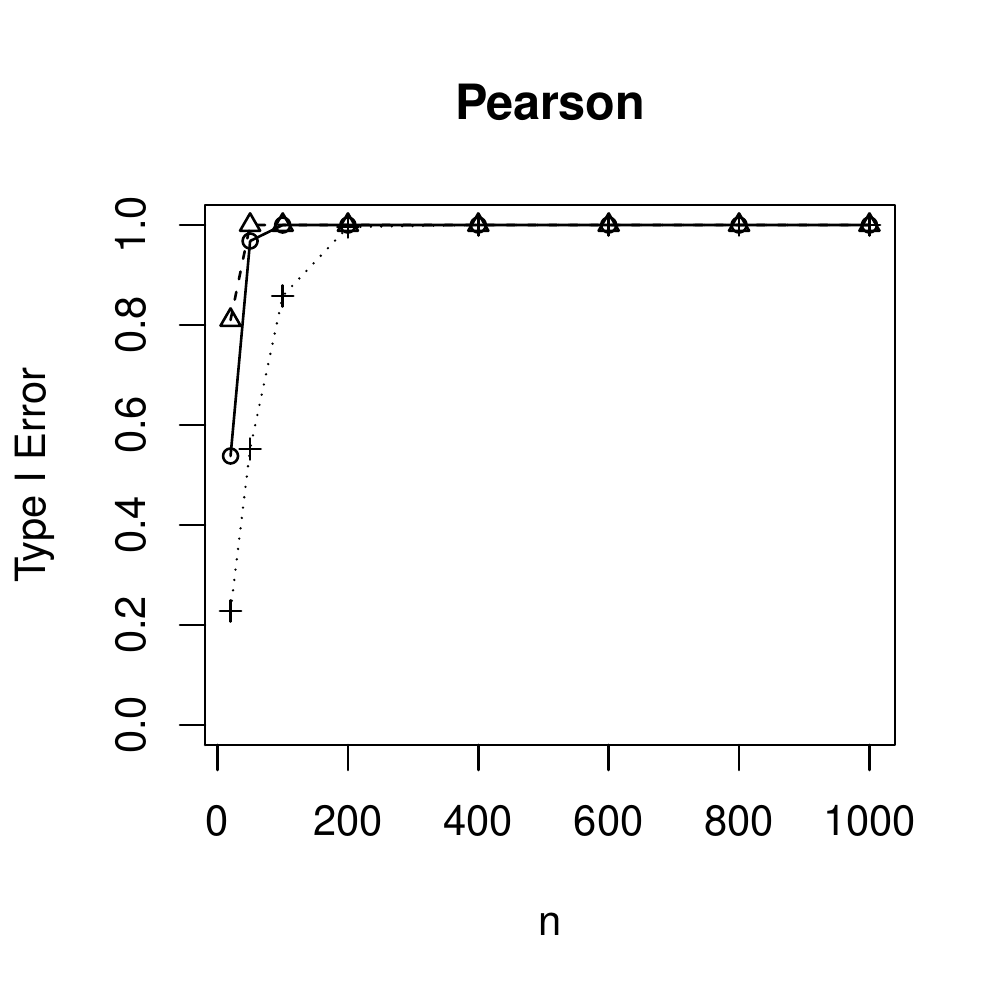}
\end{subfigure}

\begin{subfigure}[t]{0.24\textwidth}
\includegraphics[width=\textwidth,trim=0in 0.1in 0.4in 0.6in,clip]{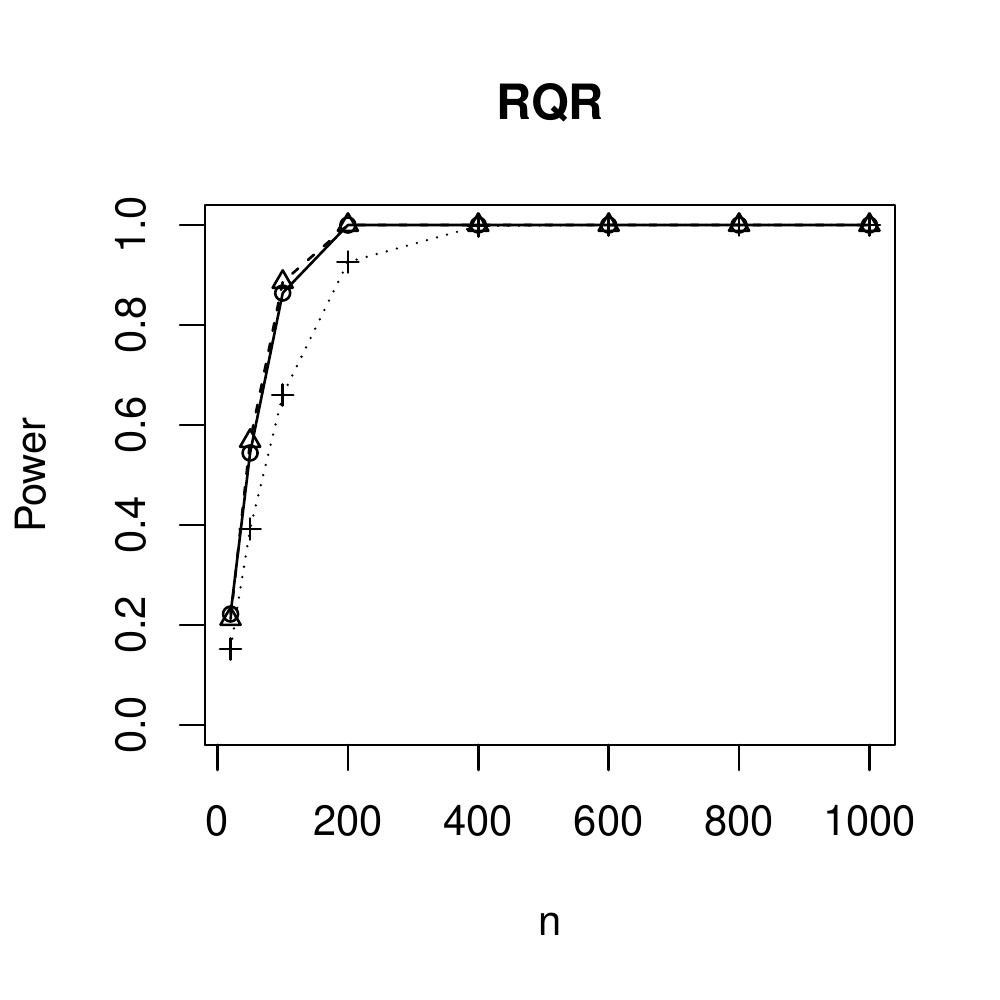}
\end{subfigure}
\begin{subfigure}[t]{0.24\textwidth}
\includegraphics[width=\textwidth,trim=0in 0.1in 0.4in 0.6in,clip]{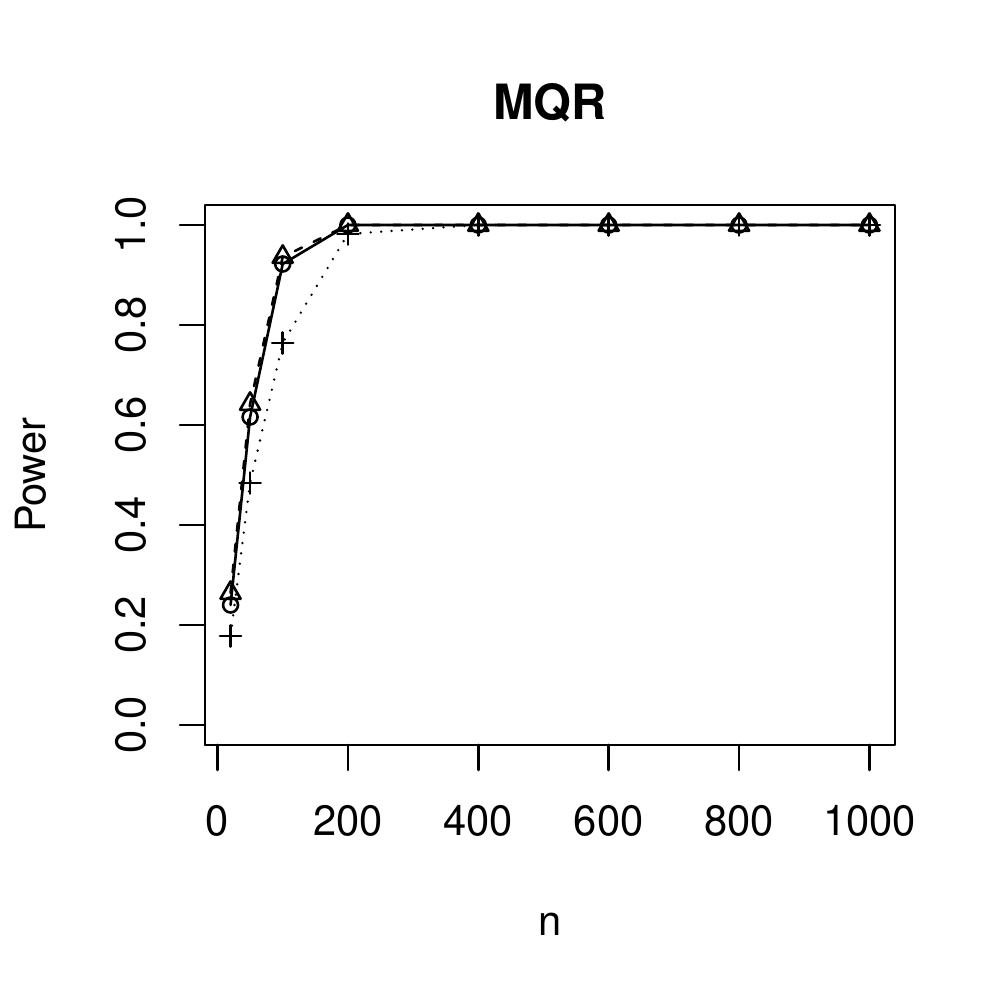} 
\end{subfigure}
\begin{subfigure}[t]{0.24\textwidth}
\includegraphics[width=\textwidth,trim=0in 0.1in 0.4in 0.6in,clip]{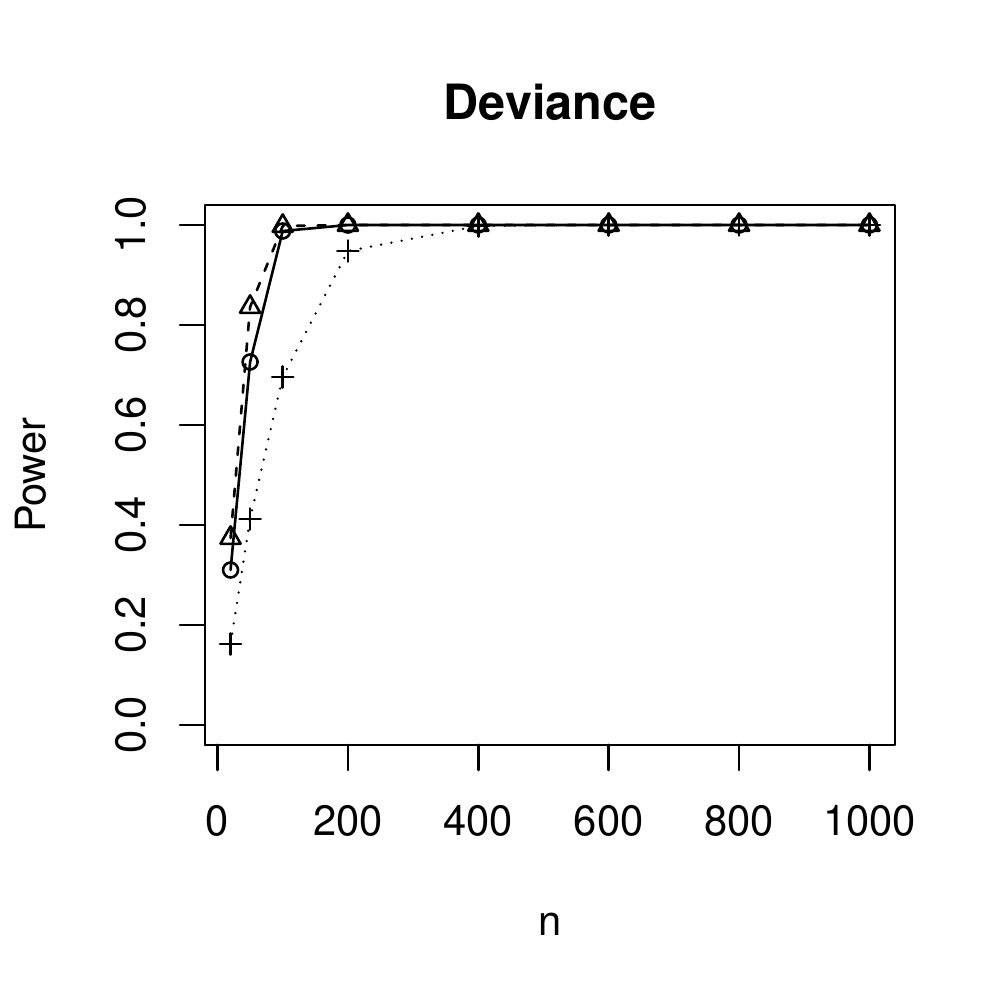} 
\end{subfigure}
\begin{subfigure}[t]{0.24\textwidth}
\includegraphics[width=\textwidth,trim=0in 0.1in 0.4in 0.6in,clip]{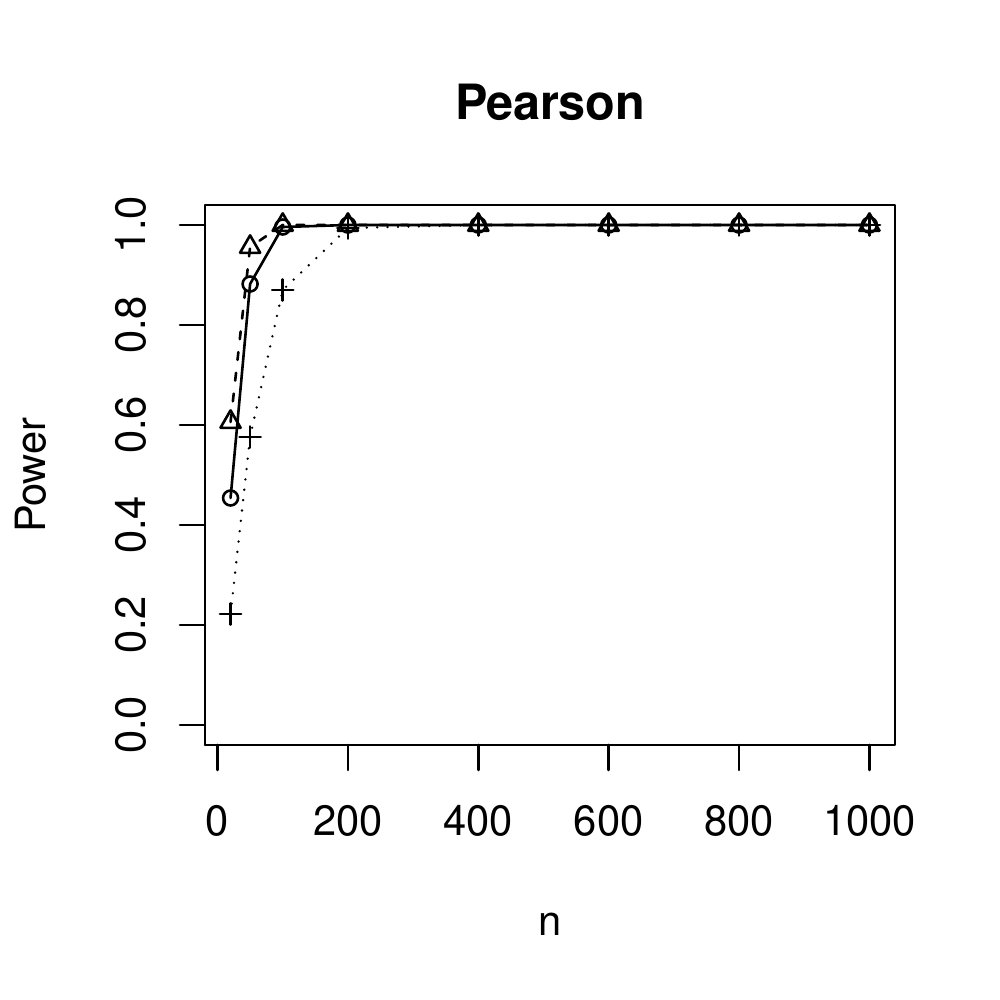} 
\end{subfigure}
\caption{\footnotesize Comparison of the type I errors and powers of the SW tests for the NRPPs, NMPPs, and deviance and Pearson residuals. Response variable is simulated from the true model at varying sample sizes and the over-dispersion parameters of $k=1$ (\protect\solidline \protect\markcircle), 2 (\protect\dashedline \protect\marktriangle) and 10 (\protect\dottedline $+$). True model: NB model with mean $\exp(\beta_0+\beta_1 x)$. Wrong model: Poisson model with mean $\exp(\beta_0+\beta_1 x)$.} \label{fig:poweranalysis_NBPois}
\end{figure}

In the power analysis, we increased the level of over-dispersion in the data by setting the dispersion parameter as $k=1, 2$ and 10. Figure \ref{fig:poweranalysis_NBPois} shows that the type I error rates of the SW test for the NRPPs remain at the nominal level 0.05 for all scenarios. In contrast, the type I error rates of the SW tests for the NMPPs, and deviance and Pearson residuals greatly exceed 0.05. Furthermore, the SW tests for the NRPPs are able to maintain high power in all scenarios as long as the sample size is sufficient. 

%\begin{landscape}
%\end{landscape}

\subsection{Detection of Zero-Inflation}\label{sec:4.3}
Finally, we conduct simulations to investigate the performance of the NRPPs in detecting zero-inflation in the data. We first simulate a covariate $x\sim \mathit{Uniform}(-1, 2)$ of size $n=1000$. Then, the response is simulated from a ZIP model as described in Section \ref{sec:zip}. The expected mean $\lambda_{i}$ of the Poisson component is set as $\lambda_i=\exp (\beta_0+\beta_1x_i)$, with $\beta_0=1$ and $\beta_1=2$, and percentage of excessive zeros $p_{i}$ is fixed at $30\%$ for all observations.  A Poisson model with the same expected mean $\lambda_i$ is fitted as a wrong model. 

\begin{figure}[htp]
\begin{subfigure}[t]{0.30\textwidth}
\subcaption{\centering \textbf{Residual plot}}
\includegraphics[width=1.8in, height=1.3in,trim=0in 1.3in 0.1in 2.0in,clip]{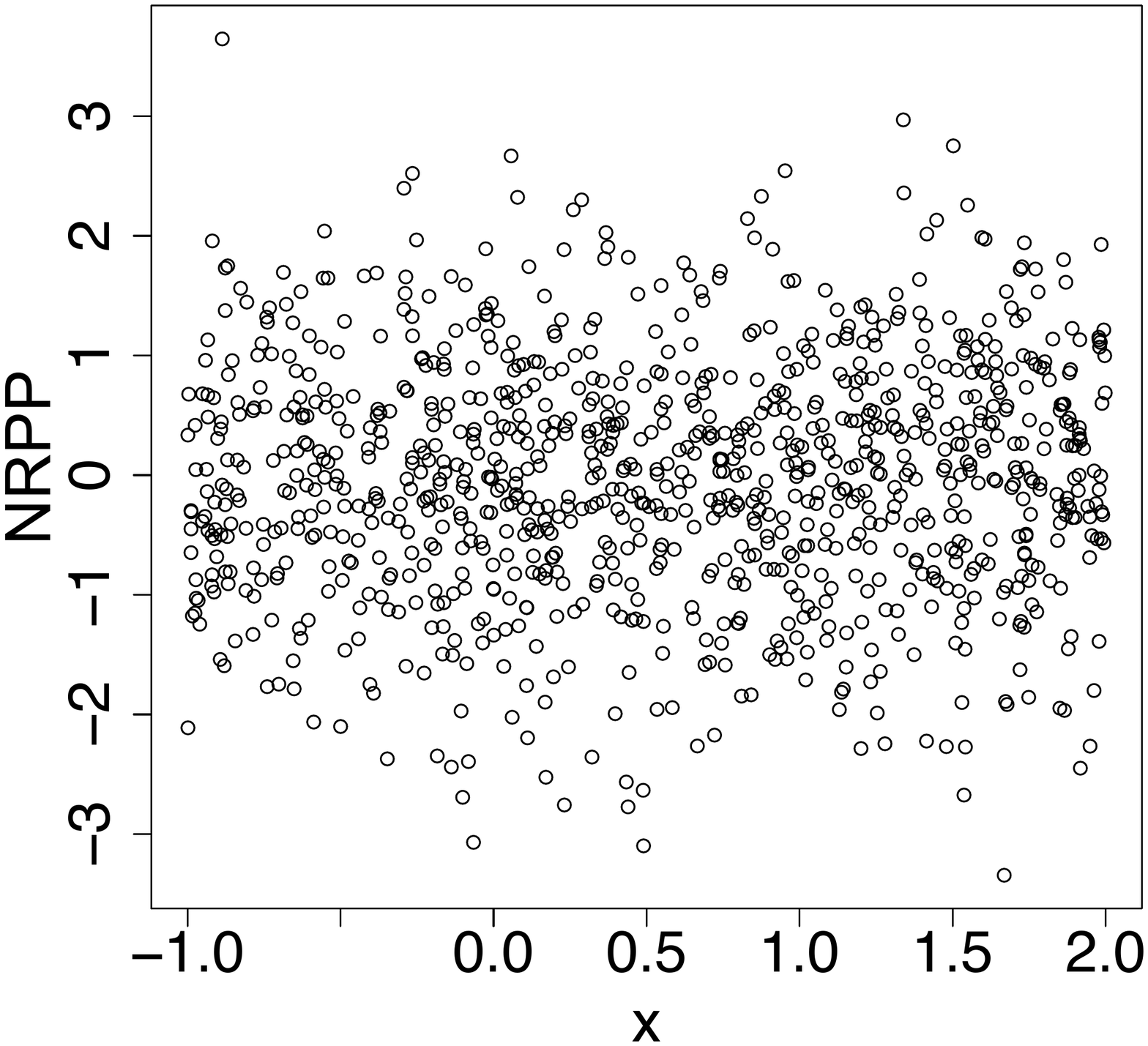}
\end{subfigure}\hspace{1em}
\begin{subfigure}[t]{0.30\textwidth}
\subcaption{\centering \textbf{QQ plot}}
\includegraphics[width=1.8in, height=1.3in,trim=0in 0.1in 0.1in 0.6in,clip]{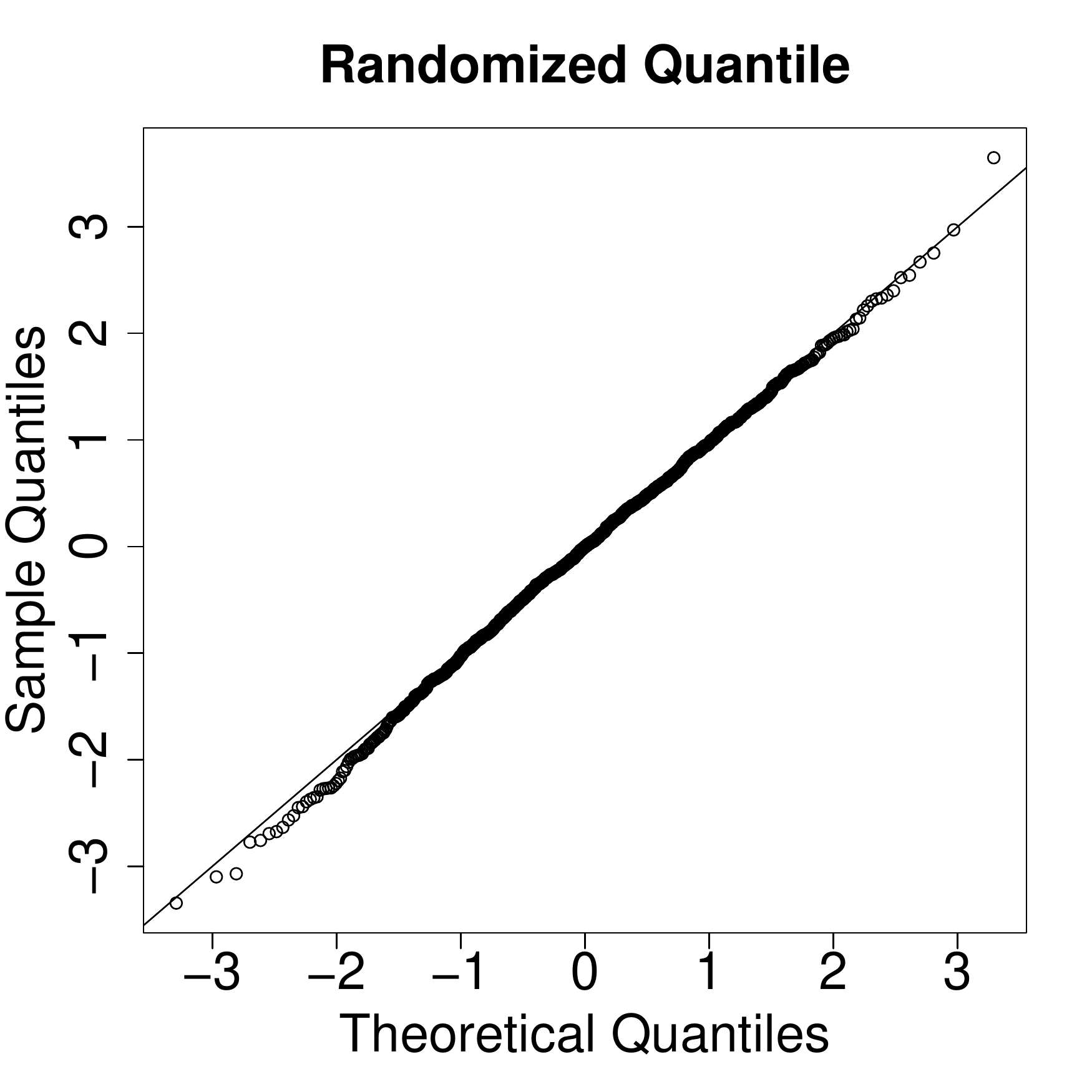}
\end{subfigure}\hspace{1em}
\begin{subfigure}[t]{0.30\textwidth}
\subcaption{\centering \textbf{SW p-values}}
\includegraphics[width=1.8in, height=1.3in,trim=0in 0.1in 0.1in 0.6in,clip]{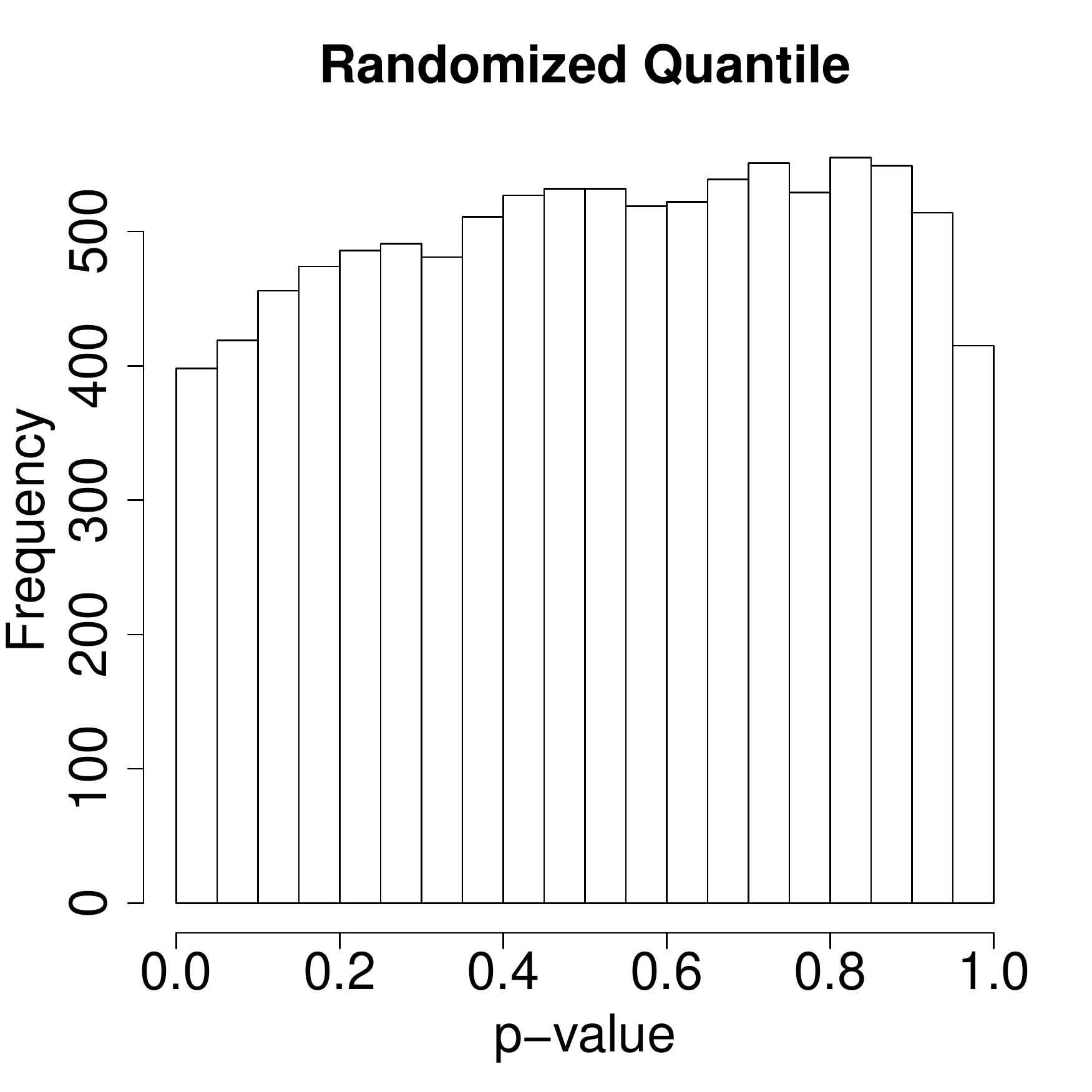}
\end{subfigure}\hspace{1em}

\begin{subfigure}[t]{0.30\textwidth}
\includegraphics[width=1.8in, height=1.3in,trim=0in 1.3in 0.1in 2.0in,clip]{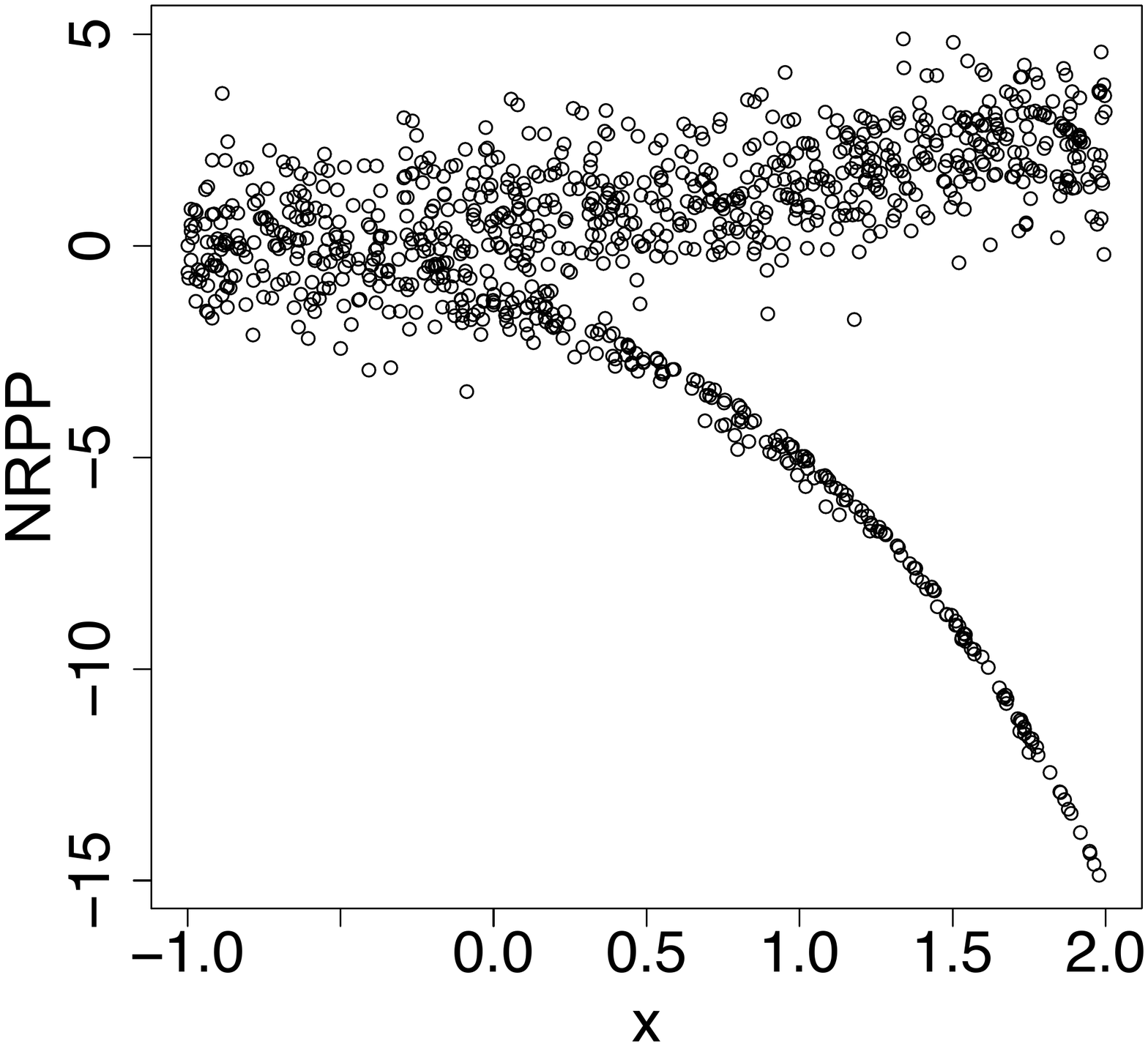}
\end{subfigure}\hspace{1em}
\begin{subfigure}[t]{0.30\textwidth}
\includegraphics[width=1.8in, height=1.3in,trim=0in 0.1in 0.1in 0.6in,clip]{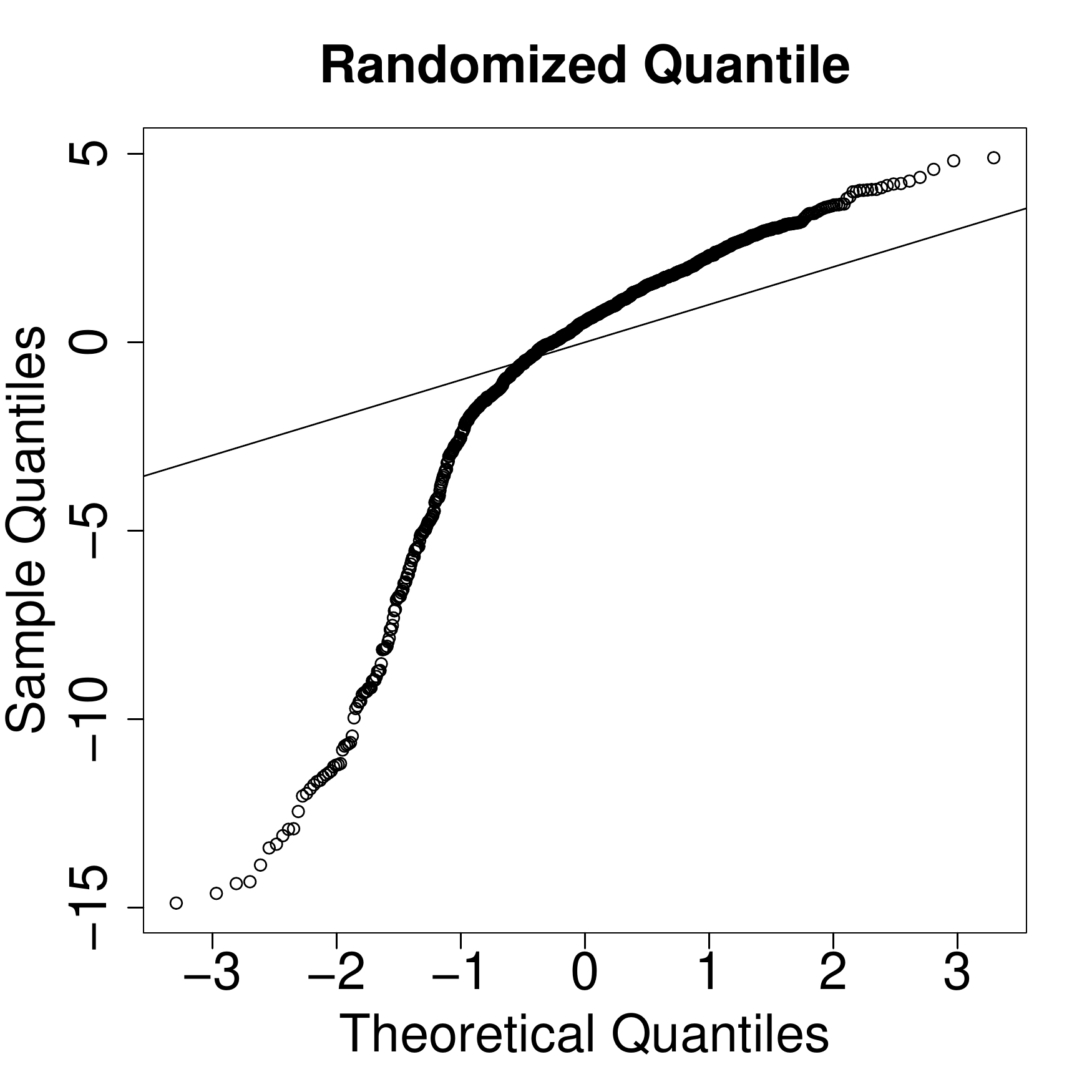} 
\end{subfigure}\hspace{1em}
\begin{subfigure}[t]{0.30\textwidth}
\includegraphics[width=1.8in, height=1.3in,trim=0in 0.1in 0.1in 0.6in,clip]{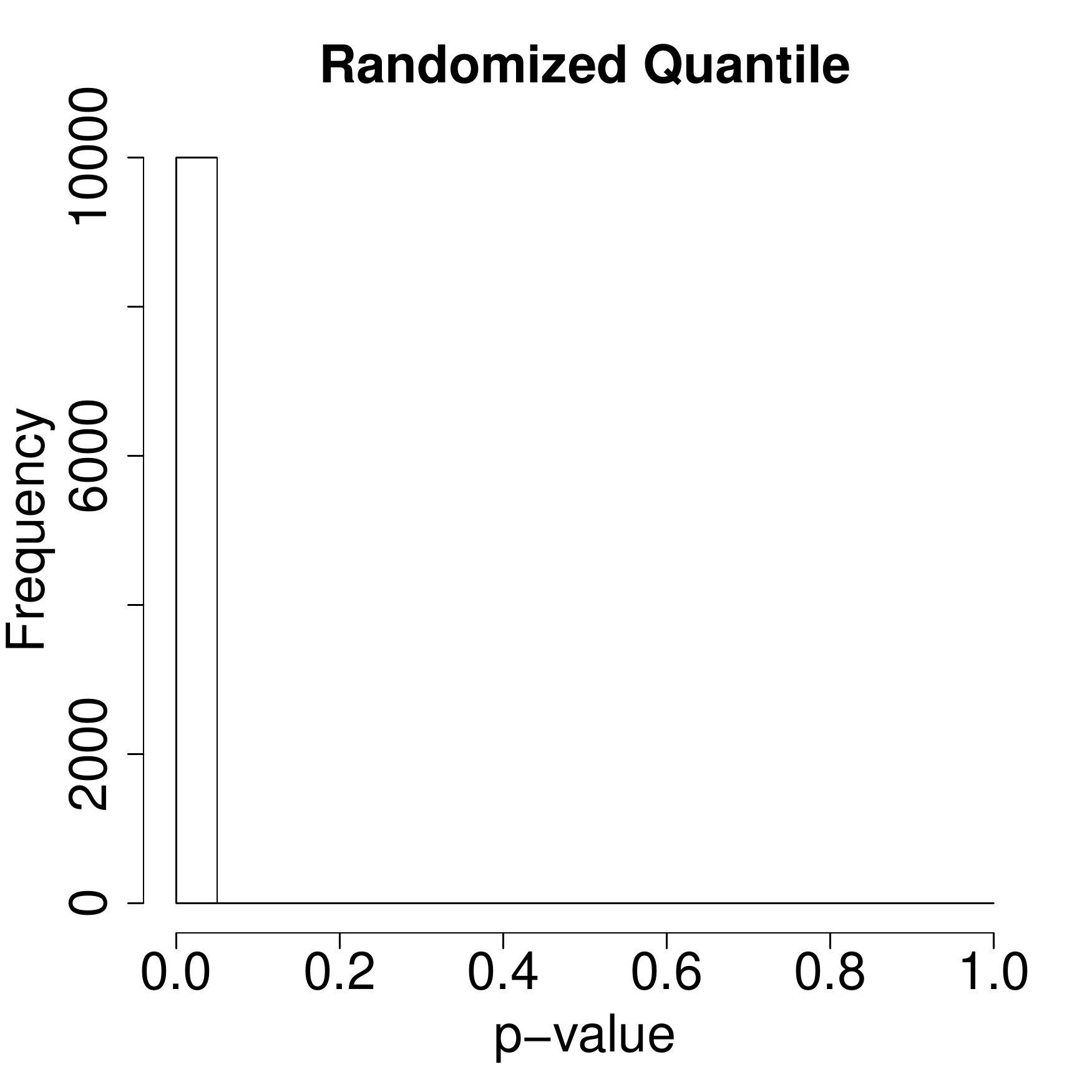} 
\end{subfigure}\hspace{1em}
\caption{\footnotesize Performance of the NRPPs in detecting zero-inflation of a sample dataset of size $n=1000$. The panels in the first row present the NRPPs for the fitted true model: ZIP model. The panels in the second row present the NRPPs for the fitted wrong Poisson model.  
%The first two columns display the scatter plots and QQ plots of the NRPPs, respectively. The third column present the histograms of the SW p-values for the NRPPs over 500 simulated datasets from the true model. 
\label{fig:zipresid}}
\end{figure}

The panels in the top row of Figure \ref{fig:zipresid} display the residual plot against the covariate, QQ plot and histogram of 500 SW p-values of the NRPPs under the true model.  The results meet the ideal expectations under the true model: Residual plot is mostly bounded between -3 and 3 as standard normal variates without any unusual patterns; QQ plot aligns with the diagonal line, and the histogram of the SW p-values are nearly uniform. The panels in the bottom row of Figure \ref{fig:zipresid} present the corresponding plots of the NRPPs resulted from fitting the wrong model. In this residual plot, a clear separation of the NRPPs is observed from the residuals associated with the zero responses, which may be typical for zero-inflated data. This QQ plot shows observable deviations from the diagonal line with a nonlinear trend present in the lower tail probably due to excessively small residuals associated with the zero responses. This histogram of SW p-values is highly distributed near 0, indicating that the wrong model will be rejected most of the times with a small threshold.  %Hence, the overall GOF test via SW tests for the NRPPs demonstrates that the type I error rates stay at the nominal level (0.05) and great statistical power in detecting zero-inflation in the dataset. 

\begin{figure}[H] 
\begin{subfigure}[t]{0.24\textwidth}
\subcaption{\centering \textbf{~~~~~NRPP}}
\includegraphics[width=\textwidth,trim=0in 0.1in 0.4in 0.6in,clip]{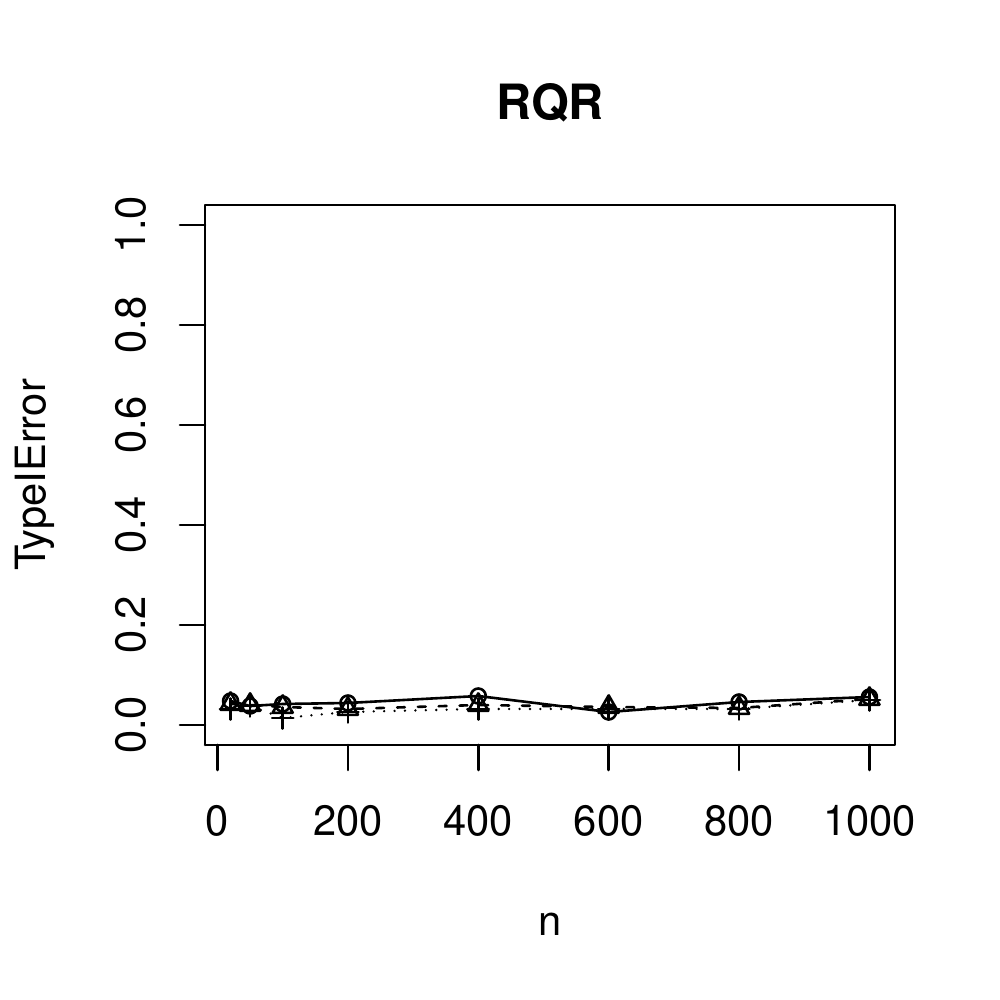}
\end{subfigure}
\begin{subfigure}[t]{0.24\textwidth}
\subcaption{\centering \textbf{~~~~~NMPP}}
\includegraphics[width=\textwidth,trim=0in 0.1in 0.4in 0.6in,clip]{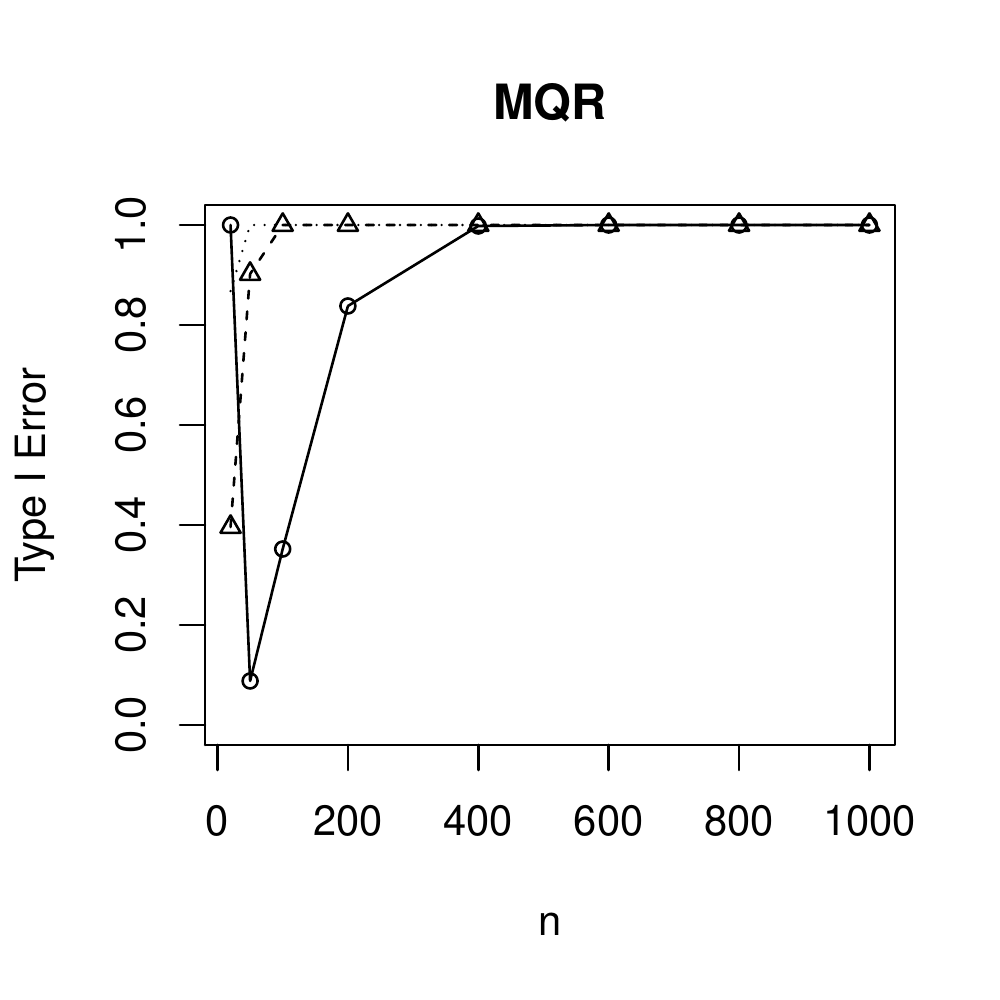}
\end{subfigure}
\begin{subfigure}[t]{0.24\textwidth}
\subcaption{\centering \textbf{~~~~~Deviance}}
\includegraphics[width=\textwidth,trim=0in 0.1in 0.4in 0.6in,clip]{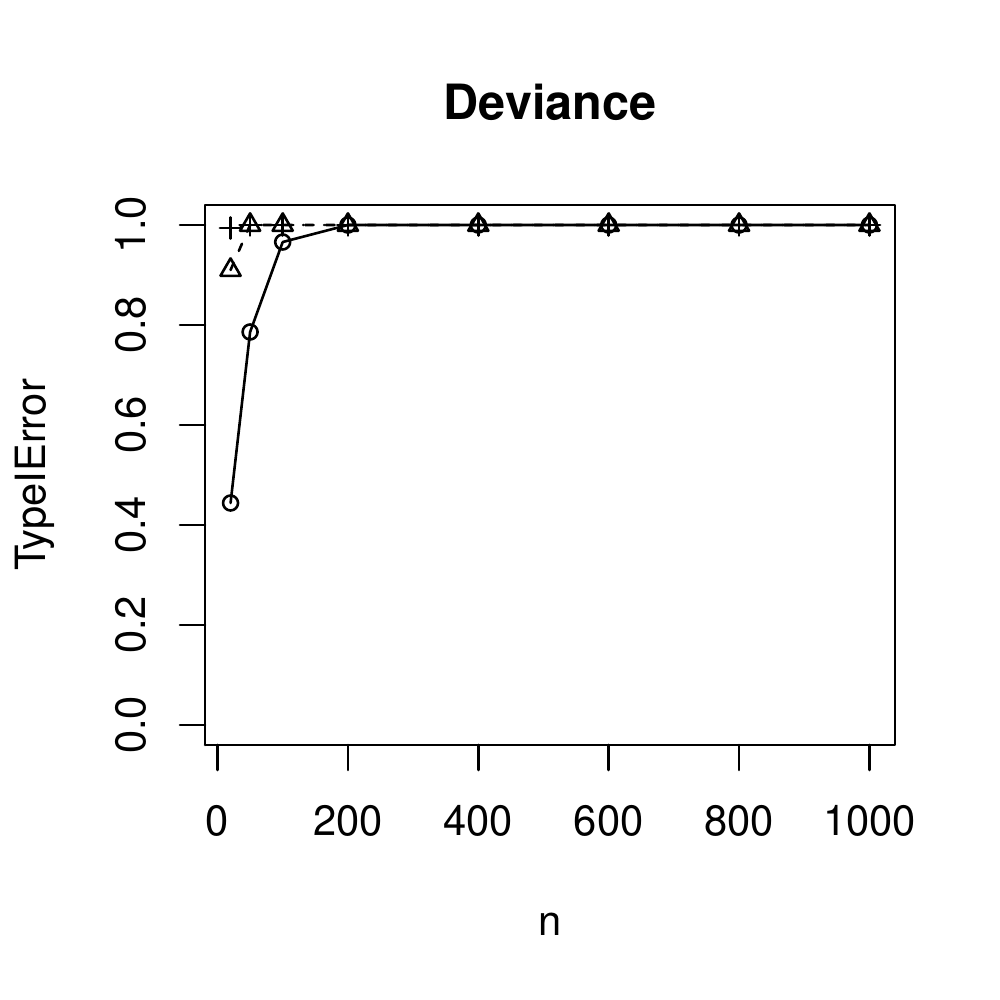}
\end{subfigure}
\begin{subfigure}[t]{0.24\textwidth}
 \subcaption{\centering \textbf{~~~~~Pearson}}
\includegraphics[width=\textwidth,trim=0in 0.1in 0.4in 0.6in,clip]{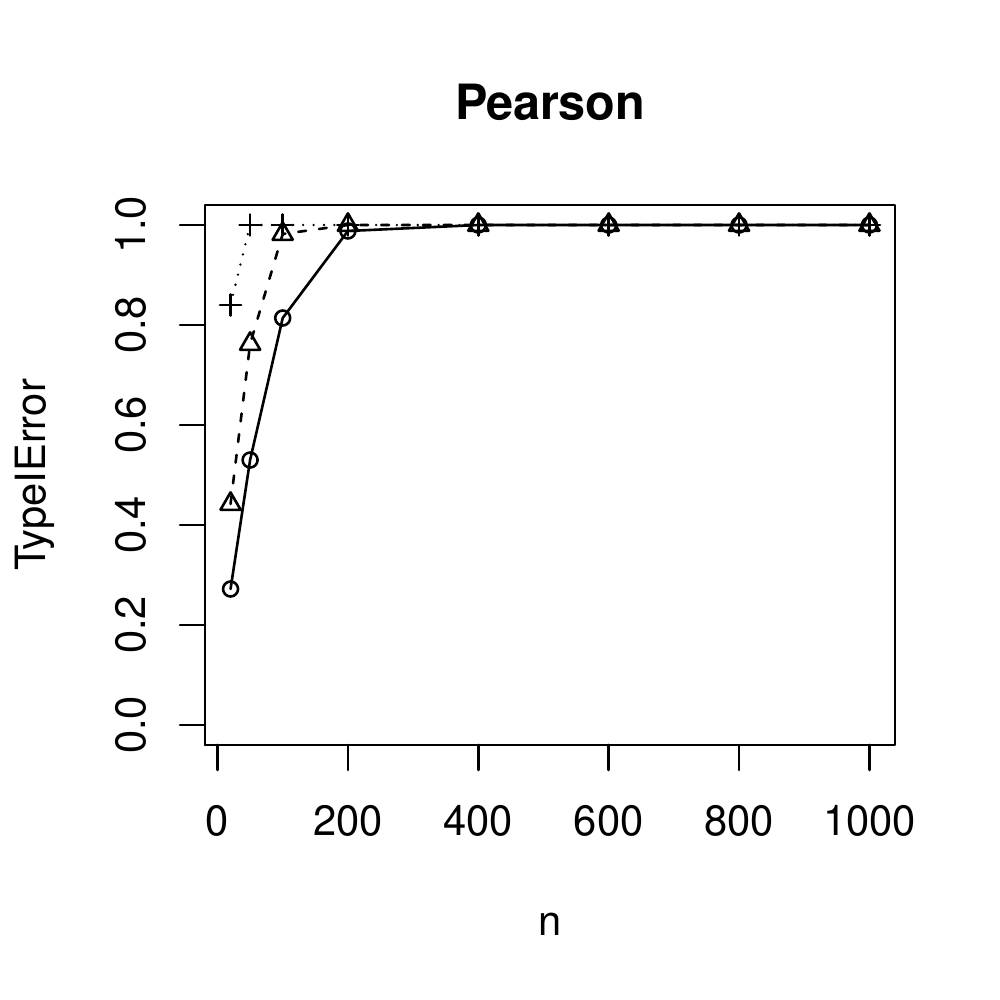}
\end{subfigure}

\begin{subfigure}[t]{0.24\textwidth}
\includegraphics[width=\textwidth,trim=0in 0.1in 0.4in 0.6in,clip]{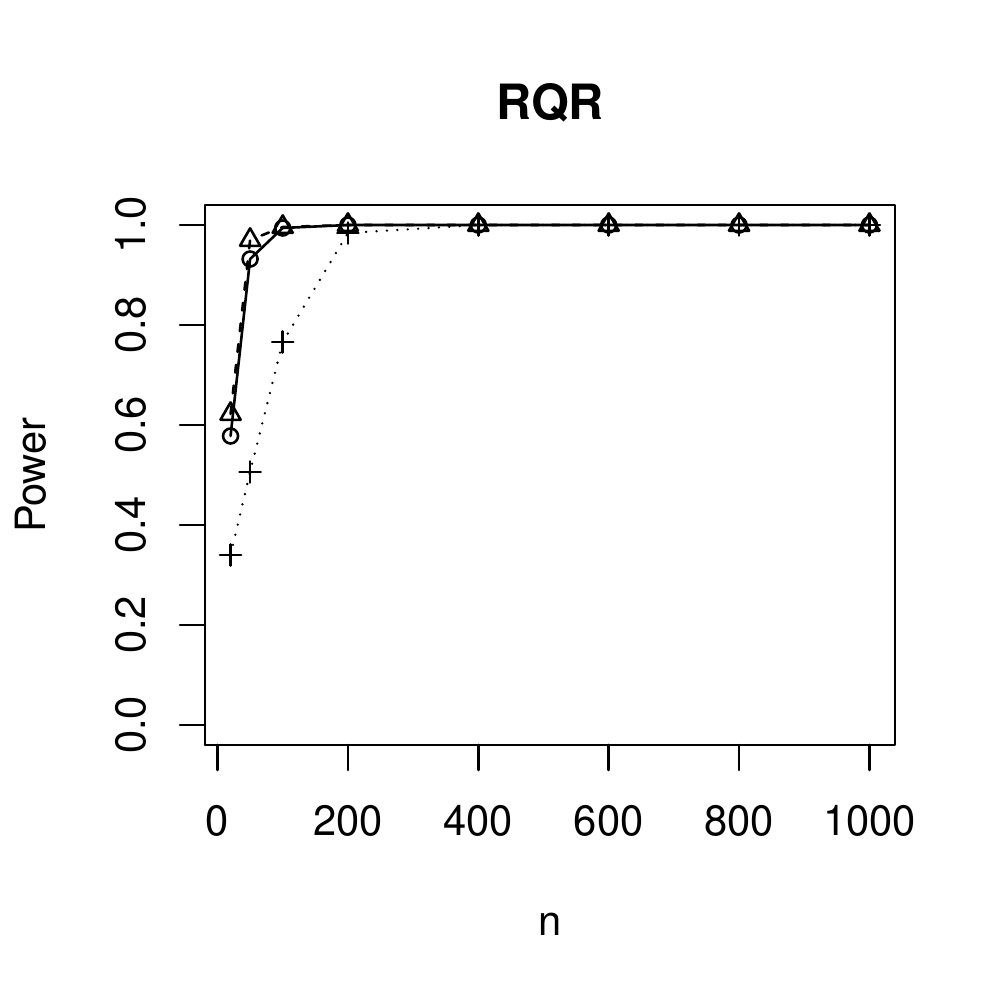}
\end{subfigure}
\begin{subfigure}[t]{0.24\textwidth}
\includegraphics[width=\textwidth,trim=0in 0.1in 0.4in 0.6in,clip]{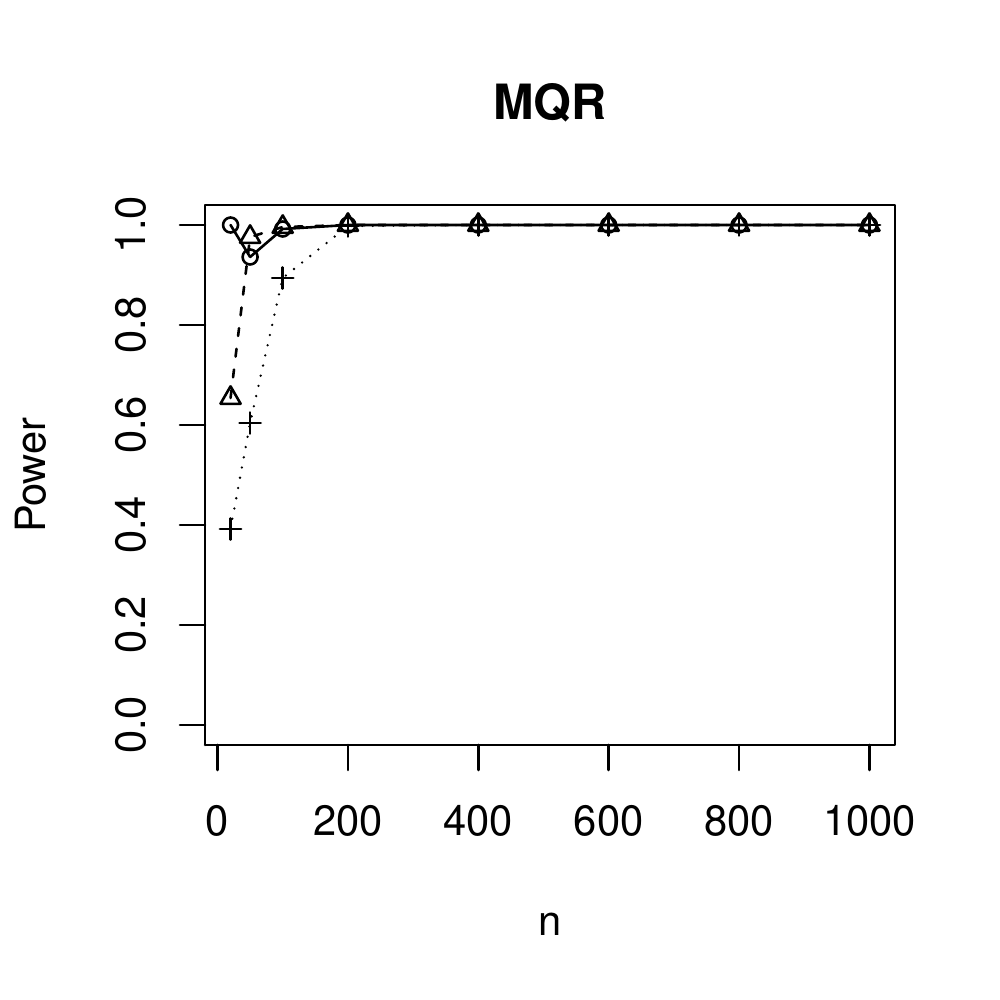} 
\end{subfigure}
\begin{subfigure}[t]{0.24\textwidth}
\includegraphics[width=\textwidth,trim=0in 0.1in 0.4in 0.6in,clip]{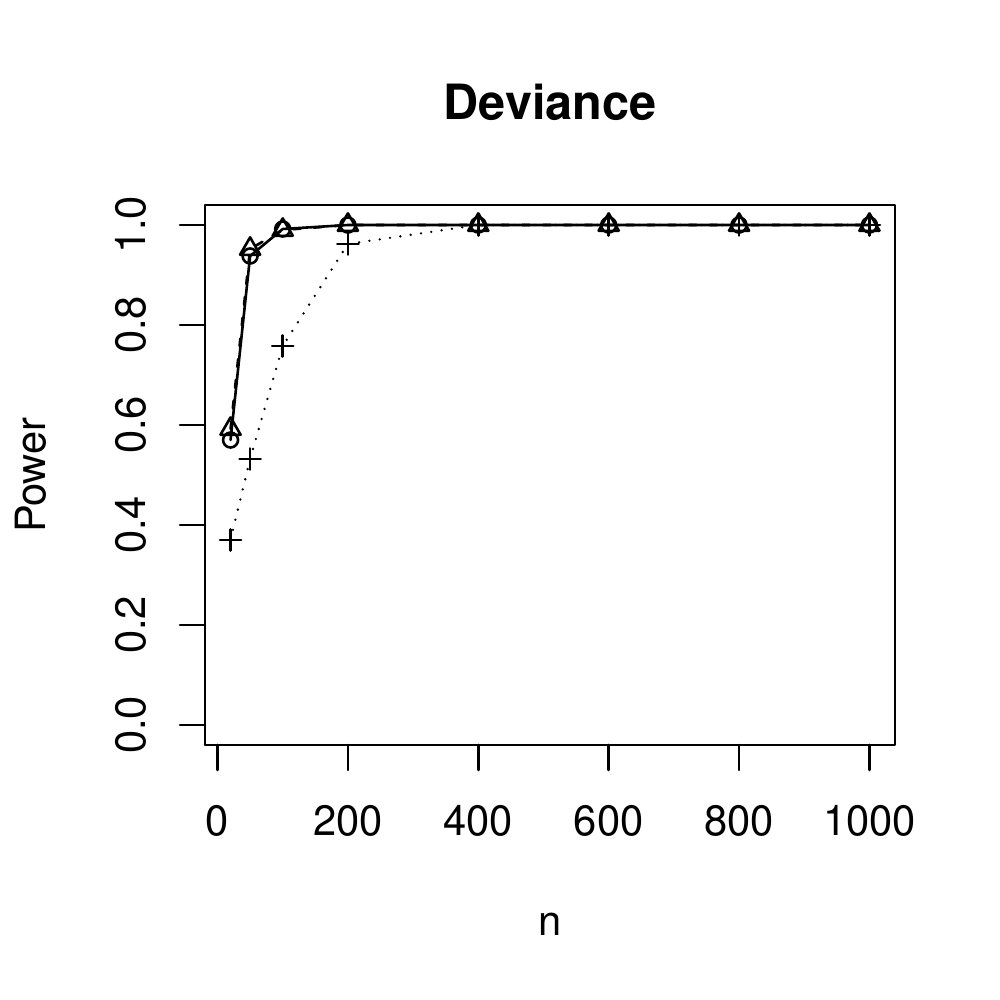} 
\end{subfigure}
\begin{subfigure}[t]{0.24\textwidth}
\includegraphics[width=\textwidth,trim=0in 0.1in 0.4in 0.6in,clip]{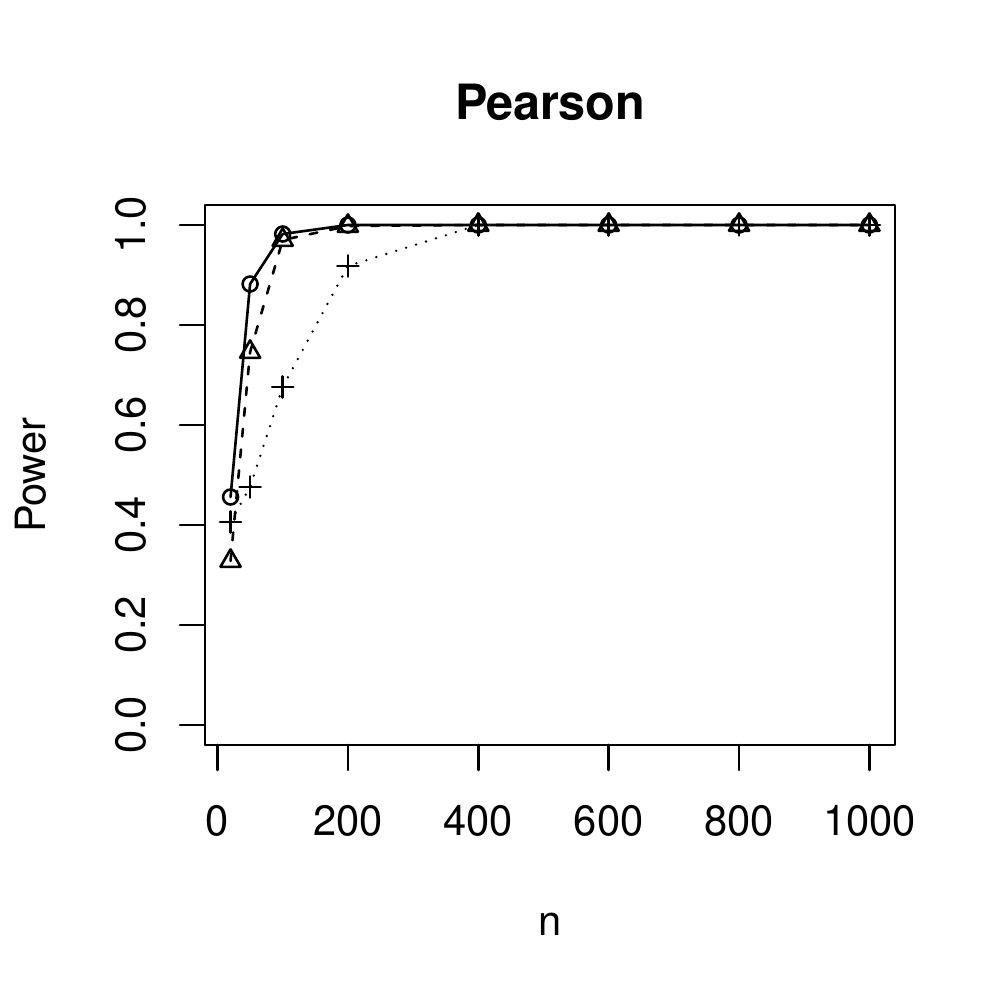} 
\end{subfigure}
\vspace*{-0.1in}
\caption{\footnotesize Comparison of the type I errors and powers of the SW tests for the NRPPs, NMPPs, and deviance and Pearson residuals. Response variable is simulated from the true model at varying sample sizes and percentages of excessive zeroes of $p=10\% $ (\protect\solidline \protect\markcircle), $30\%$ (\protect\dashedline \protect\marktriangle) and $50\%$ (\protect\dottedline $+$). True model: ZIP model with mean $\exp(\beta_0+\beta_1 x)$. Wrong model: Poisson model with mean $\exp(\beta_0+\beta_1 x)$. } \label{fig:poweranalysis_ZIP}
\end{figure}

In the power analysis, we set the probability of generating excessive zeros, $p_{i}$, as 0.1, 0.3 and 0.5 for various sample sizes. Figure \ref{fig:poweranalysis_ZIP} shows that the type I error rates of the NRPPs remain at the nominal level 0.05 for all scenarios. In contrast, the type I error rates of the NMPPs, and deviance and Pearson residuals greatly exceed the 0.05 threshold. In all scenarios, the NRPPs demonstrate high power even in the presence of small sample sizes.

%\begin{landscape}
%\end{landscape}

\vspace*{-10pt}

%%%%%%%%%%%%%%%%%%%%%%%%%%%%%%%%%%%%%%%%%%%%%%%%%%%%
\section{Application}\label{sec:application}
%%%%%%%%%%%%%%%%%%%%%%%%%%%%%%%%%%%%%%%%%%%%%%%%%%%%

In this section, we apply the NRPP approach to examine the GOF of four non-normal regression models fitted to the National Medical Expenditure Survey (NMES) \citep{Deb1997}, a large dataset on 4406 individuals surveying the demand of health care amongst the elderly in the United States. The response variable considered in this study is the number of emergency department (ED) visits. The covariates considered include demographic characteristics (e.g., age, race, sex, marital status, education and region), socioeconomic variables (e.g., family income, employment status, supplementary private insurance status and public insurance status) and health measures (e.g., self-perceived health, the number of chronic conditions and a measure of disability status).  

Over 81\% of the patient-year records were zero, implying that the majority of patients did not make any ED visits during the year of the study. The number of nonzero visits ranged from 1 to 12, with only $5\%$ having more than one visit in the study year. Due to the high evidence of over-dispersion and/or excessive zero counts in this dataset, we consider fitting Poisson, NB, ZIP and ZINB regression models (described in Section \ref{sec:zip}). Using backward elimination with a $5\%$ statistical significance level, all final models included the following covariates: The number of chronic conditions, self-perceived health (excellent vs. poor; average vs. poor), limited activities of daily living (yes vs. no) and the number of years of education. In addition to those covariates, black race was significantly associated with increased ED use for the Poisson and ZIP models, but not for the NB and ZINB models. Table \ref{tab:NBfit} contains the regression results of these models. It is observed that the standard errors of the estimated regression coefficients for the NB and ZINB models are all larger relative to Poisson and ZIP models, indicating that the choice of model distribution has a significant impact on assessing covariate effects.
This discrepancy highlights the importance of examining the model GOF; that is, the validity of statistical inferences depend on the correctness of a fitted model. 

%The binary components of the ZIP and ZINB models only included the intercept terms since no covariates were statistically significant at the $5\%$ level after backward elimination. The intercept terms were also not significantly different from zeros with the p-values for ZIP and ZINB being 0.109 and 0.828, respectively. 

\begin{table}[h] 
\centering
\caption{Estimated regression coefficients for the Poisson, NB, ZIP and ZINB models fitted to the National Medical Expenditure Survey. Standard errors are given in parentheses. } 
{\footnotesize%\scriptsize
\begin{tabular}{l|ccccc} 
\hline
Variables & Poisson &NB &ZIP &ZINB\\ \hline
Black vs. others &~0.188(0.085)*~~~& $-$ &~0.300(0.097)*~~~& $-$ \\
Chronic conditions &~0.221(0.020)** &~0.217(0.026)**&~0.216(0.023)**&~0.217(0.027)**\\
Self-perceived health&&&&\\
\, \, Excellent vs. poor &-1.093(0.190)**&-1.089(0.216)**&-1.028(0.207)**&-1.089(0.216)**\\
\, \, Average vs. poor &-0.505(0.074)**&-0.478(0.100)**&-0.451(0.088)**&-0.478(0.101)**\\
Limited daily activities &~0.453(0.070)**&~0.464(0.087)**&~0.426(0.077)**&~0.464(0.087)**\\
Years of education&~-0.017(0.008)*~~~&~-0.023(0.010)*~~~&~-0.019(0.009)*~~~&~-0.023(0.100)*~~~\\
\hline 
\multicolumn{5}{l}{\textsuperscript{a}Significance at the 5$\%$ and 1$\%$ level is indicated with $*$ and $**$, respectively. }
\end{tabular}}
\label{tab:NBfit}
\end{table}

\begin{figure}[p]
%\centering 
\begin{subfigure}[t]{0.3\textwidth}
\centering 
\subcaption{\centering \textbf{~~~~Poisson}}
\includegraphics[width=1.8in, height=1.5in,trim=0in 0.2in 0in 0.6in,clip]{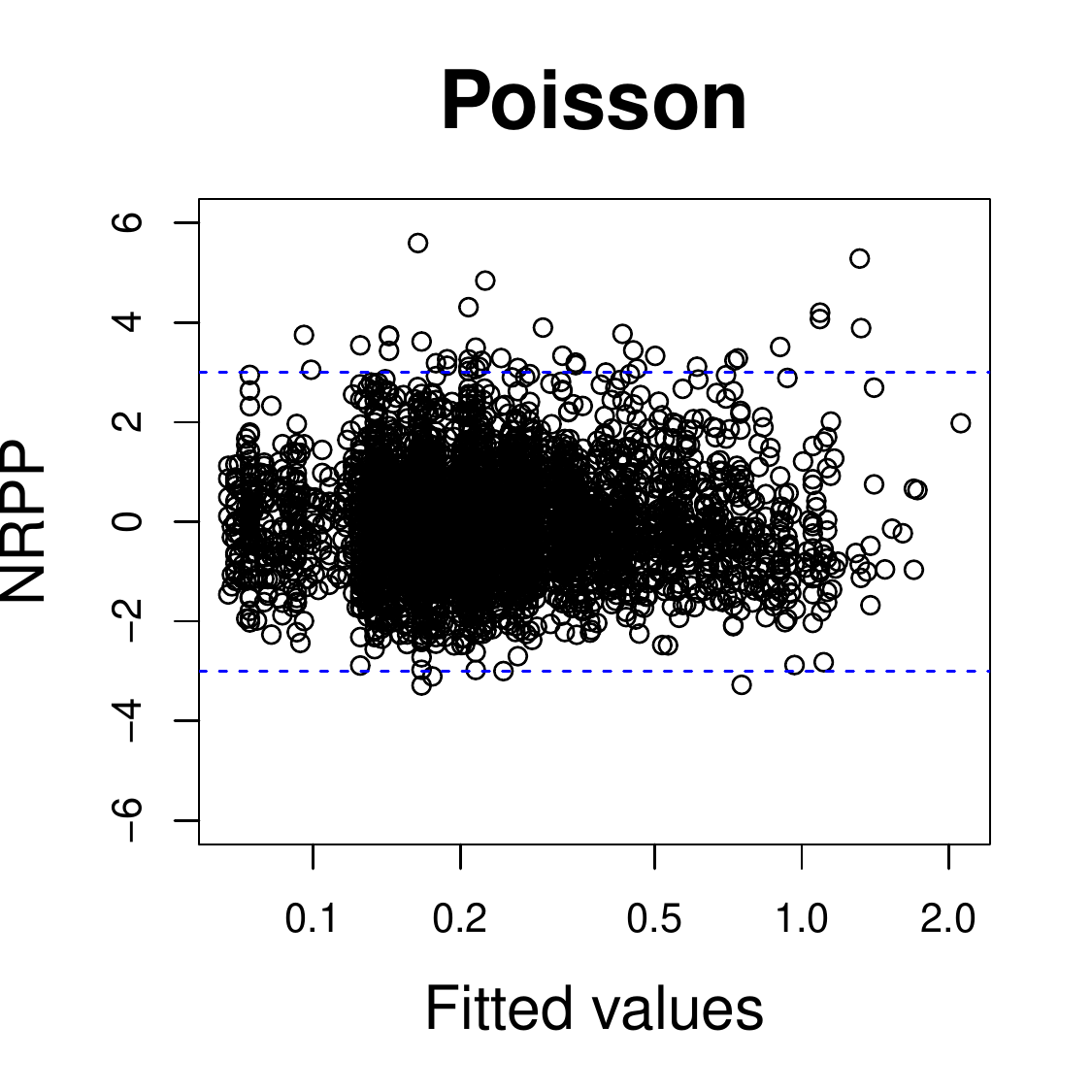}
\end{subfigure}\hspace{0.5em} 
\begin{subfigure}[t]{0.3\textwidth}
\centering 
\subcaption{\centering \textbf{~~~~~Poisson}}
\includegraphics[width=1.8in, height=1.5in,trim=0in 0.1in 0.1in 0.6in,clip]{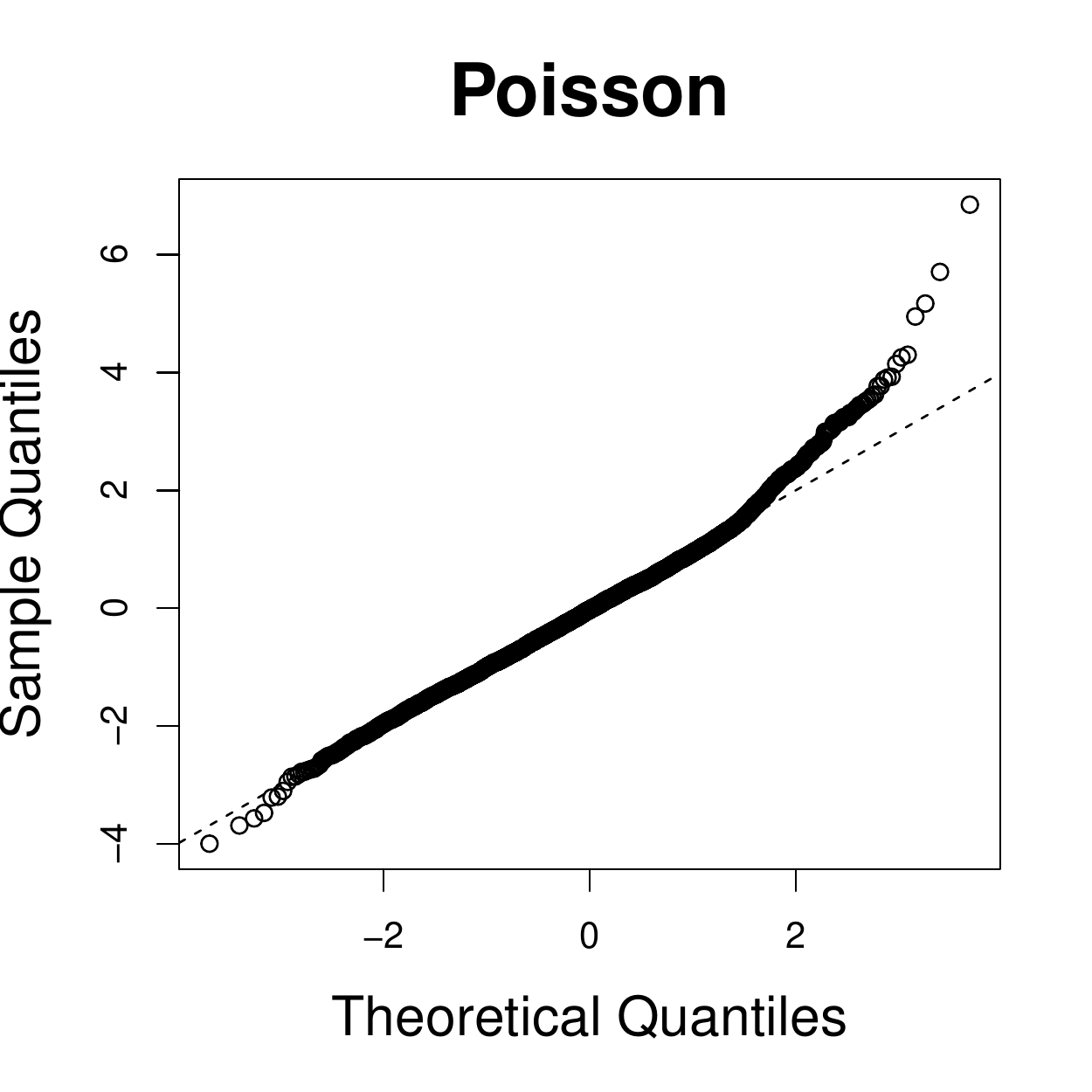}
\end{subfigure}\hspace{0.5em} 
\begin{subfigure}[t]{0.3\textwidth}
\centering 
\subcaption{\centering \textbf{Poisson}}
\includegraphics[width=1.8in, height=1.5in,trim=0in 0.1in 0.1in 0.6in,clip]{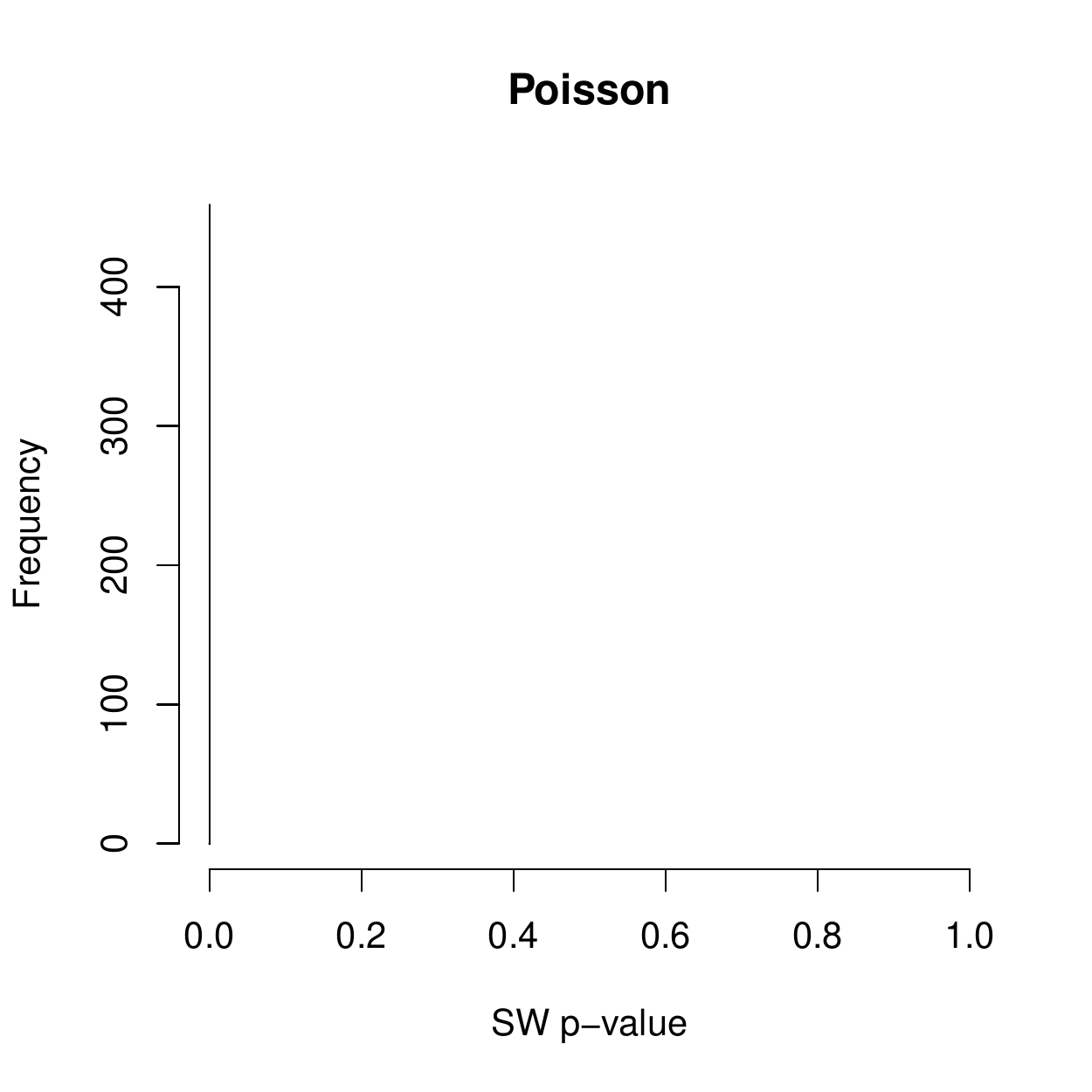}
\end{subfigure}\hspace{0.5em} 

\vspace{10pt}

\begin{subfigure}[t]{0.3\textwidth}
\centering 
\subcaption{\centering \textbf{~~~ZIP}}
\includegraphics[width=1.8in, height=1.5in,trim=0in 0.2in 0in 0.6in,clip]{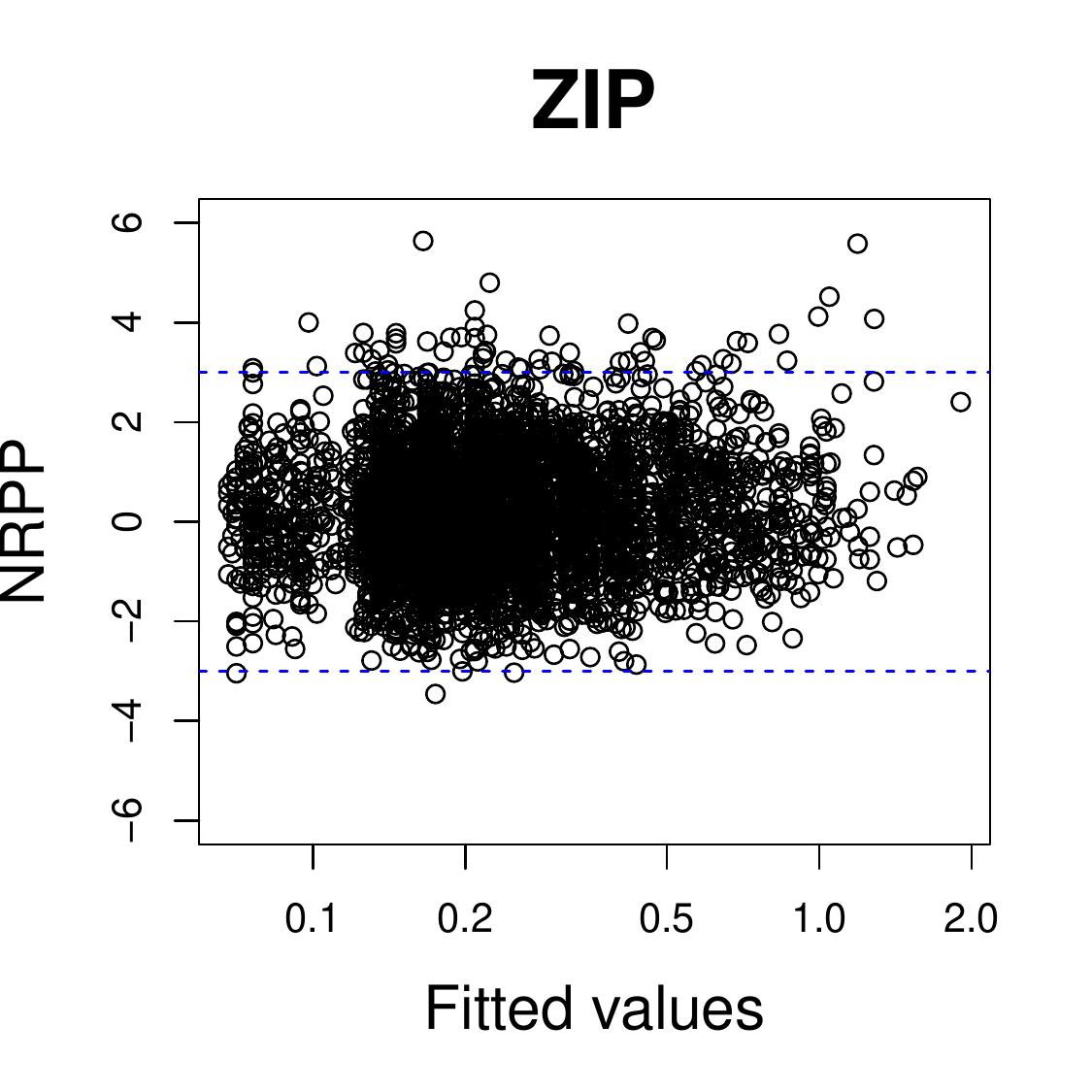}
\end{subfigure}\hspace{0.5em}
\begin{subfigure}[t]{0.3\textwidth} 
\centering
\subcaption{\centering \textbf{~~~ZIP}}
\includegraphics[width=1.8in, height=1.5in,trim=0in 0.1in 0.1in 0.6in,clip]{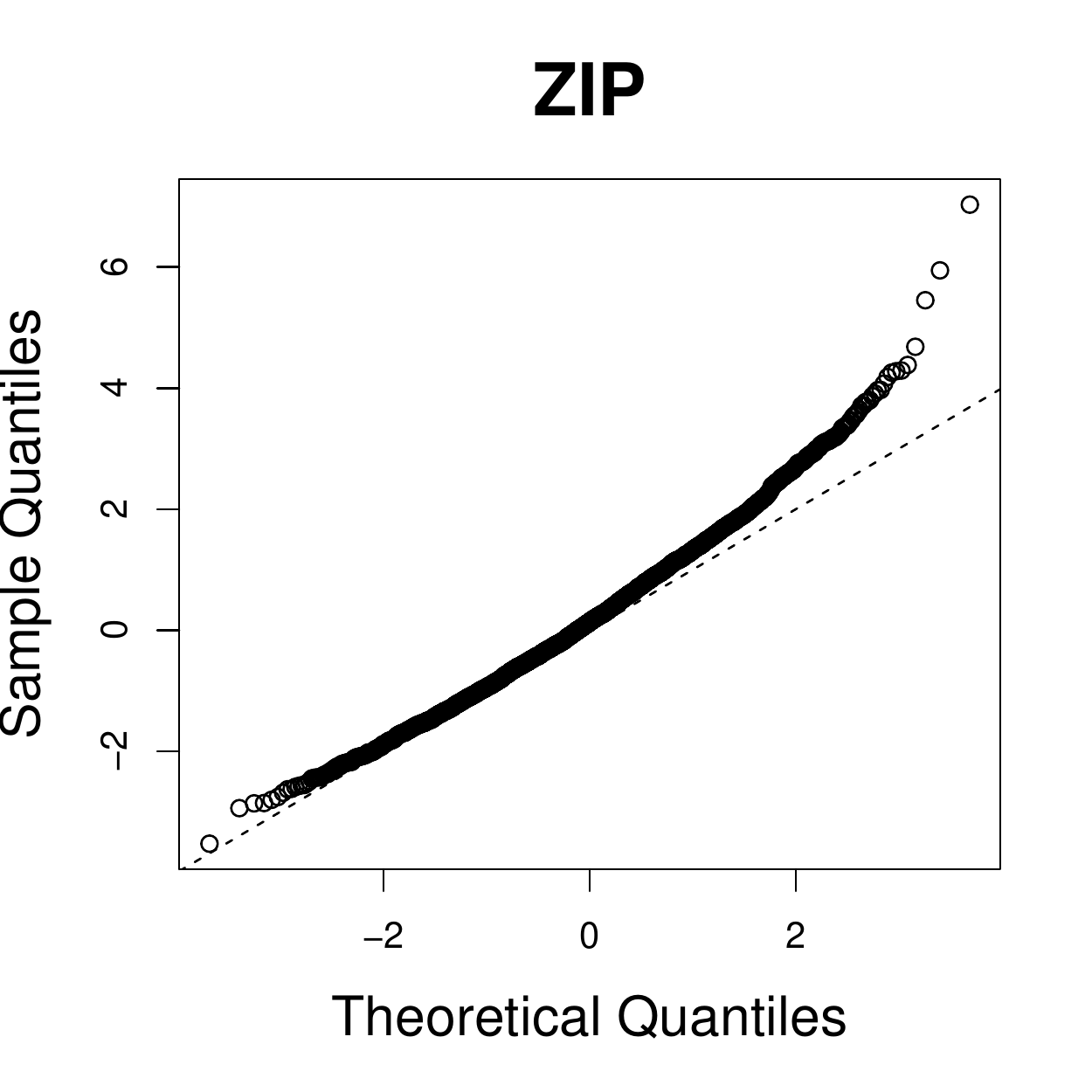}
\end{subfigure}\hspace{0.5em} 
\begin{subfigure}[t]{0.3\textwidth}
\centering 
\subcaption{\centering \textbf{~~~ZIP}}
\includegraphics[width=1.8in, height=1.5in,trim=0in 0.1in 0.1in 0.6in,clip]{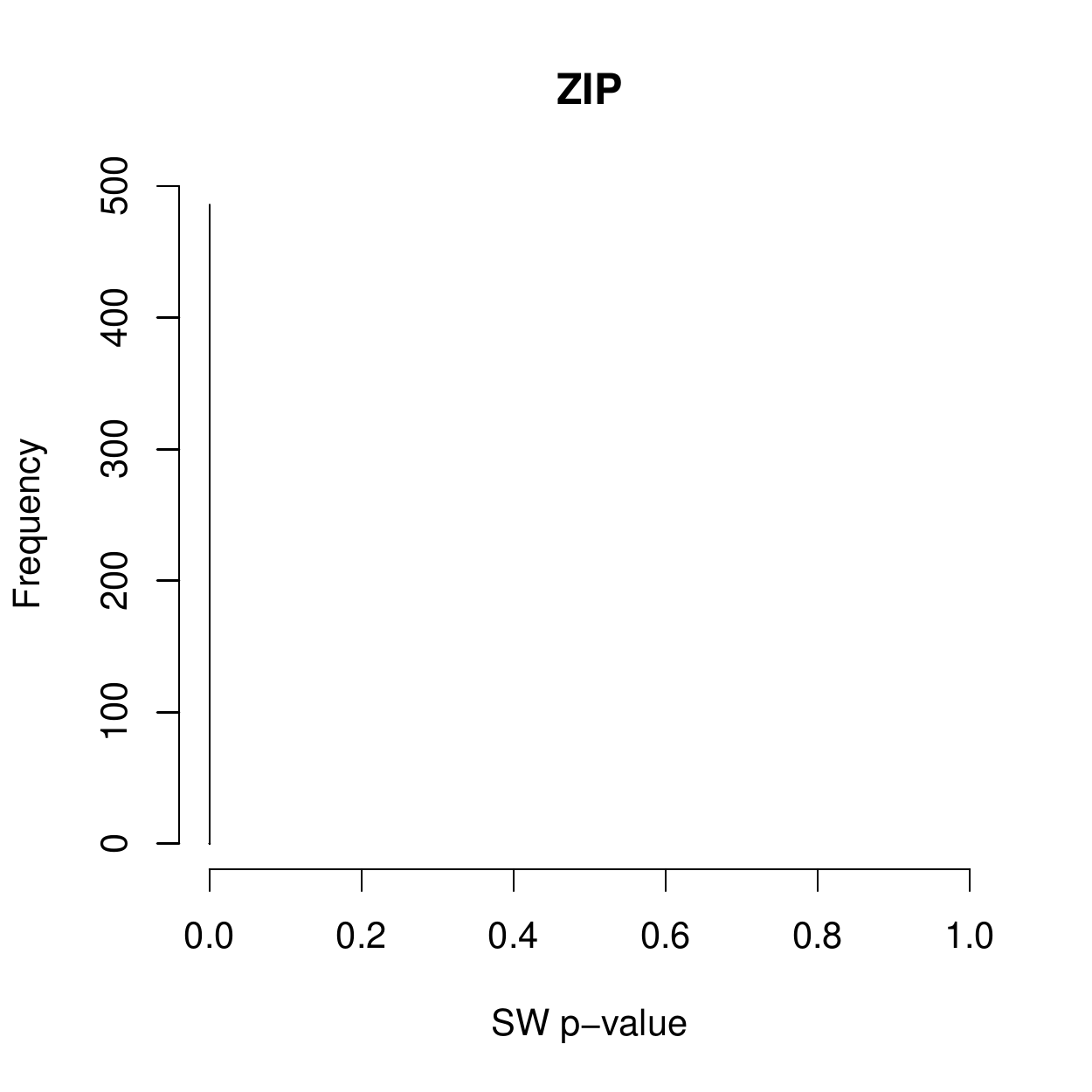}
\end{subfigure}\hspace{0.5em}

\vspace{10pt}

\begin{subfigure}[t]{0.3\textwidth}
\centering 
\subcaption{\centering \textbf{~~~NB}}
\includegraphics[width=1.8in, height=1.5in,trim=0in 0.2in 0in 0.6in,clip]{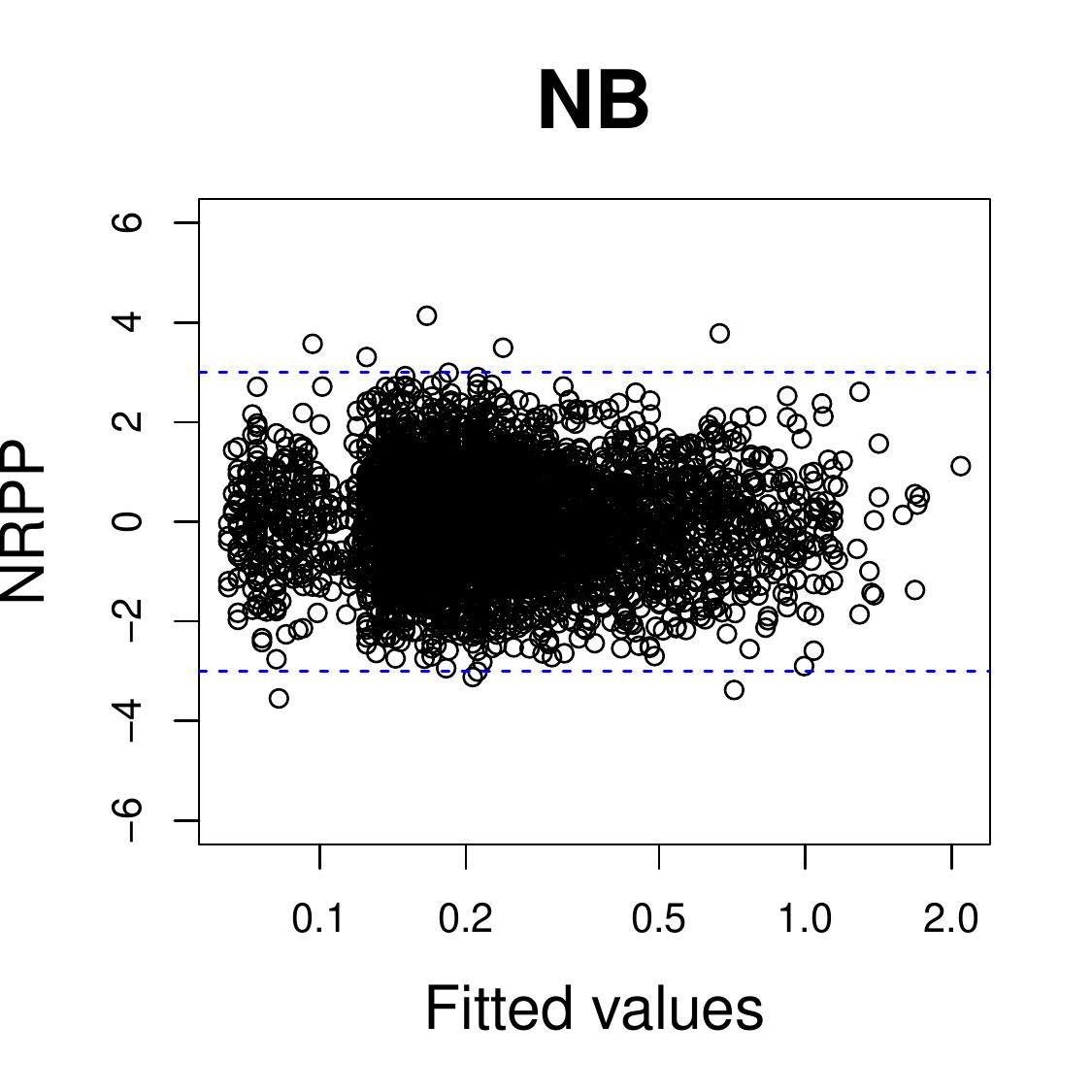}
\end{subfigure}\hspace{0.5em}
\begin{subfigure}[t]{0.3\textwidth}
\centering
\subcaption{\centering \textbf{~~~NB}}
\includegraphics[width=1.8in, height=1.5in,trim=0in 0.1in 0.1in 0.6in,clip]{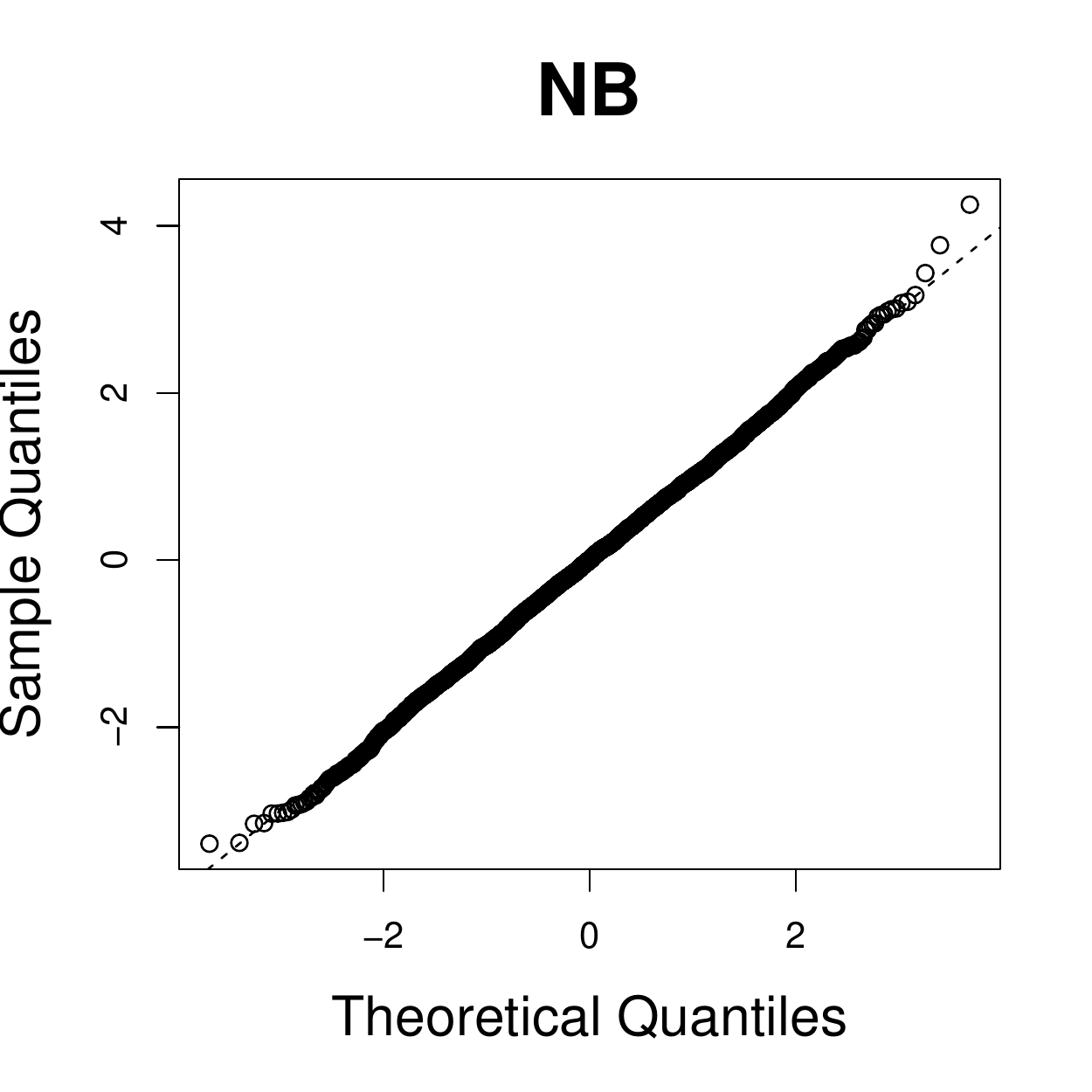}
\end{subfigure}\hspace{0.5em} 
\begin{subfigure}[t]{0.3\textwidth}
\centering 
\subcaption{\centering \textbf{~~~NB}}
\includegraphics[width=1.8in, height=1.5in,trim=0in 0.1in 0.1in 0.6in,clip]{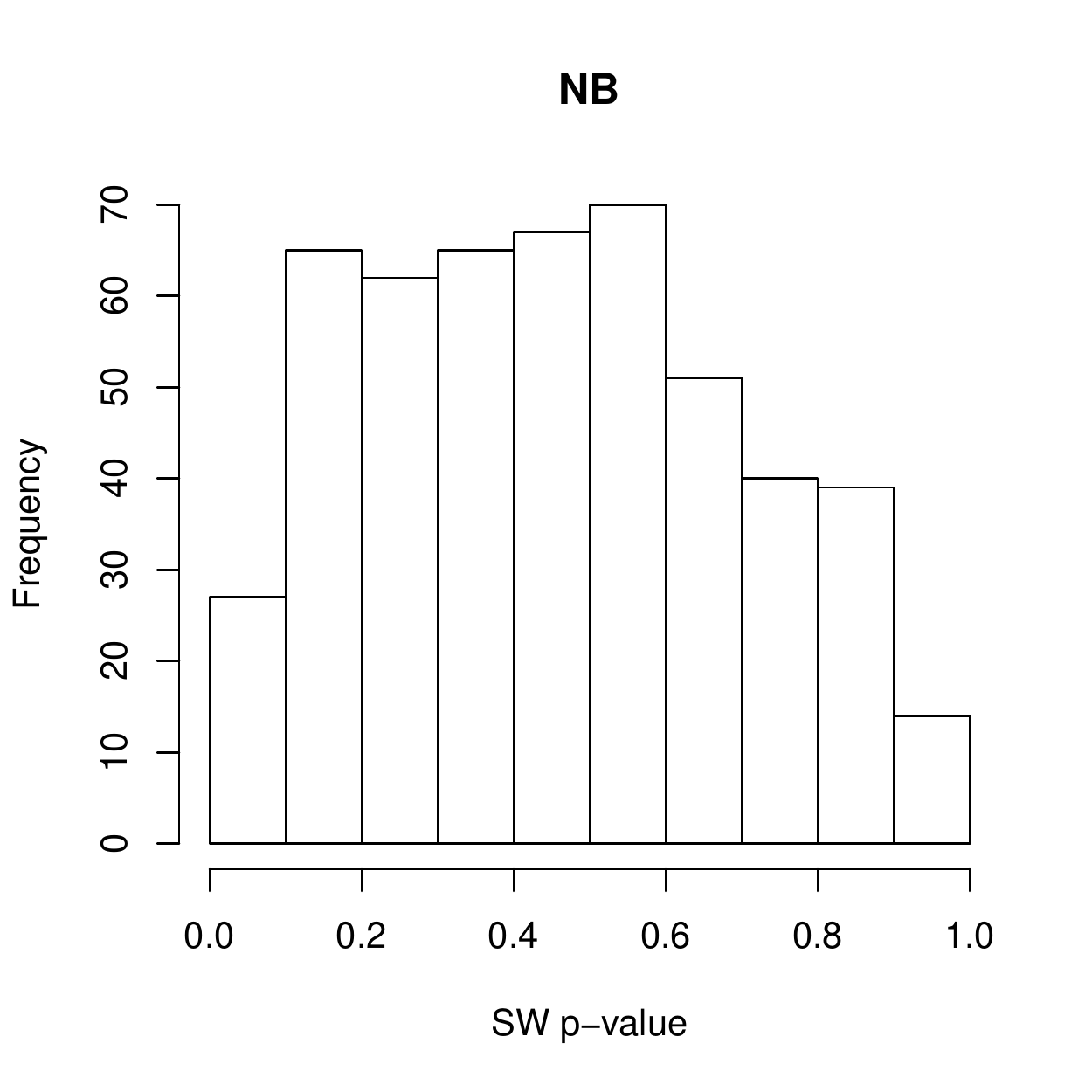}
\end{subfigure}\hspace{0.5em} 

\vspace{10pt}

\begin{subfigure}[t]{0.3\textwidth}
\centering 
\subcaption{\centering \textbf{~~~ZINB}}
\includegraphics[width=1.8in, height=1.5in,trim=0in 0.2in 0in 0.6in,clip]{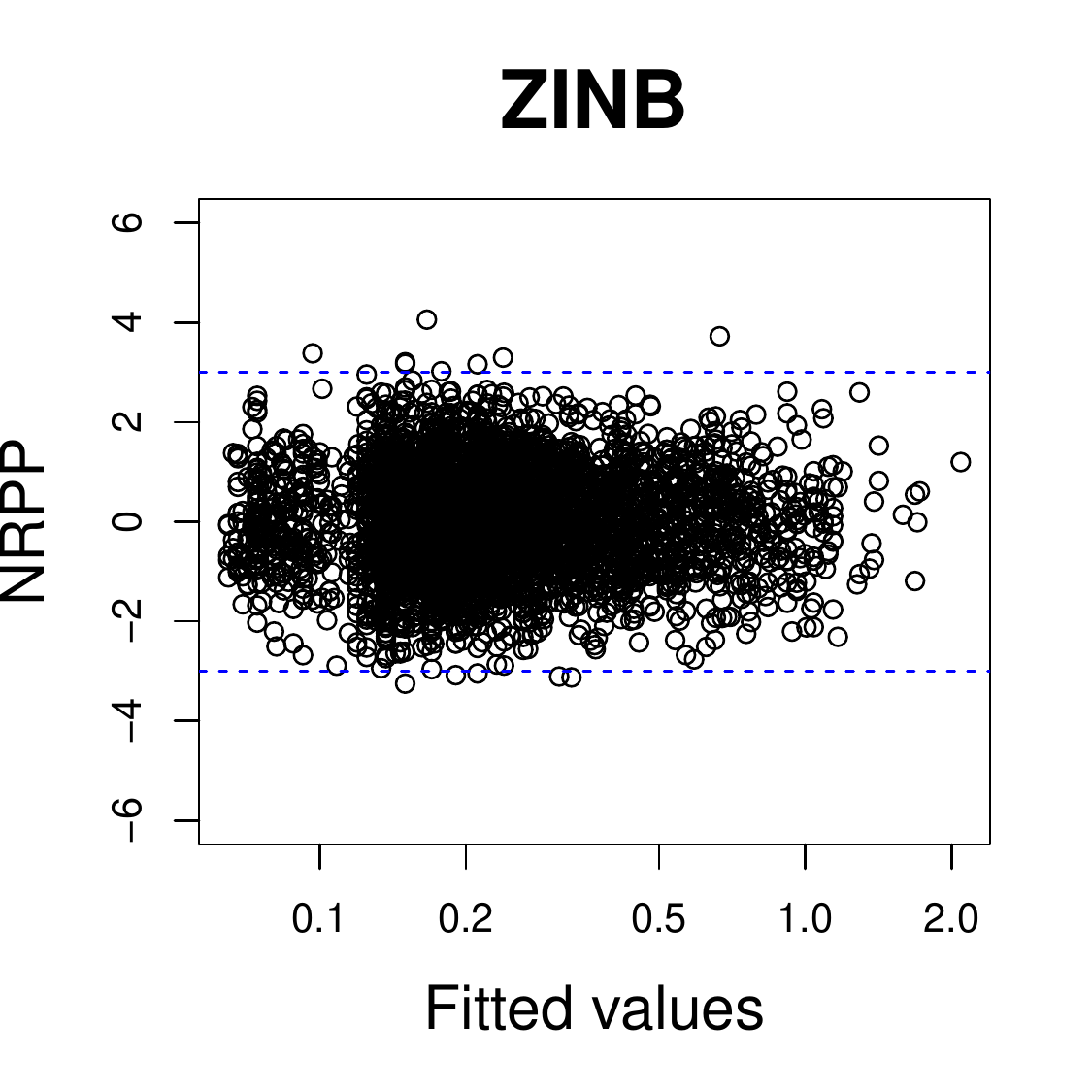}
\end{subfigure}
\begin{subfigure}[t]{0.3\textwidth}
\centering
\subcaption{\centering \textbf{~~~ZINB}}
\includegraphics[width=1.8in, height=1.5in,trim=0in 0.1in 0.1in 0.6in,clip]{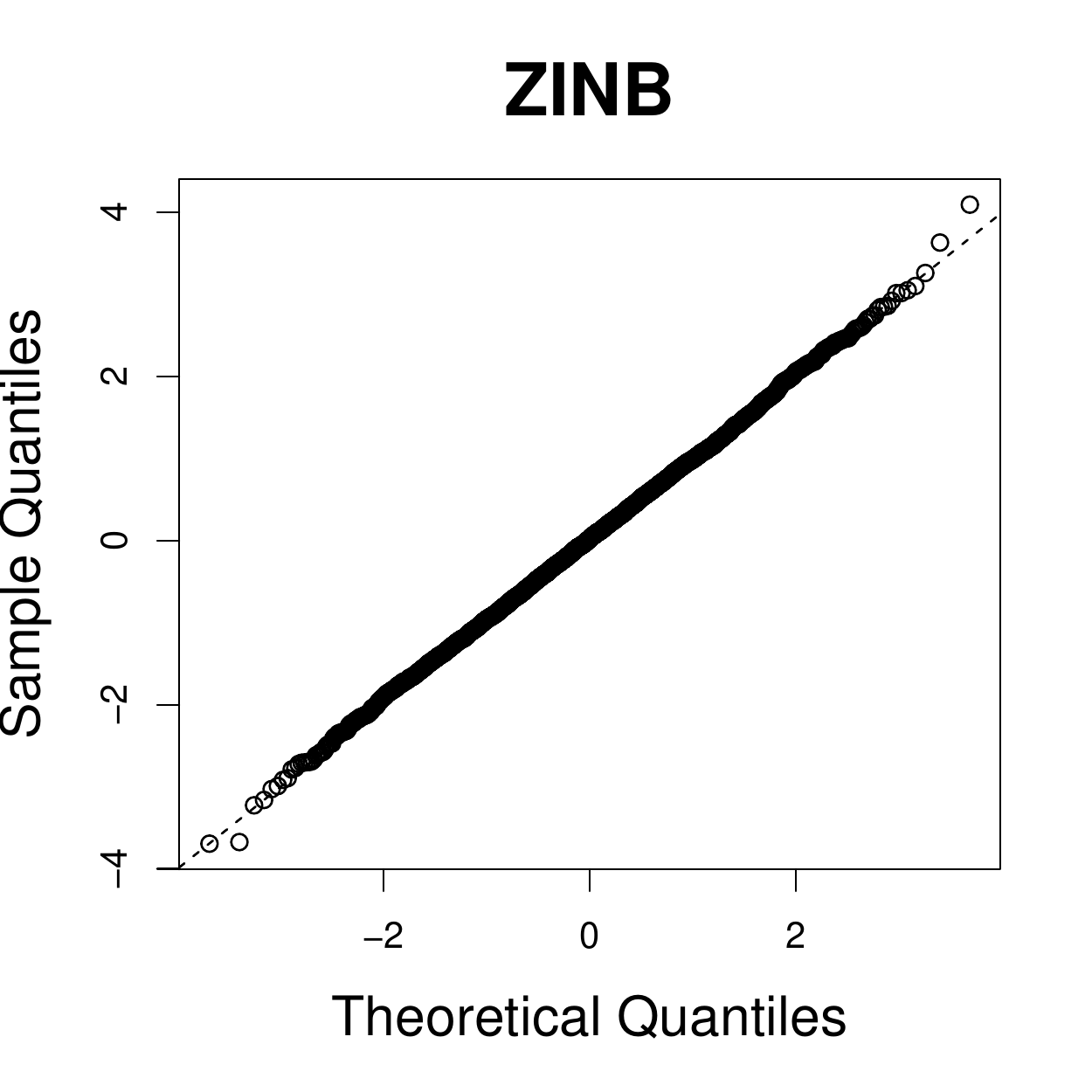}
\end{subfigure}
\hspace{0.5em} 
\begin{subfigure}[t]{0.3\textwidth}
\centering 
\subcaption{\centering \textbf{~~~ZINB}}
\includegraphics[width=1.8in, height=1.5in,trim=0in 0.1in 0.1in 0.6in,clip]{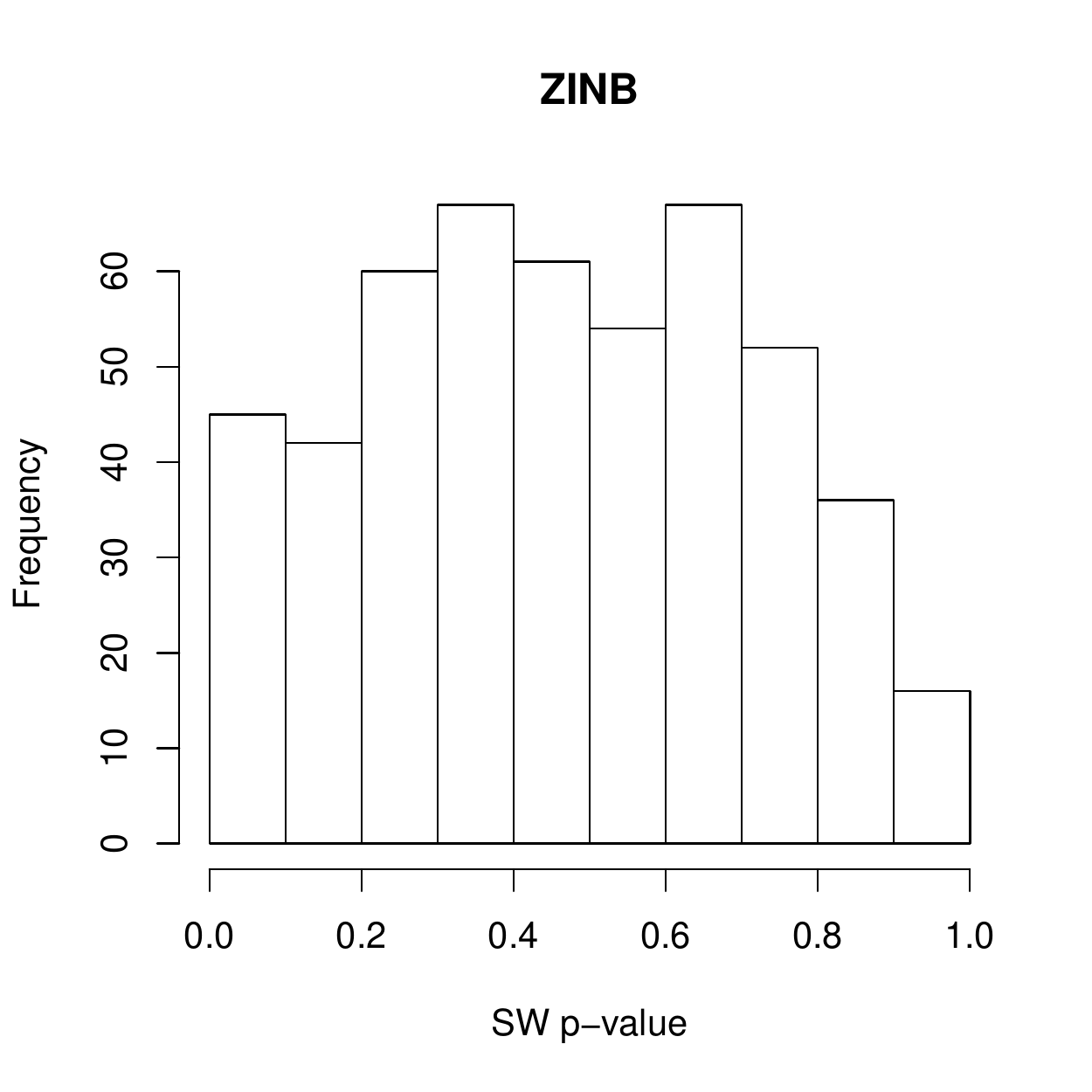}
\end{subfigure}\hspace{0.5em} 
\caption{\footnotesize NRPPs for the Poisson, NB, ZIP and ZINB models fitted to the National Medical Expenditure Survey. The panels in the first two columns present the scatter plots and QQ plots of the NRPPs versus the fitted values, respectively.  The third column presents the histograms of the p-values of the SW normality test for 1000 replicated NRPPs.} 
\label{fig:ER}
\end{figure}

To compare the competing models, the Akaike Information Criterion (AIC) is used in which smaller values indicate models with better out-of-sample prediction. The AIC scores for the Poisson, NB, ZIP, and ZINB are 5648, 5352, 5418 and 5354, respectively; this suggests that the NB and ZINB models provide an almost equivalent fit to the data and are superior to their counterpart Poisson models. Although AIC can be used to compare the GOF of competing models, it cannot check the goodness-of-fit of the models for assessing the need for additional complexity and validating the distribution assumption of the response variable. Model diagnosis with residuals is therefore imperative to address these concerns. 

The panels in the first column of Figure \ref{fig:ER} present the scatter plots of the NRPPs versus the fitted values for each model. It is evident that the NB and ZINB models fit the dataset fairly well with NRPPs ranging mostly between -3 and 3, and no specific pattern present.  In contrast, for the Poisson and ZIP models, there are many NRPPs greater than 3 and some of them are as large as 6. The QQ plots of the NRPPs, as presented in the panels of the second column of Figure \ref{fig:ER}, deviate evidently away from the straight line in the upward end. The residual plots and QQ plots both indicate that the tails of Poisson and ZIP are too light for the dataset and distributions with heavier tails to accommodate over-dispersed large counts are necessary for the dataset.  The residual and QQ plots of NB and ZINB models show that these two models with heavier right tails  appear adequate for the dataset. 

In addition, we  quantitatively check the GOF of the four models for this dataset through applying SW normality test to their NRPPs. One concern in using the NRPP method is the fluctuation in the residuals introduced by the randomization for producing continuously distributed residuals. To address this randomization, we replicated 1000 realizations of the NRPPs based on the same dataset. The panels in the third column of Figure~\ref{fig:ER} display the histograms of 1000 replicated p-values of the SW tests.  The randomization introduced little variation for the SW p-values of the NRPPs for the fitted Poisson and ZIP models as almost all their p-values were close to 0, confirming the inadequacies of both Poisson models. Conversely, the SW p-values of the NRPPs for the fitted NB and ZINB models varied between 0 and 1 with about $96\%$ of the p-values being above 0.05, confirming the adequacies of both NB and ZINB models. Hence, randomization does not compromise the statistical power of the NRPPs in this application. Nevertheless, when a model fits a dataset well, there are some fluctuations for the SW p-values with a roughly $5\%$ chance that the p-value will be below 0.05. This leads to our recommendation that a large number of replicated realizations of the NRPPs should be produced to ensure that the observed discrepancies are not made by the randomization in producing the NRPPs. Although this offers a solution to alleviate the impact of the randomness in the NRPPs, it may be still desired to have a ``non-random'' overall GOF test p-value for NRPPs. It is interesting to investigate the distribution of the mean or other summaries of replicated SW test p-values under the true model; this distribution may not be uniform.

%\begin{figure}[h]
%\centerline{\includegraphics[width=4in, height=3in]{emervisits/emervisits_ShaprioWilkPV.pdf}} 
%\caption{\footnotesize Frequency of the p-values of the Shapiro-Wilk normality test for the 1000 replicated RQRs in the real application. } \label{fig:emervisitsPV}
%\end{figure}

%%%%%%%%%%%%%%%%%%%%%%%%%%%%%%%%%%%%%%%%%%%%%%%%%%%%
\section{Conclusion and Future Work}\label{sec:conclusion}
%%%%%%%%%%%%%%%%%%%%%%%%%%%%%%%%%%%%%%%%%%%%%%%%%%%%
Diagnosing regression models is extremely important in statistical inference. Pearson and deviance residuals and their corresponding $\chi^2$ tests are commonly used in practice. However, the use of these tools in non-normal regression often lacks justification because both Pearson and deviance residuals are typically not normally distributed, and their $\chi^{2}$ tests are not well-calibrated under the true model. We have demonstrated that the NRPPs are normally distributed under the true models and the overall GOF test by checking the normality of NRPPs with Shapiro-Wilk (SW) tests is well-calibrated. We have also demonstrated the versatility and power of NRPPs  in detecting a variety of model complexities, including non-linearity, zero-inflation, and over-dispersion. % As we expected, the statistical power of the NRPPs for detecting model misspecification tends to be low for small sample sizes with minor deviations from the fitted models to the true models. Therefore, the NRPPs have substantive appeal in diagnosing non-normal regression models with moderate or large sample sizes and deviations between the fitted and true models. 

%The NRPP method uses a uniform random number $u_i$ to convert a set of discrete non-randomized residuals into continuous values in order to be examined with a unified approach. Although the randomness in $u_i$ may produce special patterns in the NRPPs, we note that the chance that the pure random numbers will make any observable pattern decreases as the sample size $n$ increases. Additionally, the influence from the randomness in $u_i$ will decrease as the number of possible values of $y_i$ increases due to the shrinkage in the continuity gaps in the PMF. Nevertheless, as suggested by \cite{dunn1996randomized}, multiple realizations of the NRPPs are needed to ensure that any pattern shown in the NRPPs is not caused by the randomness in $u_i$. 

We have also conducted simulations to investigate the behaviours of the NRPPs with datasets generated from a logistic regression model with quadratic covariate effect. Results are not included but available upon request. The residual plots of the Pearson and deviance residuals against the covariate cluster on two separated curves according to the two possible response values 0 and 1. In contrast, the NRPPs under the true model are randomly scattered without a special pattern, whereas the residual plots of NRPPs  under a wrong model assuming to have a linear covariate effect exhibit a clear quadratic pattern. The residual plot of the NRPPs is, therefore, an excellent graphical tool for detecting the non-linearity in the logistic regression model.  Nevertheless, the power of the overall GOF test by merely testing the normality of NRPPs of the logistic regression is not very high.  Development of more specialized tests based on NRPPs or RPPs for detecting various model mis-specifications for logistic regression retains an interesting topic. 

In many clinical and public health research, correlated data (i.e., longitudinal, spatial or multilevel data) involving both structural and stochastic features are often collected due to measuring unobserved heterogeneity between clusters after conditioning on the covariates. Bayesian hierarchical model with random effects (latent variables) are typically adopted to model the complex dependence structure in these types of data. The randomization as used in RPPs is also necessary to produce truly uniformly distributed predictive p-values for diagnosing complex Bayesian hierarchical models.  The randomization technique will result in modifying the CV mid-point predictive p-values defined in \cite{Marshall03, marshall2007identifying,li_estimating_2017} to CV randomized predictive p-values (CVRPP).  Computing actual CVRPPs is time-consuming because  Markov chain samplings are required to repeat for each CV posterior distribution in which an observation is excluded as a test case. There have been many approximating methods \citep{Marshall03, marshall2007identifying,li_estimating_2017, vehtari_practical_2017-3} proposed to compute CV quantities with the Markov chain samples from the posterior distribution based on the full dataset, avoiding the actual LOOCV being implemented. It is interesting to investigate the performance of CVRPPs computed with these approximating methods in complex Bayesian hierarchical models.

%Programs to extend normal and non-normal regression models to clustered or longitudinal data are widely available (e.g., \texttt{lme4} and \texttt{mgcv} packages in the R software, and \texttt{glimmix} and \texttt{nlmixed} procedures in the SAS software); however, model diagnosis for the non-normal mixed effects models are still underdeveloped. %Further development of extending the NRPP method to examine the GOF of the mixed effects models is underway for contemporary statistical application areas. 

%\section{References}
%\small
%\singlespacing
\bibliographystyle{chicago}
\bibliography{rpp,rqr}

\appendix
\section{Proof of the Uniformity of RPPs (Theorem \ref{thm:1})}\label{sec:proof}
\noindent \textbf{Proof:} First, we note that the normality of  $\phi^{-1} (F^*(Y_i,U_i)) $ can be derived from the uniformity of RPPs based on Theorem \ref{thm:0}, and hence, it suffices to prove the uniformity of RPPs. 
Suppose all the possible values (with positive mass) for $Y_i$ given $X_i = x_i$ are $k^{(1)}, k^{(2)}, \ldots$. Let $P(Y_i = k^{(j)})=p^{(j)}$, and $F^{(j)} = (F_{i}(k^{(j)}-), F_{i}(k\sj))$; note that $\lambda(F\sj)=p\sj$ where $\lambda(B)$ denotes the ordinary length (ie, Lebesgue measure) of $B$. We note that $\cup_{j=1}^\infty F\sj = (0,1) \backslash \{F_{i}(k^{(1)},F_{i}(k^{(2)},\ldots\} $ and the collection of sets $\{F\sj| j = 1,2,\ldots\}$ are mutually exclusive. Conditional on $Y_i = k\sj$, $F^*(Y_i, U_i)$ is uniformly distributed on $F\sj$ because $U_i$ is uniformly distributed on $(0,1)$. Therefore, for any interval (or Borel set) $B \subseteq (0,1)$, 
$P(\Fstar \in B | Y_i = k\sj)=\frac{\lambda(F\sj \cap B)}{p\sj}.$ By the law of total probability, we have
\begin{eqnarray*}
P(\Fstar \in B) &=& \sum_{j=1}^\infty P(\Fstar \in B | Y_i = k\sj) \times P(Y_i = k\sj)  \\
&=& \sum_{j=1}^\infty \frac{\lambda(F\sj \cap B)}{p\sj} \times p\sj 
= \sum_{j=1}^\infty\lambda(F\sj \cap B) \\
&=& \lambda(\cup_{j=1}^\infty F\sj \cap B) = \lambda (B)
\end{eqnarray*}
\noindent \textbf{End of proof}.

\end{document}